\documentclass[amsmath,aps,prl,twocolumn,superscriptaddress]{revtex4-1}

\usepackage[top=1.5cm, bottom=1.5cm, left=1.5cm, right=1.5cm, bindingoffset = 0pt, footskip = 0.75cm, marginparwidth = 0pt, marginparsep=0pt]{geometry}

\usepackage{graphicx}
\usepackage{float}
\usepackage{caption}
\usepackage{subfiles}

\usepackage[export]{adjustbox}

\graphicspath{{Figures/}}
\usepackage{amsfonts}

\usepackage{subfigure}
\usepackage[utf8]{inputenc}
\usepackage{amsmath}
\usepackage{amssymb}
\usepackage{appendix}
\usepackage{color}
\usepackage{hyperref}
\usepackage{units}
\usepackage[capitalise]{cleveref}

\usepackage{comment}

\usepackage{makecell}
\usepackage{blindtext} % Package to generate dummy text throughout this template 
\usepackage{setspace}

\usepackage[T1]{fontenc} % Use 8-bit encoding that has 256 glyphs
\usepackage{microtype} % Slightly tweak font spacing for aesthetics
\usepackage[english]{babel} % Language hyphenation and typographical rules

\usepackage{float}
\usepackage[export]{adjustbox}
\usepackage[section]{placeins}
\usepackage{framed}
\usepackage{booktabs}
\usepackage{varwidth}

\crefname{equation}{Eq.}{Eqs.}
\Crefname{equation}{Equation}{Equations}
\crefname{figure}{Fig.}{Figs.}
\Crefname{figure}{Figure}{Figures}
\crefname{section}{Sec.}{Secs.}
\Crefname{section}{Section}{Sections}

\usepackage{array}
\newcolumntype{L}{>{$}l<{$}}
\newcolumntype{C}{>{$}c<{$}}
\newcommand{\braket}[1]{\left\langle #1 \right\rangle}

\newcommand{\e}{e}

\newcommand{\dd}{d}

\newcommand{\half}{\frac{1}{2}}

\newcommand{\hphi}{\hat{\phi}}

\newcommand{\ha}{\hat{a}}

\newcommand{\hH}{\hat{H}}

\newcommand{\fc}{\text{fc}}
\newcommand{\pa}{\text{PA}}

\newcommand{\dg}{^\dagger}
\def\*#1{\mathbf{#1}}

\newcommand{\sm}{Supplementary Materials}

\usepackage[T1]{fontenc}
\usepackage{array,tabularx}
\usepackage[most]{tcolorbox}
\usepackage[export]{adjustbox}
\usepackage{float}
\usepackage[export]{adjustbox}
\usepackage[section]{placeins}
\usepackage{framed}
\usepackage{booktabs}
\usepackage{varwidth}
\usepackage{array}
\usepackage{wrapfig}
\usepackage[Symbol]{upgreek}

\newcolumntype{Y}{>{\raggedleft\arraybackslash}X}

\tcbset{tab1/.style={fonttitle=\bfseries\large,fontupper=\normalsize\sffamily,
colback=yellow!10!white,colframe=gray!75!black,colbacktitle=blue!40!white,
coltitle=black,center title,freelance,frame code={
\foreach \n in {north east,north west,south east,south west}
{\path [fill=gray!75!black] (interior.\n) circle (3mm); };},}}

\tcbset{tab2/.style={enhanced,fonttitle=\bfseries,fontupper=\normalsize\sffamily,
colback=yellow!10!white,colframe=gray!50!black,colbacktitle=blue!40!white,
coltitle=black,center title}}
\usepackage[Symbol]{upgreek}

\begin{document}
\title{Broadband Squeezed Microwaves and Amplification with \\a Josephson Traveling-Wave Parametric Amplifier}

\author{Jack Y. Qiu}
\affiliation{Research Laboratory of Electronics, Massachusetts Institute of Technology, Cambridge, MA 02139, USA}
\affiliation{Department of Electrical Engineering and Computer Science, Massachusetts Institute of Technology, Cambridge, MA 02139, USA}

\author{Arne Grimsmo}
\affiliation{ARC Centre of Excellence for Engineered Quantum Systems, School of Physics, The University of Sydney, Sydney, NSW 2006, Australia.}
\altaffiliation[present address: ]{AWS Quantum}
\author{Kaidong Peng}
\affiliation{Research Laboratory of Electronics, Massachusetts Institute of Technology, Cambridge, MA 02139, USA}
\affiliation{Department of Electrical Engineering and Computer Science, Massachusetts Institute of Technology, Cambridge, MA 02139, USA}
\author{Bharath Kannan}
\affiliation{Research Laboratory of Electronics, Massachusetts Institute of Technology, Cambridge, MA 02139, USA}
\affiliation{Department of Electrical Engineering and Computer Science, Massachusetts Institute of Technology, Cambridge, MA 02139, USA}
\author{Benjamin Lienhard}
\affiliation{Research Laboratory of Electronics, Massachusetts Institute of Technology, Cambridge, MA 02139, USA}
\affiliation{Department of Electrical Engineering and Computer Science, Massachusetts Institute of Technology, Cambridge, MA 02139, USA}
\author{Youngkyu Sung}
\affiliation{Research Laboratory of Electronics, Massachusetts Institute of Technology, Cambridge, MA 02139, USA}
\affiliation{Department of Electrical Engineering and Computer Science, Massachusetts Institute of Technology, Cambridge, MA 02139, USA}
\author{Philip Krantz}
\affiliation{Research Laboratory of Electronics, Massachusetts Institute of Technology, Cambridge, MA 02139, USA}
\author{Vladimir Bolkhovsky}
\affiliation{MIT Lincoln Laboratory, 244 Wood Street, Lexington, MA 02420, USA}
\author{Greg Calusine}
\affiliation{MIT Lincoln Laboratory, 244 Wood Street, Lexington, MA 02420, USA}
\author{David Kim}
\affiliation{MIT Lincoln Laboratory, 244 Wood Street, Lexington, MA 02420, USA}
\author{Alex Melville}
\affiliation{MIT Lincoln Laboratory, 244 Wood Street, Lexington, MA 02420, USA}
\author{Bethany M. Niedzielski}
\affiliation{MIT Lincoln Laboratory, 244 Wood Street, Lexington, MA 02420, USA}
\author{Jonilyn Yoder}
\affiliation{MIT Lincoln Laboratory, 244 Wood Street, Lexington, MA 02420, USA}
\author{Mollie E. Schwartz}
\affiliation{MIT Lincoln Laboratory, 244 Wood Street, Lexington, MA 02420, USA}
\author{Terry P. Orlando}
\affiliation{Research Laboratory of Electronics, Massachusetts Institute of Technology, Cambridge, MA 02139, USA}
\affiliation{Department of Electrical Engineering and Computer Science, Massachusetts Institute of Technology, Cambridge, MA 02139, USA}
\author{Irfan Siddiqi}
\affiliation{Quantum Nanoelectronics Laboratory, Berkeley, CA 94720, USA}
\author{Simon Gustavsson}
\affiliation{Research Laboratory of Electronics, Massachusetts Institute of Technology, Cambridge, MA 02139, USA}
\author{Kevin P. O'Brien}
\affiliation{Research Laboratory of Electronics, Massachusetts Institute of Technology, Cambridge, MA 02139, USA}
\affiliation{Department of Electrical Engineering and Computer Science, Massachusetts Institute of Technology, Cambridge, MA 02139, USA}
\author{William D. Oliver}
\affiliation{Research Laboratory of Electronics, Massachusetts Institute of Technology, Cambridge, MA 02139, USA}
\affiliation{Department of Electrical Engineering and Computer Science, Massachusetts Institute of Technology, Cambridge, MA 02139, USA}
\affiliation{MIT Lincoln Laboratory, 244 Wood Street, Lexington, MA 02420, USA}
\affiliation{Department of Physics, Massachusetts Institute of Technology, Cambridge, MA 02139, USA}
\maketitle
%

%\onecolumngrid 
{\bfseries 
Squeezing of the electromagnetic vacuum is an essential metrological technique used to reduce quantum noise in applications spanning gravitational wave detection, biological microscopy, and quantum information science. In superconducting circuits, the resonator-based Josephson-junction parametric amplifiers conventionally used to generate squeezed microwaves are constrained by a narrow bandwidth and low dynamic range. In this work, we develop a dual-pump, broadband Josephson traveling-wave parametric amplifier that combines a phase-sensitive extinction ratio of 56 dB with single-mode squeezing on par with the best resonator-based squeezers. We also demonstrate two-mode squeezing at microwave frequencies with bandwidth in the gigahertz range that is almost two orders of magnitude wider than that of contemporary resonator-based squeezers. Our amplifier is capable of simultaneously creating entangled microwave photon pairs with large frequency separation, with potential applications including high-fidelity qubit readout, quantum illumination and teleportation.
}

Heisenberg's uncertainty principle establishes the attainable measurement precision, the ``standard quantum limit (SQL),'' for isotropically-distributed vacuum fluctuations in the quadratures of the electromagnetic (EM) field~\cite{Wallraff2004,Caves,Bienfait2016}. Squeezing the EM field at a single frequency --- single-mode squeezing --- decreases the fluctuations of one quadrature below that of the vacuum at the expense of larger fluctuations in the other quadrature, thereby enabling a phase-sensitive means to beat the SQL. Squeezing can also generate quantum entanglement between observables at two distinct frequencies, producing two-mode squeezed states. Since its first experimental demonstration in 1985~\cite{Slusher1985}, squeezing has become a resource for applications in quantum optics~\cite{Toyli2016}, quantum information~\cite{Aoki2009}, and precision measurement~\cite{LIGO2011}.

\begin{figure*}[t]
\centering
\includegraphics[width=0.96\textwidth]{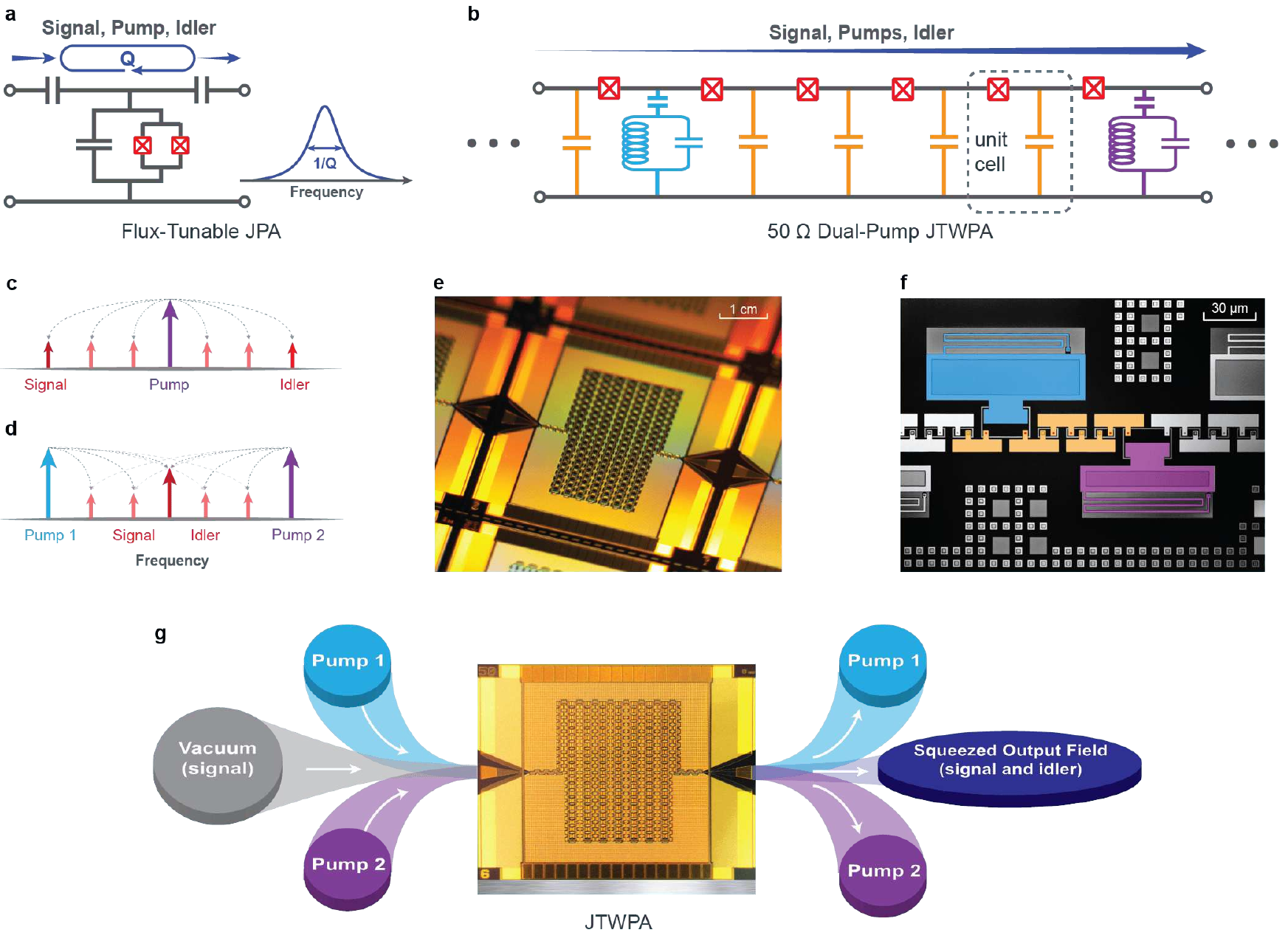}
\captionsetup{labelformat=empty, belowskip=0pt}
\caption{\textbf{Figure 1 $|$ Josephson traveling-wave parametric amplifier dispersion engineered for a bichromatic pump.} \textbf{a.} Circuit schematic of a conventional JPA. The Q-enhancement of the field produces a narrowband frequency response. \textbf{b.} A repeating section of the dual-pump JTWPA. We can identify the L-C ladder that forms a \unit[50]{$\Omega$} transmission line from lumped elements and the two phase-matching resonators for dispersion engineering. \textbf{c.} Degenerate four-wave mixing. \textbf{d.} Non-degenerate four-wave mixing. The picture shows the special case when the signal and idler are at the same frequency $\omega_{\rm c}$ at the center between the two pumps. Pairs of two-mode squeezed photons (signal and idler) are created at frequencies symmetric about the center frequency $\omega_{\rm c}$. When the two photons are frequency-degenerate at $\omega_{\rm c}$, this is referred to as single-mode squeezing. \textbf{e.} Micrograph of a 5 mm $\times$ 5 mm JTWPA chip. \textbf{f.} Zoomed-in view of the structure showing the low-frequency lumped-element phase matching resonator (blue), capacitors to ground $C_g$ (orange), high-frequency lumped-element phase matching resonator (purple), and Josephson junctions (red). The color-coded elements correspond to the circuit schematic in panel (b). \textbf{g.} The JTWPA in the presence of a bichromatic pump transforms the vacuum field at the input into a squeezed field at the output through non-degenerate four-wave mixing.}
\label{fig:mmgeneral}
\end{figure*}

The Josephson parametric amplifier (JPA) is a conventional approach to generate squeezed microwave photons (\cref{fig:mmgeneral}a). JPA squeezers use a narrow-band resonator and its Q-enhanced circulating field to increase the interaction between photons and a single or few Josephson junctions.
Josephson junctions are superconducting circuit elements with an inherently strong inductive nonlinearity with respect to the current traversing them. This is the nonlinearity that enables parametric amplification.
However, the relatively large circulating field in JPAs strongly drives the non-linearity of individual junctions, leading to unwanted higher-order nonlinear processes and saturation that impact squeezing performance~\cite{Boutin2017, Malnou_2018, Murch2013, Menzel_2012, Bienfait_2017, Krantz_2013}. Moreover, photon number fluctuations in the pump tone could lead to additional noise that reduces squeezing performance~\cite{Renger2021}.

Several alternative approaches have been developed that address some of these limitations.
For example, the impedance engineering of resonator-based JPAs has increased the bandwidth to the \unit[0.5-0.8]{GHz} range~\cite{Roy2015, Mutus2014}, but these devices still have a dynamic range limited to \unit[-110 to -100]{dBm} and sub-gigahertz bandwidth. 
%within such resonator-based JPAs. 
%
Alternative approaches using superconducting nonlinear asymmetric inductive elements (SNAILs) for both resonant~\cite{Sivak2019, Frattini2018, Sivak2020} and traveling-wave~\cite{esposito2021, perelshtein2021} parametric amplification feature a higher dynamic range in the \unit[-100 to -90]{dBm} range. However, both architectures require a magnetic field bias, making them subject to magnetic field noise.
Furthermore, the resonant version remains narrowband, and one traveling-wave approach~\cite{perelshtein2021} requires additional shunt resistors, which introduce dissipation and unwanted noise. To date, both approaches have been limited to \unit[2-3]{dB} single-mode and two-mode squeezing.

% An alternative approach using superconducting nonlinear asymmetric inductive elements (SNAILs) for both resonant~\cite{Sivak2019, Frattini2018, Sivak2020} and traveling wave~\cite{esposito2021, perelshtein2021} parametric amplification feature an improved dynamic range in the \unit[-100 to -90]{dBm} range, but the resonant version remains narrowband\textcolor{red}{; both architectures require a magnetic field bias and sometimes additional shunt resistors, which induce additional unwanted noise and dissipation}, and both have thus far been limited to \unit[2-3]{dB} single-mode and two-mode squeezing. 
%
High kinetic inductance wiring has been used in place of Josephson junctions to realize the nonlinearity needed for both resonant~\cite{parker2021} and traveling wave parametric amplification~\cite{Malnou2021, Bockstiegel2014} with higher dynamic range. However, the relatively weak nonlinearity of the wiring translates to a much larger requisite pump power to operate the devices, and the traveling wave paramps have larger gain ripple due to impedance variations on the long (up to \unit[2]{m}) lines. Furthermore, although a single-mode quadrature noise (variance) reduction has been demonstrated in narrowband resonant nanowire devices, their degree of squeezing in dB has yet to be quantified using a calibrated noise source~\cite{parker2021}. Squeezing always involves two modes, a ``signal'' and an ``idler''. We note that there are finite bandwidths associated with measurement in experimental settings. To clarify the terminology used in the paper and draw comparison with other previous works, we define ``two-mode'' as when the signal and idler are non-degenerate and their mode separation is much larger than the measurement bandwidth $|\omega_{\rm s}-\omega_{\rm i}| \gg B_{\rm meas}$, and ``single-mode'' as when the signal and idler are both nominally degenerate and within the measurement bandwidth $|\omega_{\rm s}-\omega_{\rm i}| \leq B_{\rm meas}$.

\begin{figure*}
\centering
\includegraphics[width=0.96\textwidth]{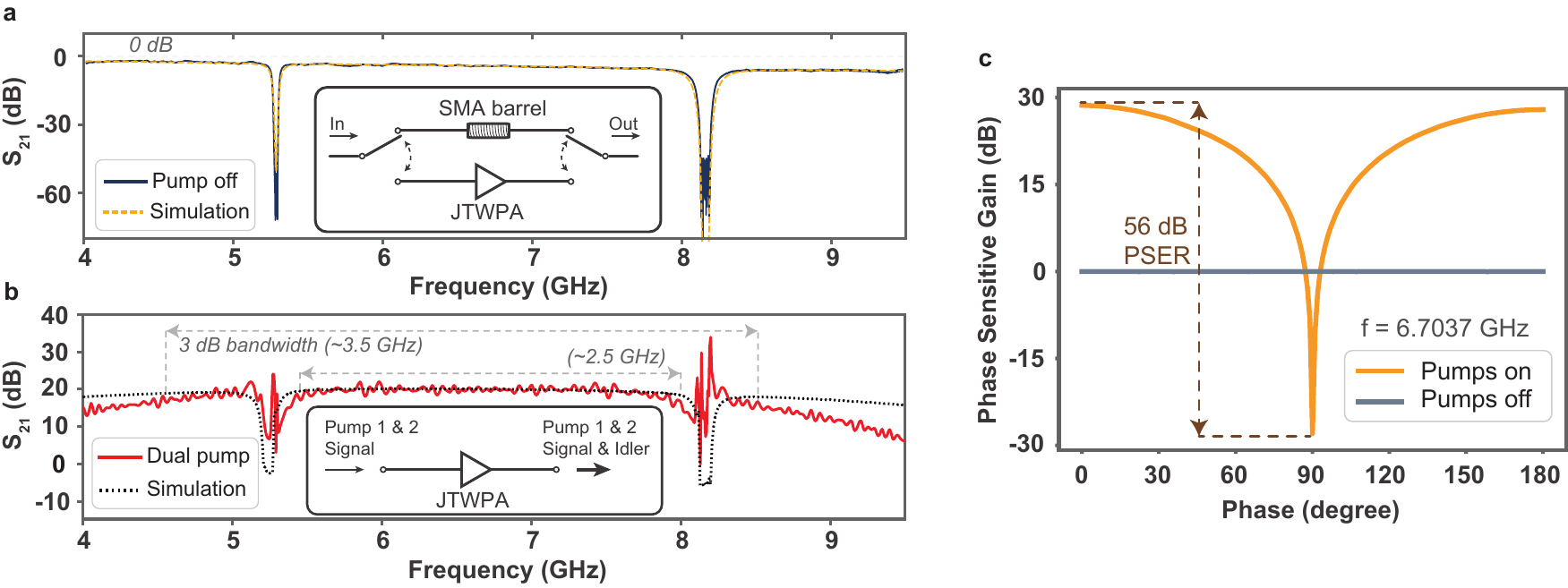}
\captionsetup{labelformat=empty}
\caption{\textbf{Figure 2 $|$ Amplification characteristics.} \textbf{a.} Undriven JTWPA transmission $S_{21}$ normalized with respect to a through line with an SMA barrel that accounts for the JTWPA package connectors. Microwave switches route the signal through the two paths with approximately equal lengths. 
The JTWPA loss %rate 
is approximately \unit[-0.00163]{dB} per unit cell, resulting primarily from two-level systems (TLSs)~\cite{Macklin307} from the dielectric material --- silicon dioxide --- in the parallel plate capacitor $C_{\rm g}$. 
The orange line is a numerical simulation of the JTWPA transmission (see~\sm{}).
\textbf{b.} Phase-preserving gain measured using a microwave vector network analyzer (red line) and a numerical simulation of the gain profile (black dotted line). The total bandwidth between the two pumps is around \unit[2.5]{GHz}, and the total \unit[3]{dB} bandwidth across the entire gain spectrum is more than \unit[3.5]{GHz}. \textbf{c.} Experimental phase-sensitive amplification at $\omega_{\rm c} = 2 \pi \times \unit[6.7037]{GHz}$. The phase-sensitive extinction ratio (PSER) is approximately \unit[56]{dB}.}
\label{fig:pppg}
\end{figure*}

In this work, we demonstrate a broadband single-mode and two-mode microwave squeezer using a dispersion-engineered, dual-pump Josephson traveling-wave parametric amplifier (JTWPA). 
%with a bichromatic pump configuration. 
As shown in \cref{fig:mmgeneral}b, the JTWPA contains a repeating structure called a unit cell, comprising a Josephson junction (red) --- a nonlinear inductor --- and a shunt capacitor (orange). Because their physical dimensions (tens of microns) are small compared to the operating wavelength (tens of millimeters) in the GHz regime, the junctions and capacitors are essentially lumped elements, constituting an effective inductance ($L$) and capacitance ($C$) per unit length. With the proper choice of $L$ and $C$, the lumped LC-ladder network forms a broadband \unit[50]{$\Omega$} transmission line, circumventing the bandwidth constraint of the JPA~\cite{Macklin307} and thereby enabling broadband operation. The use of many junctions -- here we use more than 3000 -- in a traveling-wave architecture accommodates larger pump currents before any individual junction becomes saturated~\cite{OBrien2014}, resulting in a substantially higher dynamic-range device. Therefore, with proper phase matching, the JTWPA has the potential to generate substantial squeezing and emit broadband entangled microwave photons through its wave-mixing processes.

Like a centrosymmetric crystal, the JTWPA junction nonlinearity features a spatial-inversion symmetry (in the absence of a DC current) that results in $\chi^{(3)}$-type nonlinear electromagnetic interactions. These support both degenerate-pump four-wave mixing (DFWM) and non-degenerate-pump four-wave mixing (NDFWM). 

As shown in \cref{fig:mmgeneral}c, the DFWM process --- $2\omega_{\rm p} = \omega_{\rm s} + \omega_{\rm i}$ --- converts two frequency-degenerate pump photons ($\omega_{\rm p}$) into an entangled pair of signal ($\omega_{\rm s}$) and idler ($\omega_{\rm i}$) photons. When $\omega_{\rm s} \neq \omega_{\rm p}$, energy conservation places the idler photon at a different frequency than the signal photon. This leads to two-mode squeezed photons and entanglement. However, DFWM has two drawbacks when considering single-mode squeezing, $\omega_{\rm s} = \omega_{\rm p}$. First, the signal and idler frequencies coincide with the strong pump, resulting in self-phase modulation that leads to unwanted phase mismatch, which cannot be compensated through dispersion~\cite{OBrien2014}. Second, it is challenging to later separate the signal and idler photons from the ``background'' pump photons.

In contrast, we use here (\cref{fig:mmgeneral}d) a NDFWM process -- $\omega_{\rm 1}+\omega_{\rm 2} = \omega_{\rm s} + \omega_{\rm i}$ --- that generates both single-mode and two-mode squeezed states far from the pump frequencies $\omega_1$ and $\omega_2$. To do this, we introduce a new JTWPA that uses two pumps and dispersion-engineering to achieve the desired NDFWM interaction.

The dual-pump JTWPA is fabricated in a niobium trilayer process on 200-mm silicon wafers. It exhibits a meandering geometry of its nonlinear transmission line with 3141 Josephson junctions and shunt capacitors (\cref{fig:mmgeneral}e). These are parallel-plate capacitors with silicon dioxide as their dielectric material. In addition, the JTWPA features two sets of interleaved phase-matching resonators, one (purple) at $\omega_{\rm r1} = 2 \pi \times 5.2~\mathrm{GHz}$ and the other (blue) at $\omega_{\rm r2} = 2 \pi \times 8.2~\mathrm{GHz}$ (\cref{fig:mmgeneral}f). The phase-matching resonators comprise lumped-element parallel-plate capacitors with niobium pentoxide dielectric and meandering geometric inductors. As shown in~\cref{fig:pppg}(a), the undriven JTWPA transmission $S_{21}$ is normalized with respect to the RF background of the experimental setup, utilizing a pair of microwave switches for signal routing (inset). The transmission characterization informs us of important JTWPA parameters, including the frequency-dependent loss, and the frequencies and linewidths of the phase-matching resonators, which guide us in choosing the pump frequencies.

Pumping the JTWPA at two angular frequencies $\omega_{1,2}$ generates parametric amplification that satisfies the energy conservation relation $\omega_{\rm s} + \omega_{\rm i} = \omega_1 + \omega_2$ and leads to the desired single-mode and two-mode squeezing. 
However, NDFWM also creates unwanted photons through the frequency conversion process $|\omega_{\rm s} - \omega_{\rm i'}| = |\omega_1-\omega_2|$, where $\omega_{\rm i'}$ is an extraneous idler angular frequency. 
This unwanted by-product does not participate in the desired two-mode squeezing, but rather, it is effectively noise that undermines squeezing performance. Fortunately, these unfavorable conversion processes are susceptible to phase mismatch and can be %efficiently mitigated in this device 
effectively reduced through dispersion engineering for a wide range of pump powers.

The efficiency of parametric amplification is determined by momentum conservation, i.e., phase matching~\cite{Macklin307}. To this end, we define a phase-mismatch function for the parametric amplification (\pa) process associated with NDFWM,

\begin{equation}\label{eq:deltak}
\Delta k_{\rm 12}^\pa = \left(1 {+} 2\beta_{\rm 1}^2+2\beta_{\rm 2}^2\right)\left(k_{\rm 1} {+} k_{\rm 2} {-} k_{\rm s} {-} k_{\rm i}\right)
{-} \beta_{\rm 1}^2 k_{\rm 1} {-} \beta_{\rm 2}^2 k_{\rm 2},
\end{equation}
where $k_{x} = \omega_{x}/c$ are wavevectors at frequencies $\omega_{x}$, with $x$ being the signal ($\rm s$), idler ($\rm i$), and pumps ($1$ and $2$), with $c$ being the speed of light for EM waves traveling in the JTWPA. The parameter $\beta_{\rm 1,2} \equiv I_{\rm 1,2}/4I_{\rm c}$ is a dimensionless pump amplitude scaled by the junction critical current $I_{\rm c}$, and $I_{\rm 1,2}$ are the pump currents at frequencies $\omega_{\rm 1,2}$, respectively. The linear wave vectors $k_{x}$ entering the phase-mismatch functions are determined by the unit-cell series impedance and parallel admittance to ground along the JTWPA (see \sm).

The pump-power-dependent terms in \cref{eq:deltak} --- those with the $\beta^2_{\rm 1,2}$ factors --- lead to phase mismatch that can be corrected. To achieve this, we adopt the dispersion-engineering approach of Ref.~\onlinecite{Macklin307} and extend it to two phase-matching resonators placed periodically throughout the amplifier. The resonator frequencies are chosen to be near-resonant with the desired pump frequencies. The modified admittance of the transmission line about these resonances leads to a rapid change in phase with frequency. Tuning the pump frequencies across the resonances thereby enables us to retune the pump phases periodically along the device and control the degree of phase matching. 

The precise selection of pump frequencies determines the phase matching condition and thereby enhances and suppresses different nonlinear processes. We preferentially phase match the parametric amplification process, $\omega_1+\omega_2 = \omega_{\rm s}+\omega_{\rm i}$. This is achieved if $\Delta k_{12}^\pa \simeq 0$ (see \cref{eq:deltak}), while all other processes are highly phase-mismatched. Experimentally, we sweep pump powers and frequencies in order to identify pump parameters that simultaneously maximize the dual-pump gain and minimize the single-pump gain. 
As shown in~\cref{fig:pppg}(b), with both pumps on, we obtain more than \unit[20]{dB} phase-preserving gain over more than \unit[3.5]{GHz} total bandwidth -- comparable with the single-pump JTWPA~\cite{Macklin307} and significantly broader than JPAs~\cite{Thol_2009, Malnou_2018, Castellanos_2008, Zhong_2013, Zorin2017}. The \unit[1]{dB} compression point at \unit[20]{dB} gain is \unit[-98]{dBm}, capable of amplifying more than 35,000 photons per microsecond within the microwave C-band (4 - \unit[8]{GHz}) and 20 to 30 dB higher than conventional resonator-based squeezers~\cite{Eicher2014, Bienfait_2017, Zhong_2013}. The large dynamic range enables the JTWPA to be a bright source of squeezed microwave photons.

At the center of the two pump frequencies, $\omega_{\rm c}=(\omega_1+\omega_2)/2$, the signal and idler interfere constructively or destructively, depending on their relative phase, leading to phase sensitive amplification and deamplification. We characterize such interference by injecting a probe tone at frequency $\omega_{\rm c}$ and measuring the amplifier output as a function of the probe phase $\theta_{\rm probe}$. \cref{fig:pppg}(c) shows the JTWPA output phase-sensitive gain with pumps on (orange) normalized to the case with pumps off (gray). The phase-sensitive extinction ratio (PSER), defined as the difference between the maximum phase-sensitive amplification and de-amplification, is measured to be $\unit[56]{dB}$, as far as we know, the largest value reported to date with superconducting Josephson-junction circuits~\cite{Thol_2009, Bienfait_2017, Zhong_2013}. 

Considering vacuum as the input to the JTWPA, the squeezing level -- the amount of noise reduction relative to vacuum fluctuations in decibels, $\rm{dB_{Sqz}}$ -- can be extracted based on the measurement efficiency $\eta_{\rm meas}$ of the output chain. %To accomplish this task, it 
Determining the efficiency requires an in-situ noise power calibration at the mixing chamber of a dilution refrigerator. 
%In this work
Here, we employ two independent, calibrated sources: a qubit coupled to a waveguide~\cite{Kannan2020} and a shot-noise tunnel junction~\cite{Spietz2003}. Both give consistent results, and we use these calibrated sources to extract the system noise temperature $T_{\rm sys}$ 
%and convert it into 
to calculate the measurement efficiency $\eta_{\rm meas}$~\cite{Mallet_2011}: 
\begin{equation}\label{eq:eta_meas}
\eta_{\rm meas} = \frac{\hbar \omega}{2 k_{\rm B}T_{\rm sys}},
\end{equation}
where $k_{\rm B}$ and $\hbar$ are the Boltzmann and reduced Planck constants, respectively. For example, $T_{\rm sys}$ from the output of the JTWPA at \unit[30]{mK} in the dilution refrigerator to the room temperature detectors is around $\unit[2.5]{K}$ at \unit[6.7037]{GHz}, corresponding to a measurement efficiency $\eta_{\rm meas} \approx$ 6\%. By accounting for the gain and loss in the entire measurement chain, we determine an ``input-referred'' noise at the JTWPA reference plane. See the \sm{} for details on the calibration methods and results.

\begin{figure}[tbp]
\includegraphics[width=0.475\textwidth]{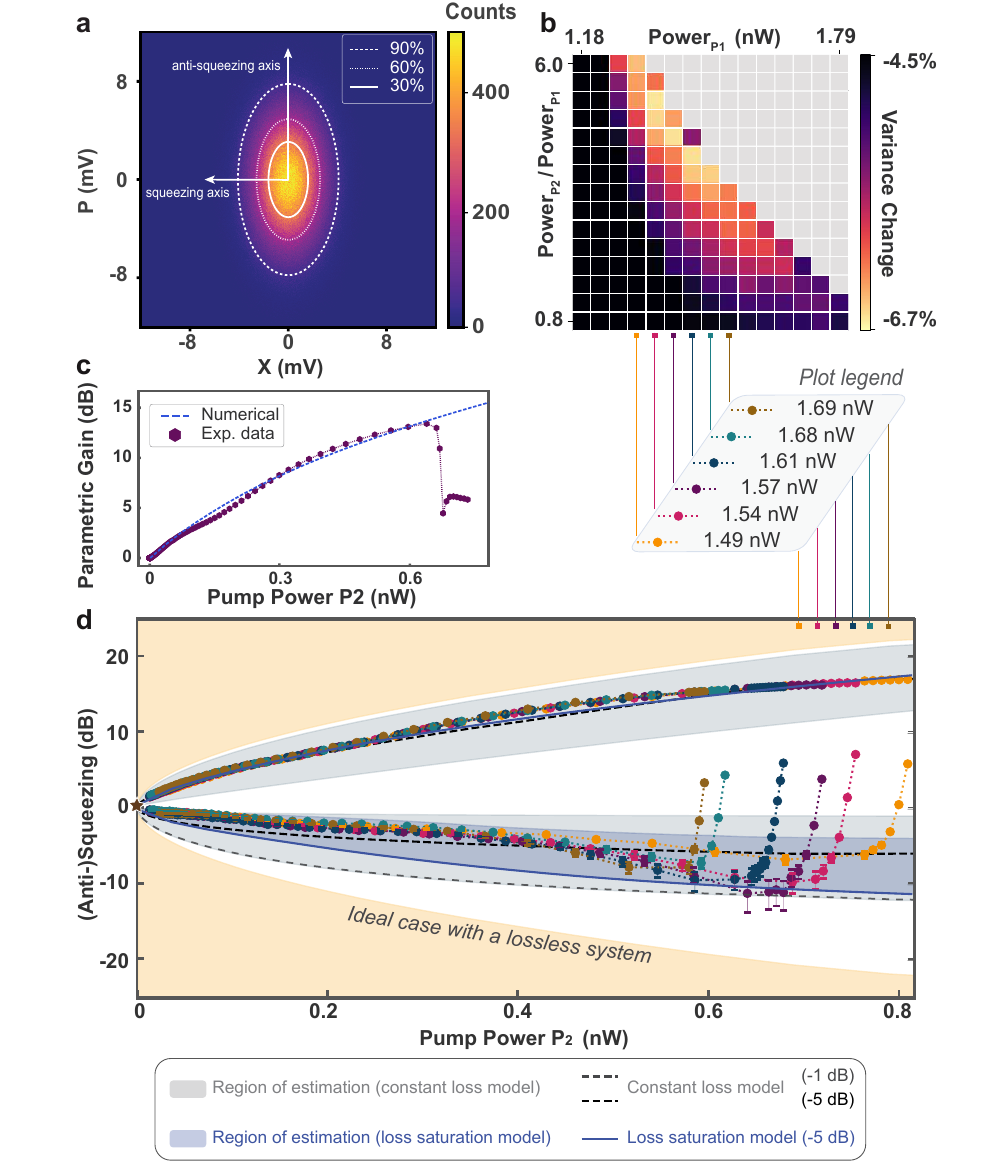}
\captionsetup{labelformat=empty, belowskip=-8pt}
\caption{\textbf{Figure 3 $|$ Single-mode squeezed vacuum.} 
\textbf{a.}  Output field histogram of an exemplary squeezed vacuum state with different confidence ellipses. The histogram comprises $6\times 10^6$ data points.  
\textbf{b.} At \unit[6.7037]{GHz}, measurement of the change in squeezing variance (relative to vacuum) %change 
versus asymmetry in the pump powers $P_1$ and $P_2$. \textcolor{black}{Colored vertical lines indicate six different values of $P_1$ in units of nW, used in the 1D measurement in panel (d). The power is referred at the input of the squeezer}.
\textbf{c.} \textcolor{black}{Experimental data of the parametric gain as a function of $P_2$ with $P_1$ fixed at \unit[1.57]{nW} (at the input of the squeezer)}. 
\textbf{d.} Measurement of squeezing and anti-squeezing versus $P_2$ with six different $P_1$ configurations (colored data). The brown star at the origin is a validation measurement to confirm there is no squeezing when pumps are turned off. The squeezing level increases as a function of $P_2$ as gain increases, but eventually degrades as the pumps become too strong and gain decreases. The shaded regions and trend lines corresponding to constant-loss and loss-saturation models are detailed in the \sm{}. The observed squeezing levels are consistent with a saturated loss of approximately -1 dB at high gain.}
\label{fig:sms}
\end{figure}

We first characterize the single-mode squeezed vacuum of the dual-pump JTWPA. To do this, we apply vacuum to the JTWPA input using a cold \unit[50]{$\Omega$} resistive load. %the 
We measure and compare the output field of the JTWPA for two cases: 1) the output with both pumps off  -- i.e., vacuum, and 2) the output with both pumps on, i.e., squeezed vacuum.
In both cases, the JTWPA output field propagates up the measurement chain to a room-temperature heterodyne detector comprising an IQ mixer that downconverts the signal into its in-phase (I) and quadrature (Q) components at \unit[50]{MHz}. These two components are then sampled using a field-programmable gate array (FPGA)-based 
%high-speed 
digitizer with a sampling rate \unit[500]{MS/s}. The components are then digitally demodulated to obtain an I-Q pair from which one can derive the amplitude and phase of the output field. 

To acquire I-Q pairs, the pumps -- and thus the squeezing -- are periodically switched on and off with a duration of \unit[10]{$\upmu$s} each. \textcolor{black}{For each \unit[10]{$\mu$s} acquisition, only the inner \unit[8]{$\mu$s} is digitally demodulated to eliminate sensitivity to any turn-on and turn-off transients. The \unit[8]{$\mu$s} signal is integrated, corresponding to a measurement bandwidth $B_{\rm meas} \approx \unit[125]{kHz}$ and yields a single I-Q pair.} We interleave the squeezer-on and squeezer-off acquisitions to reduce sensitivity to experimental drift between the measurements. When the squeezer is off, we extract an isotropic Gaussian noise distribution for the vacuum state with variance $\Delta X^2_{\rm SQZ, \ \rm off}$. When the squeezer is on, the squeezed vacuum state exhibits an elliptical Gaussian noise distribution as shown in \cref{fig:sms}(a). In total, we acquire 6 million I-Q pairs to reconstruct each histogram. We then extract the variance along the squeezing axis $\Delta X^2_{\rm SQZ, \ \rm min}$ and along the anti-squeezing axis $\Delta X^2_{\rm SQZ, \ \rm max}$. Comparing the values $\Delta X^2_{\rm SQZ, \ \rm min}$ and $\Delta X^2_{\rm SQZ, \ \rm max}$ to the vacuum level $\Delta X^2_{\rm SQZ, \ \rm off}$ along with the measurement gain and efficiency enables us to determine the degree of squeezing and anti-squeezing, respectively (see Supplementary Materials for further details on the measurement protocol). 

The squeezing process is sensitive to the power of both pumps due to the desired phase-matching condition for parametric amplification (e.g., $\Delta k_{12}^\pa \simeq 0$ in \cref{eq:deltak}) and also residual parasitic processes such as %DFWM and 
frequency conversion. 
To maximize the degree of squeezing, we perform a coarse measurement of the $\Delta X^2_{\rm SQZ, \ \rm min}$ (plotted relative to vacuum) as a function of pump powers. This enables us to identify empirically the pump powers $P_1$ and $P_2$ that correspond to higher squeezing levels. For six such near-optimal values, the six different colors in \cref{fig:sms}(d), we carry out finer scans of squeezing, anti-squeezing, and parametric gain as a function of $P_2$ for fixed $P_1$. Accounting for the measurement efficiency $\eta_{\rm meas}$ at the output, we extract a squeezing level of $\unit[-11.35^{+1.57}_{-2.49}]{dB}$ and an anti-squeezing level of $\unit[15.71^{+0.14}_{-0.15}]{dB}$ at the optimal pump conditions, comparable with the best performance demonstrated by resonator-based squeezers in superconducting circuits~\cite{Boutin2017, Bienfait_2017, Menzel_2012, Malnou_2018, Mallet_2011, Castellanos_2008, Movshovich1990, Clark2017, Zhong_2013}. 

Squeezing performance is sensitive to dissipation (loss), which acts as a noise channel. % to the squeezer. 
Within our JTWPA, loss primarily originates from defects --- modelled as two-level systems (TLSs) --- within the plasma-enhanced chemical-vapor-deposited (PE-CVD) $\rm{SiO_2}$ dielectric used in the parallel-plate shunt capacitors. Previous studies have shown a quality factor Q $\sim 10^3$ associated with this dielectric in the single-photon regime, observed at low-power and low-temperature. In this limit, the TLSs readily absorb photons from the JTWPA and cause relatively high loss.  

We observe high levels of squeezing despite the use of such lossy materials in the JTWPA. We conjecture the reason is due to TLS saturation. At sufficiently high powers (large photon numbers), the TLSs saturate and the loss is reduced~\cite{Sage2011}. 
We can understand the net impact of TLSs on squeezing performance by considering the JTWPA to be a cascade of individual squeezers. The amount of added squeezing becomes position-dependent and increases with the increased gain at the output end. 
The TLSs are also distributed along the JTWPA, and they become saturated towards the output end due to the larger number of photons associated with the higher gain. Therefore, the impact of loss on squeezing performance is reduced towards the output where the marginal squeezing is the largest~\cite{Govia2019}. As a result, we expect loss saturation at large signal gain to improve squeezing performance, as we observe in our experiment [see \cref{fig:sms}(d) at higher pump power $P_2$]. 

To verify this conjecture, we independently measure the JTWPA loss as a function of photon number %(see~\sm{} for further details) 
by varying the JTWPA temperature. The loss at small thermal photon numbers (<\unit[50]{mK}) is around \unit[-5]{dB}. This reduces to \unit[-1]{dB} for large photon numbers (>\unit[800]{mK}). These two limits are shown as dashed lines using a constant loss model. For low pump power $P_2$, our data are closer to the \unit[-5]{dB} line. At higher powers, where we see maximal squeezing, the data are more consistent with the \unit[-1]{dB} line corresponding to saturated TLSs. We then use numerical simulations to calculate the photon number in the JTWPA from its input to its output. The photon number is converted to loss from the independent loss-temperature measurement, and we plot the corresponding squeezing due to this distributed loss (solid line). It starts at \unit[-5]{dB} for low powers, and reduces toward \unit[-1]{dB} at high powers due to loss saturation. The high degree of squeezing observed in this device is consistent with the loss saturation model to within about \unit[1-2]{dB} at high powers. See \sm{} for more details. At intermediate powers, the agreement is not as good. This is likely due to our optimizing for maximum squeezing at high pump powers. Parasitic processes that are largely absent at high powers may not be completely suppressed at intermediate powers. There is ongoing research to better understand and suppress these unwanted modes~\cite{peng2022}, but this is outside the scope of the current manuscript. 

%%%%%%%%%%%% Two mode squeezing %%%%%%%%%%%%%
\begin{figure}[tbp]
\includegraphics[width=0.475\textwidth]{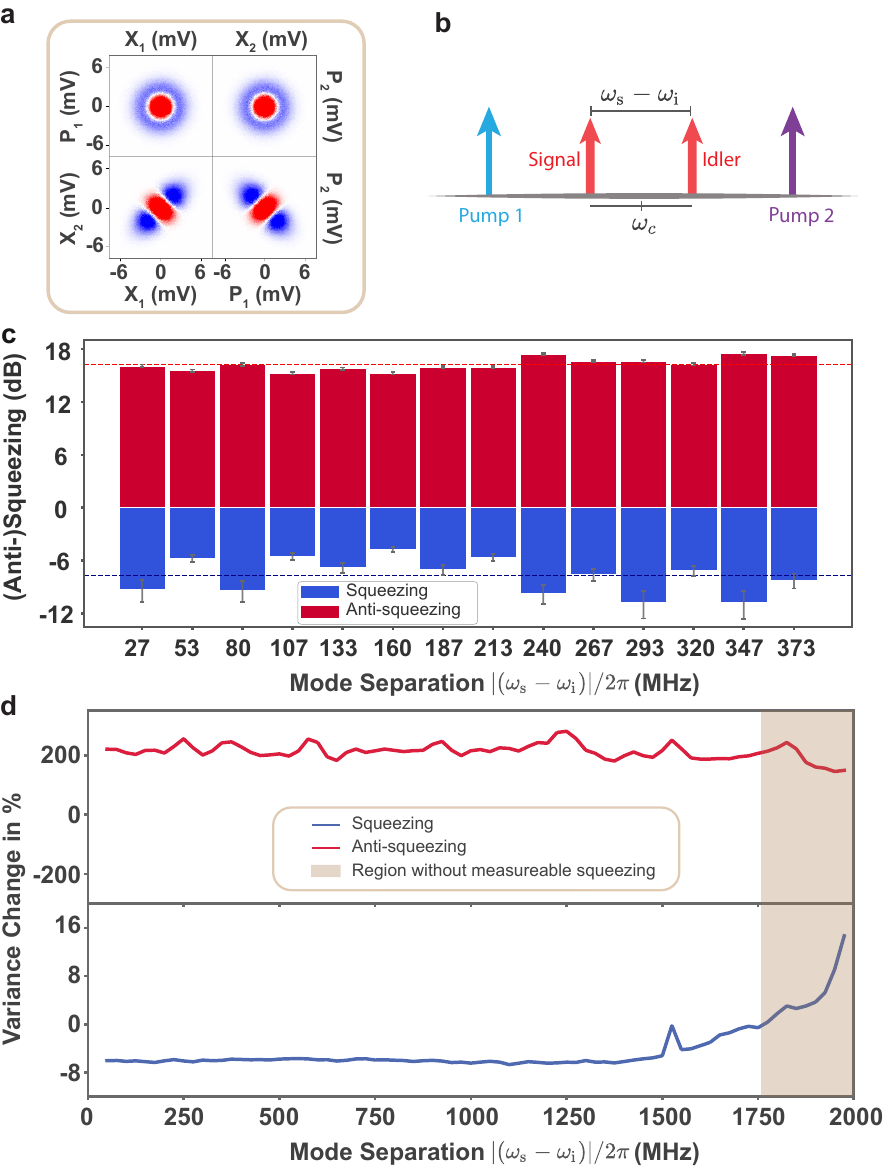}
\captionsetup{labelformat=empty, belowskip=-8pt}
\caption{\textbf{Figure 4 $|$ Broadband two-mode squeezed vacuum.} 
\textbf{a.} Difference in the output field histograms between vacuum (red) and two-mode squeezed vacuum (blue). The histograms show the X and P quadratures (equivalently, the in-phase and quadrature components) of the squeezed and vacuum states with signal and idler 320 MHz detuned from each other and centered at $\omega_{\rm c}$. 
\textbf{b.} Illustration of the frequency spectrum for the two-mode squeezing process. 
\textbf{c.} Measurement of two-mode squeezing versus frequency separation $|\omega_{\rm s} - \omega_{\rm i}|/2\pi$ between the signal and the idler. 
\textbf{d.} Percent change in variance between squeezed vacuum and vacuum for the $X_1X_2$ (or $P_1P_2$) quadrature as measured using two digitizers (see text). The beige-colored shading indicates the region where there is no measureable squeezing. The spike in the blue line plot (squeezing quadrature) around \unit[1500]{MHz} correspond to the extra mode generated by the JTWPA.}
\label{fig:tms}
\end{figure}

Using the same optimized pump configuration, we generate and characterize two-mode squeezed vacuum as a function of the frequency separation $\omega_{\rm s} - \omega_{\rm i}$ between the two modes. \textcolor{black}{We switch to a dual-readout configuration~\cite{Zhong_2013} that simultaneously demodulates the signal and idler using two separate FPGA-based digitizers, circumventing bandwidth limitations of the digitizer and other components in the experiment, such as IQ-mixers, low-frequency amplifiers, etc.} We directly measure up to a separation of \unit[373]{MHz} with the maximum squeezing of $\unit[-9.54^{+1.11}_{-1.63}]{dB}$, an average squeezing of \unit[-6.71]{dB}, and an average anti-squeezing of \unit[16.12]{dB}.
\textcolor{black}{The noise characterization method limits the measurement efficiency calibration to a frequency range $\sim \unit[500]{MHz}$, and therefore we cannot directly calibrate the degree of squeezing beyond this range. Nonetheless, squeezing is expected to continue beyond \unit[500]{MHz}~\cite{Grimsmo17}.}
% The frequency is limited solely for technical reasons by the bandwidth of the FPGA-based digitizer and other components in the experiment, such as IQ-mixers, low-frequency amplifiers, etc. As a result, the noise characterization method limits the measurement efficiency calibration to a frequency range of $\sim \unit[500]{MHz}$, and therefore we cannot directly calibrate the degree of squeezing beyond this range. Nonetheless, squeezing is expected to continue beyond \unit[500]{MHz}~\cite{Grimsmo17}. To measure squeezing at higher frequencies, we switch to a dual-readout configuration~\cite{Zhong_2013} that simultaneously demodulates the signal and idler using two separate FPGA-based digitizers.
%
As shown in \cref{fig:tms}(d), we characterize the variance change between the squeezed and the vacuum quadratures. Below 373 MHz, the results are consistent with the squeezing measured in \cref{fig:tms}(c). Above 373 MHz, the JTWPA exhibits a consistently low variance out to \unit[1500]{MHz}, beyond which we are again limited for technical reasons, in this case, by the onset of a filter roll-off. Because the signal and idler photons propagate at different frequencies, frequency-dependent variations of the loss and nonlinear processes can lead to frequency-dependent two-mode squeezing performance~\cite{Govia2019}. However, based on the flat and broadband gain profile observed in our JTWPA, we infer consistent squeezing levels out to \unit[1.5]{GHz} total signal-to-idler bandwidth, and net squeezing out to \unit[1.75]{GHz} total signal-to-idler bandwidth. \textcolor{black}{These results represent almost two-orders-of-magnitude increase in two-mode squeezing bandwidth compared to conventional resonator-based squeezers}~\cite{Eichler2011,Eicher2014, Menzel_2012, Malnou_2018, Flurin_2012, Schneider2020}.

In conclusion, we design and demonstrate a dual-pump Josephson traveling-wave parametric amplifier that exhibits both phase-preserving and phase-sensitive amplification, and both single-mode and two-mode squeezing. We measured \unit[20]{dB} parametric gain over more than \unit[3.5]{GHz} total instantaneous bandwidth (\unit[1.75]{GHz} for each the signal and the idler) with a \unit[1]{dB} compression point of \unit[-98]{dBm}. This gain performance is comparable with the single-pump JTWPA, yet it features minimal gain ripple and gain roll-off within the frequency band of interest. This advance alone holds the promise to improve readout of frequency-multiplexed signals~\cite{Heinsoo2018}. 
In addition, the favorable performance of this device enabled us to measure a \unit[56]{dB} phase-sensitive extinction ratio, \textcolor{black}{useful for qubit readout in quantum computing and phase regeneration in quantum communications. We also achieve} a single-mode squeezing level of $\unit[-11.35^{+1.57}_{-2.49}]{dB}$, and two-mode squeezing levels averaging \unit[-6.71]{dB} with a maximum value of $\unit[-9.54^{+1.11}_{-1.63}]{dB}$ measured directly over approximately 400 MHz and extending to over more than \unit[1.5]{GHz} total bandwidth (signal to idler frequency separation). \textcolor{black}{The results enable direct applications of the JTWPA in superconducting circuits, such as suppressing radiative spontaneous emission from a superconducting qubit~\cite{Murch2013} and enhancing the search for dark matter axions~\cite{Backes2021}.}
%
%, and can potentially read out multiplexed signals more efficiently~\cite{Heinsoo2018}. 

We have observed high levels of squeezing, despite the presence of dielectric loss from the $\rm{SiO_2}$ capacitors, which we attribute predominantly to distributed TLS saturation in the high-gain regions of our JTWPA.
Nonetheless, squeezing performance can be further improved by introducing a lower-loss capacitor dielectric. \textcolor{black}{} 
Performance can also be improved by exploring distributed geometries and Floquet-engineered JTWPAs that reduce the impact of unwanted parasitic processes~\cite{peng2022}.

The broad bandwidth and high degree of squeezing demonstrated in our device represents a new, resource-efficient means to generate multimode, non-classical states of light with applications spanning qubit-state readout~\cite{Barzanjeh2014, didier}, quantum illumination~\cite{Barzanjeheabb0451, LasHeras2017}, teleportation~\cite{Mallet_2011, Zhong_2013, Fedorov2021}, and quantum state preparation for continuous-variable quantum computing in the microwave regime~\cite{Grimsmo17, Fedorov2016}. In addition, the technique of using dispersion engineering to phase match different nonlinear processes can be extended to explore dynamics within superconducting Josephson metamaterials with engineered properties not otherwise found in nature.

\section{\label{sec:acknowledgement}Acknowledgement}
\noindent We thank Aditya Vignesh for valuable discussions and Joe Aumentado at NIST for providing the SNTJ. This research was funded in part by the NTT PHI Laboratory and in part by the Office of the Director of National Intelligence (ODNI), Intelligence Advanced Research Projects Activity (IARPA) under Air Force Contract No. FA8721-05-C-0002. The views and conclusions contained herein are those of the authors and should not be interpreted as necessarily representing the official policies or endorsements, either expressed or implied, of ODNI, IARPA, or the US Government.
ALG acknowledges support from the Australian Research Council, through the Centre of Excellence for Engineered Quantum Systems (EQUS) project number CE170100009 and Discovery Early Career Research Award project number DE190100380.

\setcounter{figure}{0}
\setcounter{equation}{0}
\setcounter{table}{0}
\renewcommand\theequation{S\arabic{equation}}
\renewcommand\thefigure{S\arabic{figure}}
\renewcommand\thetable{S\arabic{table}}

\setcounter{section}{0}

% \clearpage

\maketitle
\onecolumngrid
\section{Supplementary Materials}

\section{Cryogenic setup and control instrumentation}
\begin{figure}[!htbp]
\centering
\includegraphics[width=0.9\textwidth]{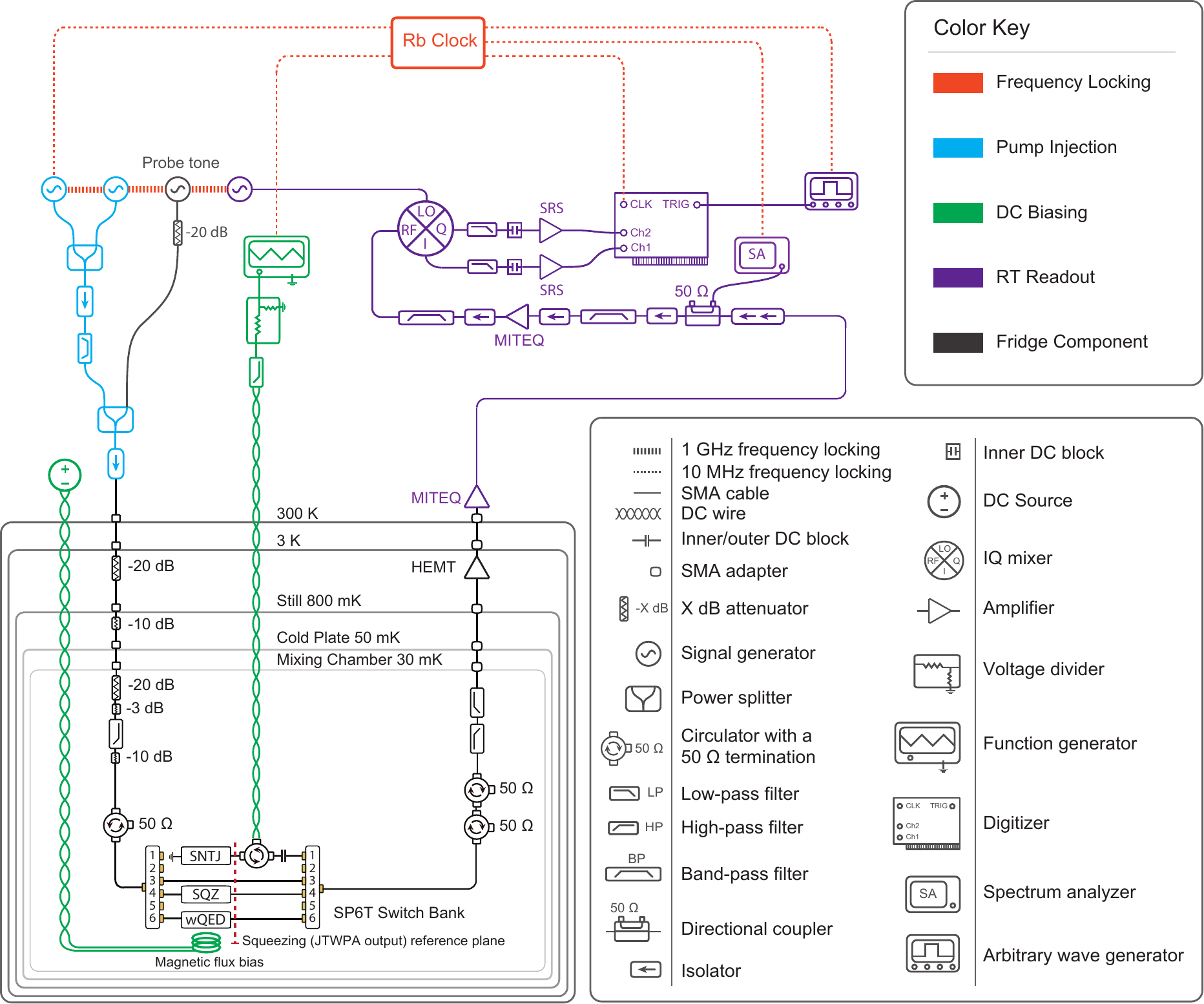}
% \captionsetup{labelformat=empty}
\caption{\textbf{$|$ Cryogenic setup and wiring diagram.} The diagram is color-coded to illustrate individual functional groups. The blue circuit shows the JTWPA pump sources with additional components including power combiner, isolator, and bandpass filter. The purple circuit represents the data acquisition setup that includes a spectrum analyzer and a digitizer with several filters and amplifiers. The green circuit is the DC biasing of shot-noise tunnel junction (SNTJ) and wQED qubit in the noise temperature characterization.}

\label{fig:cryosetup}
\end{figure}
In~\cref{tbl:hardware}, we list the major experimental components used in the experiment.
\begin{table}[!ht]
  \centering
 \begin{tabular}{|c|c|c|}
  \hline
  Component                & Manufacturer               & Type         \\
  \hline
  \hline
  Control Chassis       & Keysight    & M9019A   \\
  AWG & Keysight & M3202A \& 33250A\\
  ADC & Keysight & M3102A \\
  RF source & Rohde \& Schwarz & SGS100 \\
  Refrigerator & Leiden & CF450 \\
  DC Bias & Yokogawa & GS 200\\
  \hline
 \end{tabular}
\caption{Major experimental equipment used in the experiment.}
\label{tbl:hardware}
\end{table}

\cref{fig:cryosetup} shows the overall wiring diagram for the experiments conducted in a Leiden CF450 dilution refrigerator with a base temperature around \unit[30]{mK}. The pumps and probe signal generated by RF sources (Rhode and Schwarz SGS100A) are combined at room temperature (\unit[290]{K}) and sent via semi-rigid microwave coaxial cable to the squeezer (SQZ), a Josephson traveling-wave parametric amplifier (JTWPA). The line is attenuated by \unit[20]{dB} at the \unit[3]{K} stage, \unit[10]{dB} at the still, and \unit[33]{dB} at the mixing chamber to ensure proper thermalization of the line and attenuation of thermal photons from higher-temperature stages. In addition, coaxial cables and other components from the input line contribute around \unit[8]{dB} loss. A Cryoperm-10 shield magnetically shields the samples. We use Radiall single-pole-6-throw (SP6T) microwave switches to transmit the signal from either the squeezer, the shot-noise tunnel junction (SNTJ), or the waveguide QED (wQED) qubit to the measurement chain. The microwave signal at the output of the SP6T switch propagates through two \unit[50]{Ohm}-terminated circulators, a combination of a \unit[3]{GHz} high-pass and a \unit[12]{GHz} low-pass filter, and then into a superconducting NbTi coaxial cable that connects the \unit[30]{mK} and \unit[3]{K} stages. The NbTi cable allows high electrical and low thermal conductivity to minimize attenuation and heat transfer between different temperature stages. The signal is then amplified by a high electron mobility transistor (HEMT, Low Noise Factory LNF-LNC4\_8C) amplifier and room temperature stages for further amplification (MITEQ, AMP-5D-00101200-23-10P) and signal processing (frequency downconversion, filtering, digitization, and demodulation). The pump tones are band-pass filtered (and reflected into the \unit[50]{Ohm} termination of the room-temperature isolators) before the signal enters the IQ mixer for downconversion to avoid saturating the setup. In addition, the pump phase drift with \unit[1]{GHz} frequency locking is negligible compared to the measurement noise in the squeezing quadrature data. We also preemptively minimize any potential experimental drifts with our interleaved acquisition method described in the main text.

We use an arbitrary waveform generator (AWG Keysight 33250A) to bias the SNTJ. The AWG sends a low-frequency triangle wave with an amplitude $V_{\rm bias}$ through a \unit[993]{$\mathrm{k\Omega}$} resistor at room temperature to current bias the device in the $\upmu$A range. The current then passes through a stainless steel thermocoax to attenuate microwave and infrared noise. The resistance of the SNTJ at base temperature is measured in-situ to allow accurate extraction of the bias voltage across the junction. The frequency of the wQED qubit is controlled with a global flux line filtered at the \unit[3]{K} stage, using a DC source (Yokogawa GS200) at room temperature.

\begin{figure}[!htbp]
\centering
\includegraphics[width=0.75\textwidth]{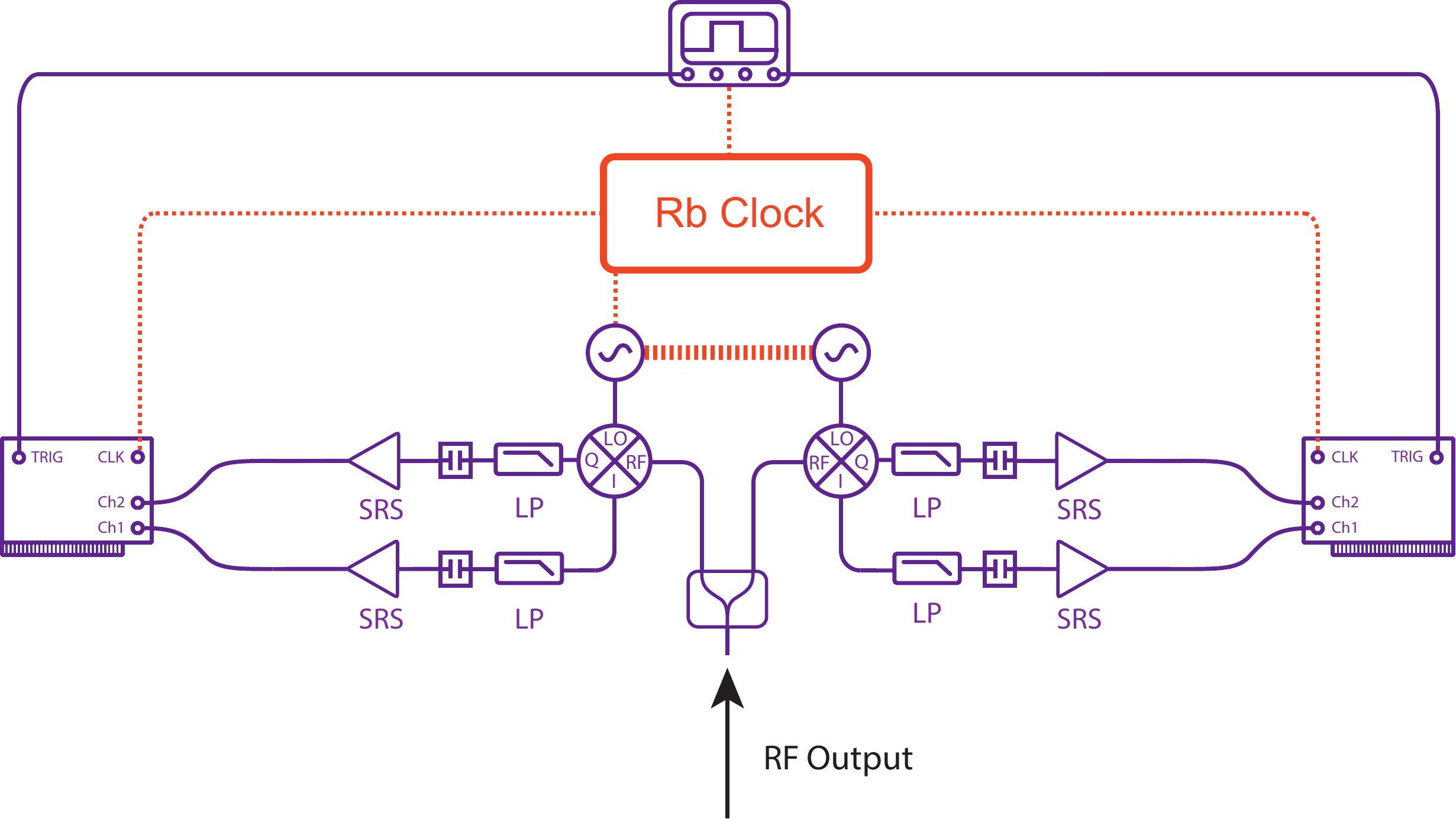}
% \captionsetup{labelformat=empty}
\caption{\textbf{$|$ Dual readout setup schematic diagram for two-mode squeezing data acquisition.} The JTWPA output from the dilution fridge is divided by a power splitter at room temperature into two identical branches of IQ down-conversion circuits, where the signal at the two modes can be simultaneously demodulated. Refer to \cref{fig:cryosetup} for more detailed schematic illustration.}
\label{fig:dualreadout}
\end{figure}

\newpage
\section*{Output Field Data Analysis}
\subsection{Single-Mode Squeezing}
We measure the output fields from the JTWPA using a room-temperature digitizer and demodulation scheme. An example of such a measurement is shown in Fig.~\ref{fig:gaussianfit}(a). This corresponds to pump 1 power $P_1 = $ \unit[1.57]{nW} and pump 2 power $P_2 = $ \unit[0.665]{nW} in Fig. 3d of the main text and reproduced in the supplementary material as Fig.~\ref{fig:sqztrend}. The measured distributions for squeezing (blue) and vacuum (red) are plotted independently in the insets, and then also together by subtracting the vacuum distribution from the squeezing distribution. Although this bias point corresponds to a high-degree of squeezing, the room-temperature measurement result is somewhat ameliorated due to several factors [e.g., see probability densities, right-hand side of  Fig.~\ref{fig:gaussianfit}(a)]. 
The reason is that we are measuring the quadratures at room temperature, rather than at the JTWPA output. Our measurement incorporates all of the loss, gain, and added amplifier noise in the measurement chain from the JTWPA output to the room temperature digitizer, and we must account for these to obtain the degree of squeezing at the JTWPA output. In addition, the digitizer measures the distributions in the voltage basis, and while this is sufficient for relative measurements between squeezing and vacuum, we also convert to the photon basis to make a standardized assessment in the photon basis.

\setlength\extrarowheight{5pt}
\setlength{\leftmargini}{0.5cm}
\begin{table*}[!htbp]
\small
\centering
\begin{tabular}{p{0.2\linewidth}p{0.4\linewidth}p{0.4\linewidth}}
\toprule
{Quantity} & {Definition} & {Value}  \\ 
\midrule \midrule
$\boldsymbol{\Delta X_{\rm SQZ, \ off}^2}$ & Vacuum variance, photon basis, stage 1 & $\dfrac{1}{2} + \bar{n}$\\ 
\midrule
$\boldsymbol{T_{\rm sys}}$ &  System noise temperature, stage 2 & measured quantity (noise temperature calibration)  \\
\midrule
$\boldsymbol{\eta_{\rm meas}}$ &  Measurement efficiency, stage 2 & $\hbar \omega/2k_{\rm B}T_{\rm sys}$\\
\midrule
$\boldsymbol{\alpha}$ &  Transfer function, stage 2  & $\Delta x^2_{\rm SQZ, \ off}/\Delta V^2_{\rm SQZ, \ off}$ (see text) \\ 

\midrule
$\boldsymbol{\Delta V^2_{\rm SQZ, \ min/max}}$ & Field variance, voltage basis, stage 3 &  measured quantity  \\
\midrule
$\boldsymbol{\Delta x^2_{\rm SQZ, \ min/max}}$ & Field variance, photon basis, stage 2  & $\alpha \ \Delta V^2_{\rm SQZ, \ min/max}$   \\
\midrule

$\boldsymbol{\Delta X^2_{\rm SQZ, \ min/max}}$ &  Field variance, photon basis, stage 1 & $\left(\Delta x^2_{\rm SQZ, \ min/max} - (1-\eta_{\rm meas}) \dfrac{1}{2} \right)/\eta_{\rm meas}$ \\

\bottomrule
\end{tabular}
% \captionsetup{labelformat=empty}
\caption{\textbf{$|$ Definition of quantities used to convert measured fields to squeezed fields at the JTWPA output.} These quantities are used to convert the measured fields in the voltage basis to the desired squeezing and anti-squeezing fields at the output of the JTWPA in the photon basis. Stages refer to the modeling of the measurement chain shown in Fig.~\ref{fig:beamsplitter}.}
\label{tab:2}
\end{table*}

\begin{figure}[!htbp]
\centering
\includegraphics[width =0.7\textwidth]{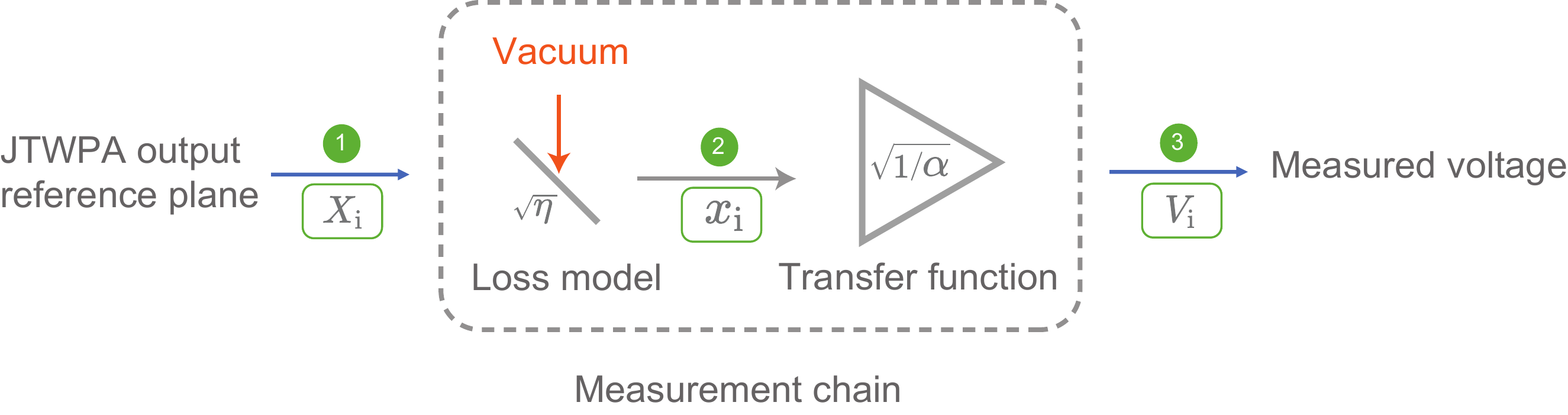}
% \captionsetup{labelformat=empty}
\caption{\textbf{$|$ Physical model connecting the field at the JTWPA output to the measured voltage.} Due to non-negligible loss and noise in the microwave setup, the output field measured by the room-temperature digitizer is different from that at the output of the JTWPA at milliKevin temperature. Therefore, to infer the squeezing levels at the JTWPA output, we use a model for the measurement chain from the output of the JTWPA (input of the model with quadrature field amplitude $X_{\rm i}$ in the photon basis) to the digitizer (output of the model with quadrature field amplitude $V_{\rm i}$ in the voltage basis). The model uses a fictitious beamsplitter that accounts for noise (loss) in the measurement chain followed by a lossless transfer function that accounts for amplifier gain and the conversion between the photon basis and the voltage basis. The losses and other Gaussian noise sources~\cite{Mallet_2011} are captured by the beamsplitter with transmissivity $\eta$, after which the quadrature field amplitude in the photon basis is denoted as $x_{\rm i}$. The transfer function $\alpha$ with a field conversion factor $\sqrt{1/\alpha}$ encompasses any linear scaling in the measurement chain, including amplifier gain (linear) and analog-to-digital conversion process of the digitizer. Green label 1, 2, and 3 mark the relative position in the model as referenced in our discussion.}
\label{fig:beamsplitter}
\end{figure}

We use the procedure following Mallet \textit{et al.} in Ref.~\onlinecite{Mallet_2011} to go from the room temperature measurement in the voltage basis to the degree of squeezing at the JTWPA output in the photon basis. The procedure is summarized in~\cref{tab:2} and goes as follows:

\begin{itemize}
    \item 
    We first determine the vacuum state variance in the photon basis at the output of the JTWPA to be $\Delta X_{\rm SQZ, \ off}^2 = 1/2 + \bar{n}$. To begin, the variance of the vacuum state in the photon basis $\Delta X_{\rm SQZ, \ off}^{2 \textrm{ (input)}}$ at the JTWPA input is $1/2+\bar{n}$, where $\bar{n} = \sum_i \bar{n}_i$ is the sum of average residual thermal photons arriving at the JTWPA from different temperature stages stages $i$ in the refrigerator. For each temperature stage $i$, the residual photon number is given by the Bose-Einstein distribution, $\bar{n}_i = A_{\rm i} / (e^{\hbar\omega/k_{\rm B}T_{\rm i}}-1)$, where $T_{\rm i}$ is the temperature of stage $i$, and the net average photon number is reduced by the collective attenuation $A_{\rm i}$ from stage $i$ to the JTWPA input. 
    \item
    Next, we determine the measurement efficiency $\eta_{\rm meas}$ from the system noise temperature $T_{\rm sys}$ determined using the noise calibration methods described in more detail in the next section. The efficiency is primarily affected by the HEMT, which we use as our first-stage amplifier with a large dynamic range, chosen to prevent gain saturation that would otherwise affect the measurement outcome. The efficiency is also affected by distributed loss in the measurement chain between the JTWPA and room-temperature digitizer. For the case shown in Fig.~\ref{fig:gaussianfit} and Fig.~\ref{fig:SMSevolv}, $\eta_{\rm meas} = 6.53^{+0.23}_{-0.22}\%$ at measurement frequency \unit[6.70]{GHz}.
    For our specific setup, we estimate an effective temperature $T < \unit[40]{mK}$ (or $\bar{n} < 0.00014$) at \unit[6.70]{GHz}, which has a negligible effect on the output field from the squeezer~\cite{Fei2018, Jin2015}. This is further validated by the noise characterization experiment using a shot noise tunnel junction (SNTJ), where we extract an average temperature $T$ = \unit[30.4]{mK} of the tunnel junction. The JTWPA input vacuum state is nearly ideal, with only a negligible thermal background, that is, $\bar{n} \ll 1/2$. Therefore, we could safely take  $\Delta X_{\rm SQZ, \ off}^{2 \textrm{ (input)}} = \bar{n}+1/2 \approx 1/2$. Nonetheless, although negligible, for completeness, we carry forward the small $\bar{n}$ to the JTWPA output. We note that this is an overestimate (worst-case), since the non-equilibrium thermal-photon portion of $\bar{n}$ -- the portion arriving from higher temperature stages in the refrigerator -- is  further attenuated by the JTWPA itself. Therefore $\bar{n}_{\textrm{output}} < \bar{n}_{\textrm{input}}$. Since the JTWPA attenuation changes with the bias point, we simply use the worst-case estimate $\bar{n}_{\rm output} = \bar{n}_{\rm input} \equiv \bar{n}$. This means we take $\Delta X_{\rm SQZ, \ off}^{2} \equiv \Delta X_{\rm SQZ, \ off}^{2 \textrm{ (output)}} = \Delta X_{\rm SQZ, \ off}^{2 \textrm{ (input)}} = (1/2 + \bar{n})$. Again, we have confirmed that $\bar{n}$ at this level has no discernible impact on our results. 
    \item
    We next determine the factor $\alpha$ that converts between the voltage basis and photon basis, $\alpha = \Delta x^2_{\rm SQZ, \ off}/\Delta V^2_{\rm SQZ, \ off} = \unit[0.129052]{\rm quanta/mV^2}$, obtained from the calculated value for $\Delta x_{\rm SQZ, \ off}^2$ and the measured value of $\Delta V_{\rm SQZ, \ off}^2$ for the vacuum state in the voltage basis~\cite{Mallet_2011}. 
    This conversion factor enables us to utilize a beamsplitter model as shown in~\cref{fig:beamsplitter} that accounts for the measurement efficiency, which for the variances we consider here, leads to:
    \begin{align}
        \Delta x_i^2 &= \eta_{\rm meas} \Delta X_i^2 + (1-\eta_{\rm meas})\frac{1}{2},
        \label{eq:variance-i}
    \end{align}
    where $\Delta X_i^2$ is the variance of the $X$ quadrature field at the beamsplitter input (i.e., the JTWPA output, the quantity we want to extract), $\Delta x_i^2$ is the variance of the $x$ quadrature field at the beamsplitter output that accounts for measurement efficiency, the factor (1/2) is the variance of vacuum introduced by the vacuum port of the beamsplitter, and $i$ corresponds to ``SQZ, off'' (vacuum), ``SQZ, min'' (squeezing), and ``SQZ, max'' (anti-squeezing). The same holds for the $P$ quadrature field. 
    
    We then use $\alpha$ in this equation to calculate the desired quadratures from the measured voltage variance~\cite{Mallet_2011}:
    \begin{align*}
        \Delta x_i^2 = \alpha \Delta V_i^2 &= \eta_{\rm meas} \Delta X_i^2 + (1-\eta_{\rm meas}) \frac{1}{2} \\
       \left( \frac{\Delta V_i^2}{\Delta V^2_{\rm SQZ, \ off}} \right)  \Delta x^2_{\rm SQZ, \ off}  &= \eta_{\rm meas} \Delta X_i^2 + (1-\eta_{\rm meas}) \frac{1}{2}
    \end{align*}
    which takes the decrease (increase) of the squeezed (anti-squeezed) voltage variance relative to the voltage variance obtained for vacuum, and uses it to scale the variance of vacuum in the photon basis. 

    \begin{align*}
        \Delta x^2_{\rm SQZ, \ off} &= \eta_{\rm meas} \Delta X^2_{\rm SQZ, \ off} + (1-\eta_{\rm meas})\frac{1}{2} \\
                                    &= \eta_{\rm meas} (\frac{1}{2} + \bar{n}) + (1-\eta_{\rm meas})\frac{1}{2}
                                    = \frac{1}{2} + \eta_{\rm meas} \bar{n}
    \end{align*}
    \item
    Finally, we obtain the desired variance at the JTWPA output by inverting~\cref{eq:variance-i}, which in turn accounts for the measurement efficiency, leading to the final entry in~\cref{tab:2}:
    \begin{align}
        \Delta X^2_{\rm SQZ, \ min/max} &= \frac{\Delta x^2_{\rm SQZ, \ min/max} - (1-\eta_{\rm meas})\frac{1}{2}}{\eta_{\rm meas}}.
    \end{align}
    This converts from $\Delta x^2_{\rm SQZ, \ min/max}$ to $\Delta X^2_{\rm SQZ, \ min/max}$, that is, the (anti-)squeezed quadratures at the JTWPA output. The same is done for the $P$ quadrature. 
\end{itemize}

Using this procedure, we can convert from the measured distributions in the voltage basis in Fig.~\ref{fig:gaussianfit}(a) to distributions in the photon basis Fig.~\ref{fig:gaussianfit}(b). For the particular bias point in Fig.~\ref{fig:gaussianfit}, we provide a few numbers. We process the output field data using the \emph{GaussianMixture} module within the \emph{sklearn.mixture} package in Python to compute the output field variance. 
The extracted variances for vacuum and squeezed states are denoted $\Delta V^2_{\rm SQZ,  \ off}$ and $\Delta V^2_{\rm SQZ, \ min/max}$ respectively. The corresponding standard deviations for the squeezed output field distributions shown in \cref{fig:gaussianfit}(a) are $\sigma_{\rm min}^{\rm sqz} = 1.90805 \times \pm 4.63\times 10^{-4}$ mV and $\sigma_{\rm max}^{\rm sqz} = 3.62295 \pm 7.64\times 10^{-4}$ mV when the squeezer is turned on. When it is off, the measured minimum and maximum standard deviations of the vacuum are identical within the fitting error bar $\sigma_{\rm min}^{\rm vac} = 1.96837 \pm 4.65 \times 10^{-4}$ mV and $\sigma_{\rm max}^{\rm vac} = 1.96837 \pm 4.04 \times 10^{-4}$ mV, which is expected for vacuum output field.
The transfer function $\alpha$ is calculated as the ratio $\Delta x^2_{\rm SQZ, \ off}/\Delta V^2_{\rm SQZ, \ off} = \unit[0.129052 \pm 5.2 \times 10^{-5}]{\rm quanta/mV^2}$.

\begin{figure}
\centering
\includegraphics[width =0.9\textwidth]{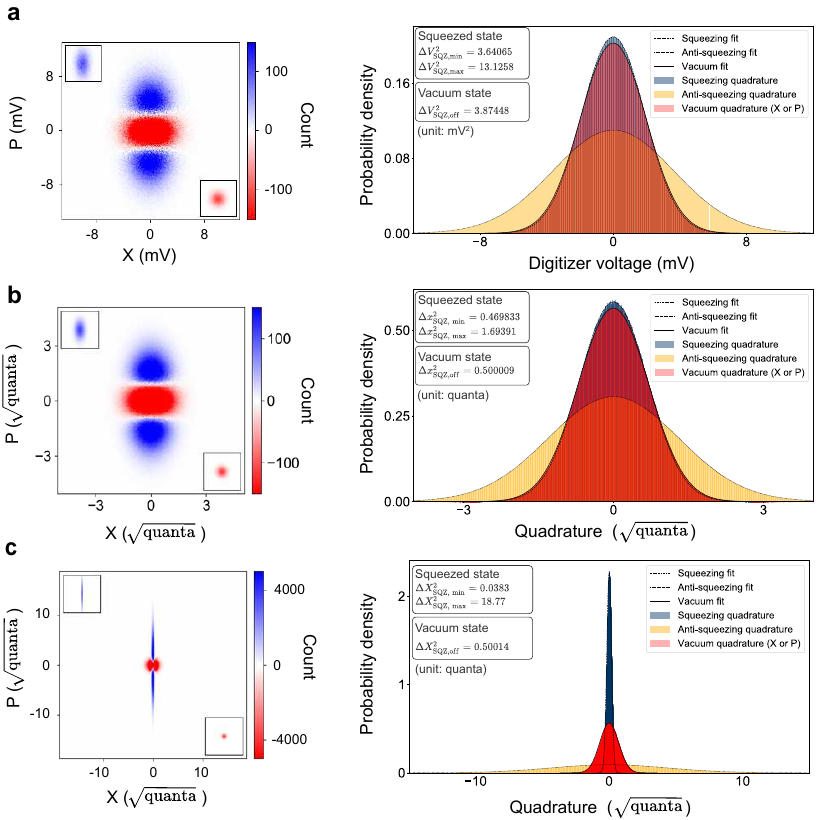}
% \captionsetup{labelformat=empty}
\caption{\textbf{$|$ Gaussian fits to quadrature data in voltage and photon bases.} \textbf{a.} Left: the histograms display the X and P quadratures for the single-mode squeezed vacuum (blue) and vacuum (red) states in the digitizer voltage basis (position 3 in~\cref{fig:beamsplitter}). Their output fields are shown individually in the insets. Right: probability density distribution of the X and P quadratures for the vacuum and squeezed states plotted together with Gaussian fits. \textbf{b.} Histograms and probability density distribution for the same vacuum and squeezed vacuum states are plotted in photon basis before the converter but after the beamsplitter (position 2 in~\cref{fig:beamsplitter}). \textbf{c.} Histograms and probability density distribution for the same vacuum and squeezed vacuum states are plotted in photon basis before the beamsplitter (position 1 in~\cref{fig:beamsplitter}). The measurement is taken at the pump configuration (pump 1 power $P_1 = $ \unit[1.57]{nW} and pump 2 power $P_2 = $ \unit[0.665]{nW} in \cref{fig:sqztrend}.}
\label{fig:gaussianfit}
\end{figure}

\begin{table}[!ht]
\centering
\begin{tabular}{|c|c|c|c|}
\hline
Stage & Parameter & Value & Error\\
\hline
\hline

3& $\sigma_{\rm min}^{\rm off}$ & $\unit[1.96837]{\rm mV}$ & $\unit[\pm 4.65 \times 10^{-4}]{mV}$\\

3& $\sigma_{\rm max}^{\rm off}$ & $\unit[1.96837]{\rm mV}$ & $\unit[\pm 4.04 \times 10^{-4}]{mV}$\\

3& $\sigma_{\rm min}^{\rm sqz}$ & $\unit[1.90805]{\rm mV}$ & $\unit[\pm 4.63 \times 10^{-4}]{mV}$\\

3& $\sigma_{\rm max}^{\rm sqz}$ & $\unit[3.62295]{\rm mV}$ & $\unit[\pm 7.64 \times 10^{-4}]{mV}$\\

3& $\Delta V^2_{\rm SQZ, min}$ & $\unit[3.64065]{\rm mV^2}$ & $\unit[\pm 1.77 \times 10^{-3}]{mV^2}$\\

3& $\Delta V^2_{\rm SQZ, max}$ & $\unit[13.1258]{\rm mV^2}$ & $\unit[\pm 5.54 \times 10^{-3}]{mV^2}$\\

3& $\Delta V^2_{\rm SQZ, off}$ & $\unit[3.87448]{\rm mV^2}$ & $\unit[\pm 1.83 \times 10^{-3}]{mV^2}$\\
\hline
\hline
2& $\Delta x^2_{\rm SQZ, \ min}$ & \unit[0.469833]{\rm quanta} & $\unit[\pm 2.28 \times 10^{-4}]{\rm quanta}$\\

2& $\Delta x^2_{\rm SQZ, \ max}$ & \unit[1.69391]{\rm quanta} & $\unit[\pm 7.15 \times 10^{-4}]{\rm quanta}$\\

2& $\Delta x^2_{\rm SQZ, \ off}$ & \unit[0.500009]{\rm quanta} & $\unit[+5\times 10^{-6} / -2\times 10^{-7}]{\rm quanta}$\\

2& $\alpha$ & $\unit[0.129052]{\rm quanta/mV^2}$& $\unit[\pm 5.2 \times 10^{-5}]{\rm quanta/mV^2}$\\

2& $\eta_{\rm meas} (\unit[6.70]{GHz})$ & \unit[6.534]{\%} & $\unit[+0.234\ /-0.218]{\%}$\\
\hline
\hline
1& $\Delta X^2_{\rm SQZ, \ min}$ & \unit[0.0383]{\rm quanta} & $\unit[+0.0160\ /-0.0159]{\rm quanta}$\\

1& $\Delta X^2_{\rm SQZ, \ max}$ & \unit[18.77]{\rm quanta} & $\unit[\pm0.63]{\rm quanta}$\\

1& $\Delta X^2_{\rm SQZ, \ off}$ & \unit[0.50014]{\rm quanta} & $\unit[+1.3 \times 10^{-4}/ -1.0 \times 10^{-4}]{\rm quanta}$\\

1& $\rm dB_{\rm SQZ}$ & \unit[-11.16]{\rm dB} & $\unit[+1.51\ /-2.33]{\rm dB}$\\

1& $\rm dB_{\rm ANTI}$ & \unit[15.74]{\rm dB} & $\unit[+0.14\ /-0.15]{\rm dB}$\\
\hline
\end{tabular}
\caption{A list of experimental parameter values and errors and their corresponding stage in the physical model in~\cref{fig:beamsplitter}.}
\label{tbl:experror}
\end{table}

Individual parameter errors lead to different variations in the overall squeezing and anti-squeezing levels. For example, errors in $\sigma_{\rm min}^{\rm off}$, $\sigma_{\rm max}^{\rm off}$, $\sigma_{\rm min}^{\rm sqz}$, $\sigma_{\rm max}^{\rm sqz}$ and $\alpha$ lead to a maximum of $\unit[\sim 0.4]{dB}$ change in the squeezing level and $\unit[\sim 0.01]{dB}$
in the anti-squeezing level. In contrast, errors in the measurement efficiency $\eta_{\rm meas}$ amount to $\unit[\sim 2]{dB}$ and $\unit[\sim 0.15]{dB}$ change in the squeezing and anti-squeezing level, respectively. Fast FPGA demodulation enables efficient collection of large quadrature datasets and thus results in small variations while $\eta_{\rm meas}$ is limited by microwave measurement losses and noises. Therefore, we primarily consider the errors associated with measurement efficiency $\eta_{\rm meas}$, the most significant error source, to estimate variations in squeezing and anti-squeezing levels.

For the output fields in \cref{fig:gaussianfit}(b), accounting for measurement efficiency, the squeezed variance $\Delta X^2_{\rm SQZ, \ min} = \unit[0.0383^{+0.0160}_{-0.0159}]{\rm quanta}$ and the anti-squeezing variance $\Delta X^2_{\rm SQZ, \ max} = \unit[18.77 \pm 0.63]{quanta}$ give $\unit[-11.16^{+1.51}_{-2.33}]{dB}$ squeezing and $\unit[15.74^{+0.14}_{-0.16}]{dB}$ anti-squeezing, relative to the vacuum state with a variance $\Delta X^2_{\rm SQZ, \ off}  = \unit[0.5001]{quanta}$. The conversion to decibels, $\rm dB_{\rm SQZ}$, is:
\begin{equation}\label{eq:sm:sqzlevel}
    \rm dB_{\rm SQZ} = 10 \log_{10}\dfrac{\Delta X^2_{\rm SQZ, \ min}}{\Delta X_{\rm SQZ,\ off}^2},
\end{equation}
while the anti-squeezing level $\rm dB_{\rm ANTI}$ is
\begin{equation}\label{eq:sm:antilevel}
    \rm dB_{\rm ANTI} = 10 \log_{10}\dfrac{\Delta X^2_{\rm SQZ, \ max}}{\Delta X_{\rm SQZ,\ off}^2}.
\end{equation}

Fig.~\ref{fig:SMSevolv} shows the evolution of squeezing as a function of pump powers for 6 of the points shown in Fig. 3d of the main text and reproduced in the supplementary material as Fig.~\ref{fig:sqztrend}. The vacuum and squeezed states from Fig.~\ref{fig:gaussianfit} are the middle panels (top and bottom) in Fig.~\ref{fig:SMSevolv} and correspond approximately to the maximal degree of squeezing observed. The left-most panels correspond to vacuum states with the pumps off. The second pair of panels from the left show a moderate degree of squeezing. The third pair of panels, as mentioned, are those from Fig.~\ref{fig:gaussianfit}. For even higher pump powers, the squeezing becomes distorted (fourth pair of panels) and even disappears (sixth pair of panels) as the junctions in the JTWPA become overpowered, the gain starts to saturate, and higher-order nonlinearities~\cite{Boutin2017} and even losses manifest.
\subsection{Two-Mode Squeezing}
\begin{figure}[!htbp]
\centering
\includegraphics[width=0.95\textwidth]{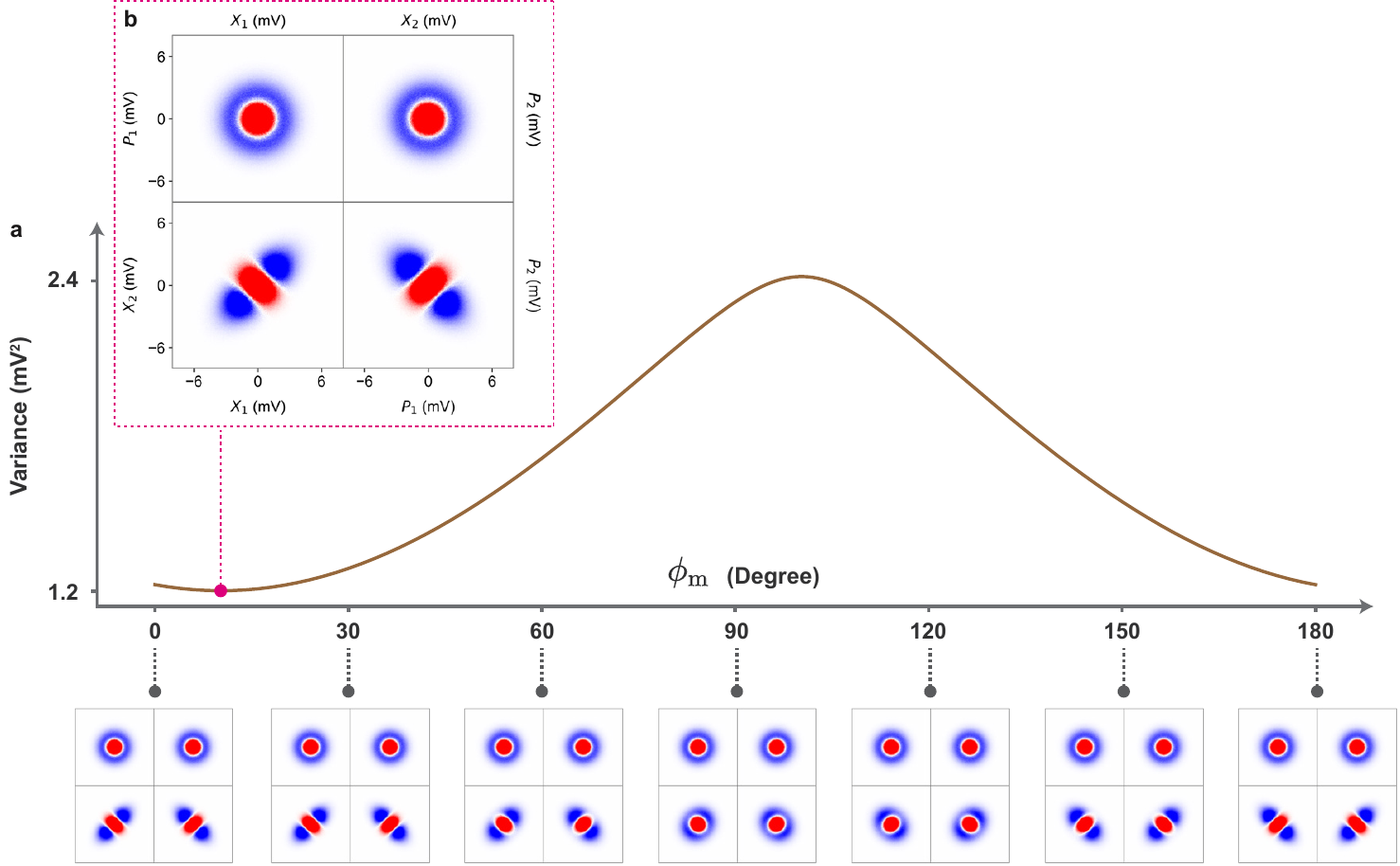}
% \captionsetup{labelformat=empty}
\caption{\textbf{| Two-mode squeezing signal-idler phase calibration at \unit[187]{MHz} mode separation.} \textbf{a.} Variance of the squeezed quadrature $X_1X_2$ or $P_1P_2$ as a function of relative phase $\phi_{\rm m}$, which we vary in data analysis. In addition, we display different output field histograms with different $\phi_{\rm m}$. Difference in the output field histograms between vacuum (red) and two-mode squeezed vacuum (blue). The histograms show the X and P quadratures (equivalently, the in-phase and quadrature components) of the squeezed and vacuum states. \textbf{b.} Output field histograms for optimal two-mode squeezing with $\phi_{\rm m} = \unit[0.17]{rad} (10^\circ$).}
\label{fig:tmsphi}
\end{figure}

To observe two-mode squeezing, we need to construct collective quadrature operators of the signal and idler defined as
\begin{equation}\label{eq:tmsX}
\begin{aligned}\hat{X}_\pm & = \hat{X}_{\rm s} \pm e^{i\phi_{\rm m}}\hat{X}_{\rm i},\\
\hat{P}_\pm &= \hat{P}_{\rm s} \pm e^{i\phi_{\rm m}}\hat{P}_{\rm i},
\end{aligned}
\end{equation}
% \begin{equation}
% \textcolor{red}{\hat{P}_\pm = \hat{P}_{\rm s} \pm e^{i\phi_{\rm m}}\hat{P}_{\rm i},}
% \label{eq:tmsP}
% \end{equation}
\noindent where $\hat{X}_{\rm s}$, $\hat{P}_{\rm s}$ and $\hat{X}_{\rm i}$, $\hat{P}_{\rm i}$ are quadrature components of the signal and the idler; $\phi_{\rm m}$ is the phase difference between the signal and the idler. In the ideal case where the signal and idler have the same phase, i.e., $\phi_{\rm m} = 0$ and $e^{i\phi_{\rm m}} = 1$, we can find the maximum squeezing. However, in practice the relative phase might not be zero due to the frequency dependency of the output line at the individual modes. Therefore, after acquiring the quadrature components of the two modes, we sweep $\phi_{\rm m}$, construct new histograms for $\hat{X}_\pm$ and $\hat{P}_\pm$ for each $\phi_{\rm m}$ as shown in~\cref{fig:tmsphi}, and extract the variance of the squeezed quadrature in the voltage basis $\Delta V^2_\pm$. to find the the minimum variance corresponding to the maximum two-mode squeezing.

\FloatBarrier
For example, at mode separation of \unit[187]{MHz}, we first set up the dual readout scheme as shown in~\cref{fig:dualreadout}. We then simultaneously demodulate the output signal at the two modes $\omega_{\rm s}/2\pi = \unit[6.6102]{GHz}$ and $\omega_{\rm i}/2\pi = \unit[6.7972]{GHz}$. The demodulation frequency $\omega_{\rm demod}/2\pi = \unit[47]{MHz}$ is the same for both, and we use two frequency-locked signal generators as local oscillators at frequencies $\omega^{\rm LO}_{\rm s}/2\pi = \omega_{\rm s}/2\pi +\omega_{\rm demod}/2\pi =  \unit[6.6572]{GHz}$ and $\omega^{\rm LO}_{\rm i}/2\pi = \omega_{\rm s}/2\pi+\omega_{\rm demod}/2\pi = \unit[6.8442]{GHz}$, respectively. After the demodulation, we obtain pairs of I-Q data for the two modes. We also correct for the power difference between the signal and idler mode that could lead to asymmetry in the output field due to any discrepancy such as attenuation between the two RF paths in the dual readout setup. To compensate for this effect, we measure the ratio $\nu_{\rm TMS}$ in vacuum state (JTWPA off) variance of the two modes and normalize that of the idler mode $\Delta V_{\rm i, off}^2 = \unit[1.424]{mV^2}$ with respect to that of the signal $\Delta V_{\rm s, off}^2 = \unit[1.119]{mV^2}$ --- $\nu_{\rm TMS} = \sqrt{\Delta V_{\rm s, off}^2/\Delta V_{\rm i, off}^2} = 0.8865$. As a result, we achieve normalized I-Q pairs with variances $\Delta \tilde{V}_{\rm s, off}^2 = \unit[1.119(3)]{mV^2}$ and $\Delta \tilde{V}_{\rm i, off}^2 = \unit[1.118(8)]{mV^2}$, now with an asymmetry of 0.04\% ($\Delta \tilde{V}_{\rm s, off}^2/\Delta \tilde{V}_{\rm i, off}^2 = 1.0004$) in vacuum state variance between the two modes; the asymmetry in the squeezed state variance of normalized data is also negligible at \unit[0.03]{\%} --- $\Delta \tilde{V}_+^2/\Delta \tilde{V}_-^2 = \unit[1.064(7)]{mV^2}/\unit[1.065(0)]{mV^2} = 0.9997$. This procedure accounts for the frequency dependence of the output line without amplification but does not compensate for asymmetry in the squeezer when it is turned on, e.g., the small ripples in the gain. Similar to the single-mode analysis, the variances of output fields are extracted using the \emph{GaussianMixture} module within the \emph{sklearn.mixture} package in Python. The histograms are plotted in~\cref{fig:tmsphi}
(b). In the same plot, we have also calibrated the relative phase $\phi_{\rm m} = \unit[0.17]{rad} (10^\circ)$ and achieved a maximum squeezing for this dataset.~\cref{fig:tmsphi}
(b) shows the signature of two-mode squeezing, in which the individual modes are in a ``thermal-like'' state with an increased variance (blue histograms in the $X_1P_1$ and $X_2P_2$ quadrants) while we have squeezing and anti-squeezing in the collective quadratures (blue histograms in the $X_1X_2$ and $P_1P_2$ quadrants).
% For the current example, we have its system noise temperature $T_{\rm sys}^{\rm TMS} = \unit[2.59]{K}$, corresponding to $\eta_{\rm meas} = \unit[6.13]{\%}$. 

To extract the squeezing level, we collect the joint distribution in the $X_1X_2$ or $P_1P_2$ quadrant (analogous to the single-mode squeezed state statistics) and perform the same analytical procedure \cref{eq:variance-i} - \cref{eq:sm:antilevel} as detailed in the previous single-mode squeezing section using the measured system noise temperatures (details of system noise characterization can be found in the next section). The results are shown in Fig. 4(c) from the main text up to around \unit[500]{MHz}, the bandwidth of our noise calibration device. Outside that bandwidth, we perform the same two-mode squeezing analysis except without the system noise temperature and report variance change between the squeezed and vacuum states as $1-\Delta \tilde{V}^2_+/\Delta \tilde{V}^2_{\rm +, off}$, where $\tilde{V}^2_{\rm +, off}$ is the variance for the two-mode vacuum state in the voltage basis. In the case of no squeezing, we have $\Delta \tilde{V}^2_+ = \Delta \tilde{V}^2_{\rm +, off}$ and variance change would be 0; in the case of squeezing, the squeezing variance drops below that of the vacuum, i.e., $\Delta \tilde{V}^2_+ <\Delta \tilde{V}^2_{\rm +, off}$, and variance change would be $<1$. This corresponds to the results in Fig. 4(d) from the main text. We note that although our noise calibration device was limited to 500 MHz bandwidth, the system noise temperature likely remains similar outside of this frequency range, as there is no apparent reason why it would suddenly change value. Therefore, we expect similar reductions in measured variance outside the calibrated band to correspond to similar levels of inferred squeezing measured within the band. However, since we did not explicitly calibrate the system noise at those frequencies, we report the measured reduction in variance.

\subsection{Squeezing Purity}
\begin{figure}[!htbp]
\centering
\includegraphics[width=0.99\textwidth]{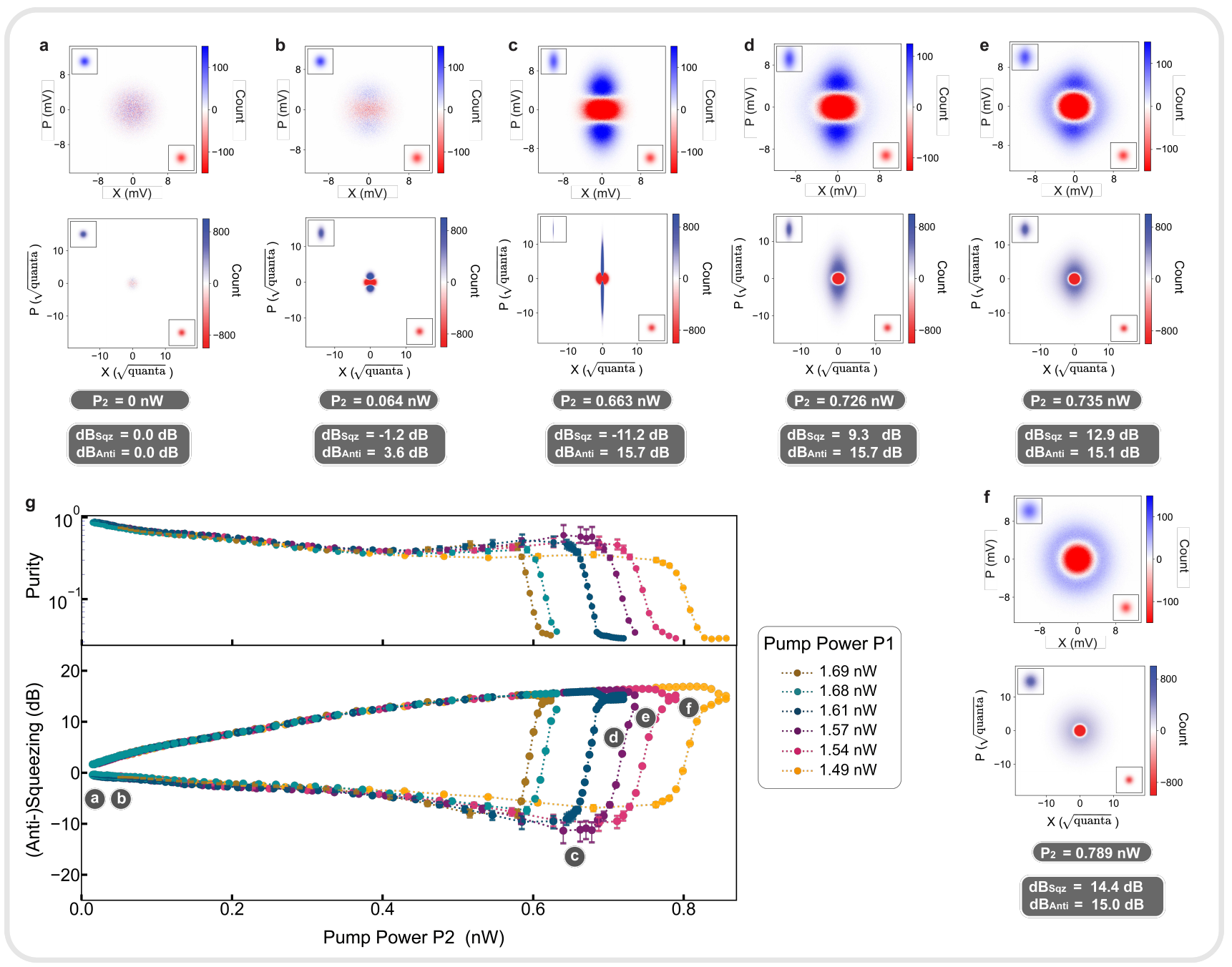}
%\captionsetup{labelformat=empty}
\caption{\textbf{$|$ Output field imaging of squeezed state evolution as pump power increases.} The top row demonstrates the output fields in voltage basis measured at room temperature, and the bottom row shows the same state in photon basis at the JTWPA. The histograms show the X and P quadratures of the squeezed and vacuum states. Panels \textbf{a} - \textbf{f} display the difference in the output field histograms between vacuum (red) and single-mode squeezed vacuum (blue), which are shown individually in the insets. The histograms give a different perspective on squeezing degradation shown in Fig. 3 in the main text and indicate different higher-order nonlinearities compared to a JPA that exhibits a distorted output field~\cite{Boutin2017}. After the junctions become saturated and lossy, they can generate excessive noise due to dissipation. The output field looks like an effective thermal state shown as the blue enlarged circular ``blob'' in panel \textbf{e} and more so in \textbf{f} (from a different trace with pump 1 power at \unit[1.54]{nW}). As the pump power continues to increase, the power dissipation leads to an increasing quadrature variance as seen in Fig. 3 from the main text and in \textbf{g} showing more details towards the high-pump-power region. The top panel in \textbf{g} shows the purity of the squeezed states as a function of pump power. As mentioned in the main text, the data are presented as mean values of 3 sets of repeated measurement (each with $6\times 10^6$ sample points). Their statistical variation is almost entirely due to the uncertainty in estimating the noise temperature (\cref{fig:NTcorrected}), which dominates the error bars shown in the plot.}
\label{fig:SMSevolv}
\end{figure}

\begin{figure}[!htbp]
\centering
\includegraphics[width=0.75\textwidth]{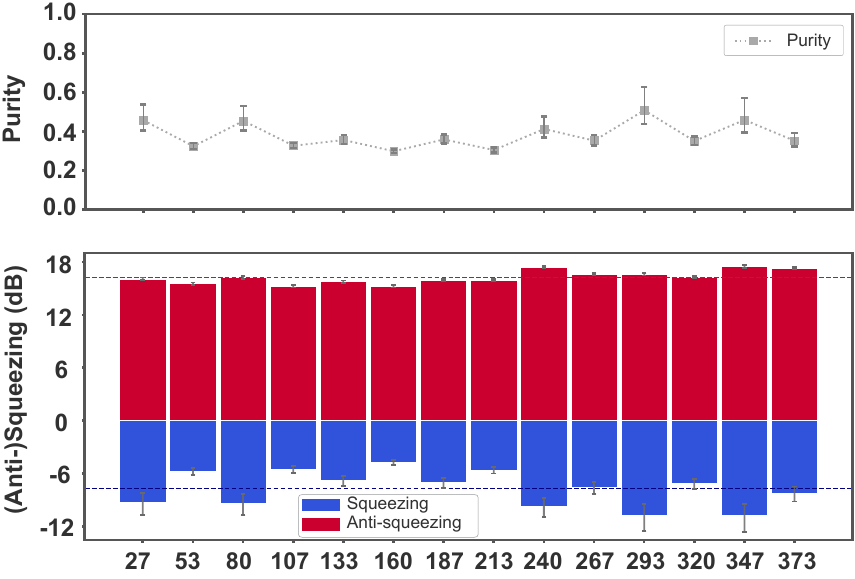}
%\captionsetup{labelformat=empty}
\caption{\textbf{$|$ Squeezing level and purity of two-mode squeezed vacuum state.} The top panel displays the corresponding squeezing purity associated with the squeezing levels shown in the main text Fig. 4. Similar to the single-mode squeezing results, the data are presented as mean values of 3 sets of repeated measurement (each with $6\times 10^6$ sample points). Their statistical variation is almost entirely due to the uncertainty in estimating the noise temperature, which dominates the error bars shown in the plot.}
\label{fig:TMSpurity}
\end{figure}

Additionally, in both~\cref{fig:SMSevolv} (g) and~\cref{fig:TMSpurity}, we have shown the purity of the squeezed states as a function of pump power and mode separation, respectively. Following the definition used in Ref.~\cite{Dassonneville2021}, purity of the squeezed state can be expressed as $\mathcal{P} = 1/\sqrt{S_-S_+}$ for a Gaussian state, where $S_-$ and $S_+$ denote squeezing and anti-squeezing factors. For single-mode squeezing, we extract the purity around the maximum squeezing level to be $\unit[0.605^{+0.201}_{-0.100}]{}$. Similarly, for two-mode squeezing, the average purity is \unit[0.379]{} and a maximum purity of $\unit[0.507^{+0.120}_{-0.070}]{}$ at \unit[293]{MHz} mode separation. The two-mode squeezing is measured under the same pump configuration for the maximal single-mode squeezing and can be further optimized. In comparison with cavity-based squeezers, the purity values have more room for improvement. The remarkably high levels of squeezing (as high as -11.3 dB for single-mode squeezing and -9.5 dB for two-mode squeezing) with only 40\%-50\% purity suggests that the JTWPA is capable of achieving even better squeezing performance, e.g., if we reduce the internal loss that likely limits the purity in this Nb-based version of the JTWPA. We are currently developing a new generation of JTWPAs that we expect to have a much lower internal loss and further suppression of spurious nonlinear processes~\cite{peng2022}.

\FloatBarrier

\FloatBarrier
\clearpage
\newpage
\section*{Noise Temperature Calibration}

The measurement efficiency of the output chain needs to be determined accurately to extract squeezing levels at the output of the JTWPA. However, direct access to the mixing chamber while the refrigerator is at milliKelvin temperatures in vacuum presents a fundamental challenge to this task. Calibrating at room temperature by passing a signal through the entire setup is insufficient, as the insertion loss for the input and the overall transmission of the output changes dramatically with temperature. Therefore, it necessitates an in-situ noise power calibration device at the mixing chamber, ideally at a relevant reference plane for the squeezer. In this work, we use a wQED device as our primary noise calibration device and a voltage-biased tunnel junction generating shot noise as our secondary method.

\subsection*{Shot-Noise Tunnel Junction (SNTJ) Noise Characterization}
A shot-noise tunnel junction~\cite{Spietz2003} is a metal-insulator-metal aluminum (Al) junction, with the Al operated in the normal state via a strong magnetic field from an in-situ neodymium magnet.

\begin{figure}[!htbp]
\centering
\includegraphics[width=0.75\textwidth]{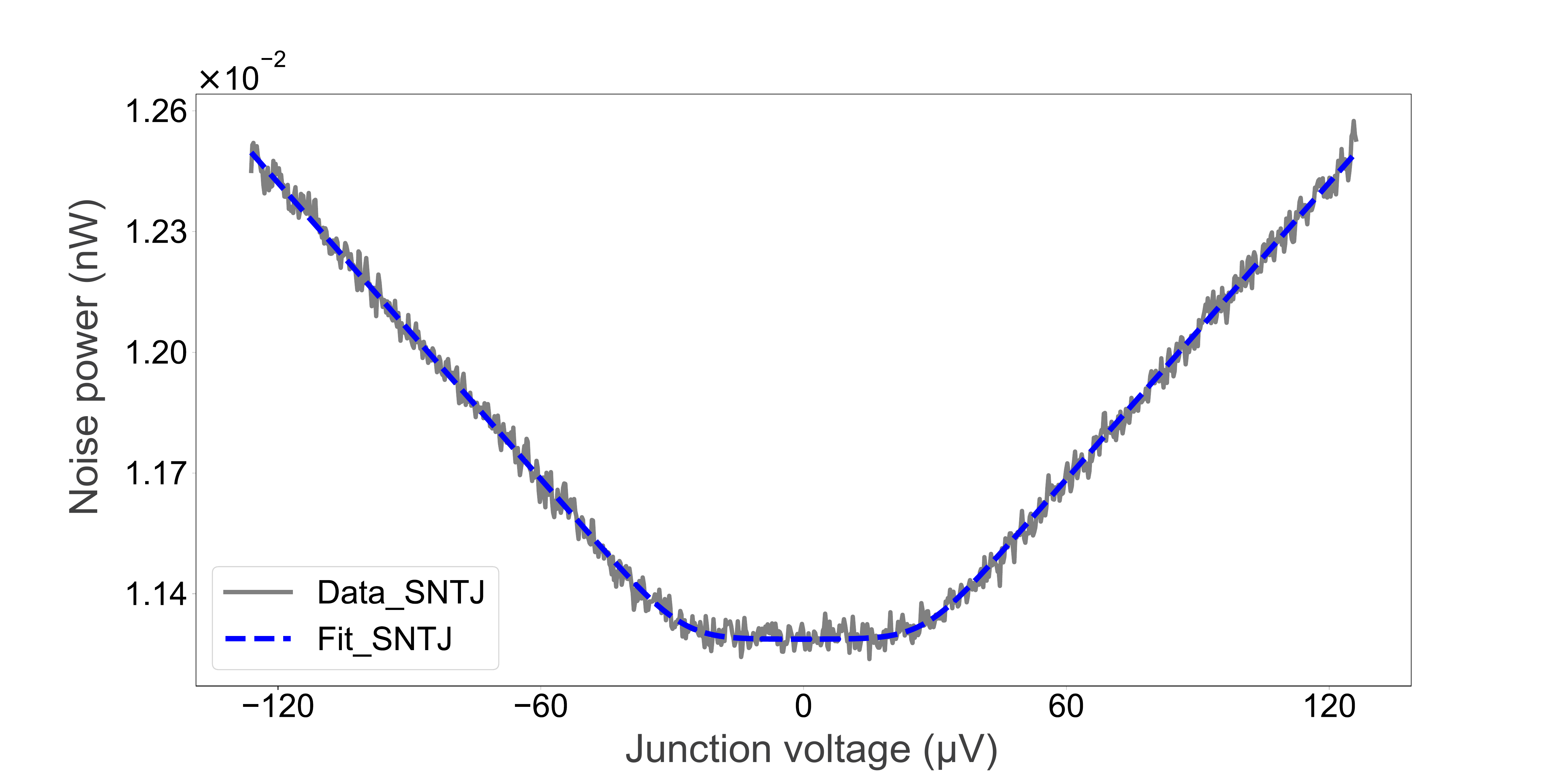}
% \captionsetup{labelformat=empty}
\caption{\textbf{$|$ ``Noise mustache curve'' generated from a SNTJ.} Experimental data plotted together with a fit using \cref{eq:noisetemp}. The horizontal axis is the voltage drop across the tunnel junction, and the vertical axis shows the noise power from the SNTJ.}
\label{fig:sntjcurve}
\end{figure}
\FloatBarrier
With a matched load, the noise power at frequency $f$ generated by a voltage-biased SNTJ at temperature $T$ is \cite{Spietz2003}
\begin{equation}
\begin{aligned}
  N =  Gk_{\rm B}B_{\rm SNTJ}\Bigg\{T_N + \frac{1}{2} \bigg[ \bigg(\frac{eV + hf}{2k_{\rm B}} \bigg)\coth\bigg(\frac{eV + hf}{2k_{\rm B}T} \bigg)+\bigg(\frac{eV - hf}{2k_{\rm B}} \bigg)\coth\bigg(\frac{eV - hf}{2k_{\rm B}T} \bigg)\Bigg\},
\end{aligned}
\label{eq:noisetemp}
\end{equation}
where $V$ is the voltage bias across the shot noise tunnel junction, $B_{\rm SNTJ}$ is the measurement bandwidth for the SNTJ noise measurement, $G$ is the system gain, and $T_{\rm N}$ is the system noise temperature.

In the limit $eV \ll hf$,~\cref{eq:noisetemp} is dominated by thermal and quantum noise. When $eV \gg hf$, the noise is dominated by the Poissonian shot noise of the electron current through the tunnel junction. Dilution refrigerators with a base temperature \unit[20-30]{mK} are sufficient to reach the quantum noise floor within the frequency range of interest here --- 4 - \unit[8]{GHz} represented by the plateau in the vicinity of \unit[0]{V} junction voltage. From the fit, we can extract both the system noise temperature $T_N$ as well as the temperature of the noise source $T$.
\newpage
\subsection*{Waveguide Quantum Electrodynamics (wQED) System Power Calibration}
\begin{figure}[!htbp]
\centering
\includegraphics[width=0.98\textwidth]{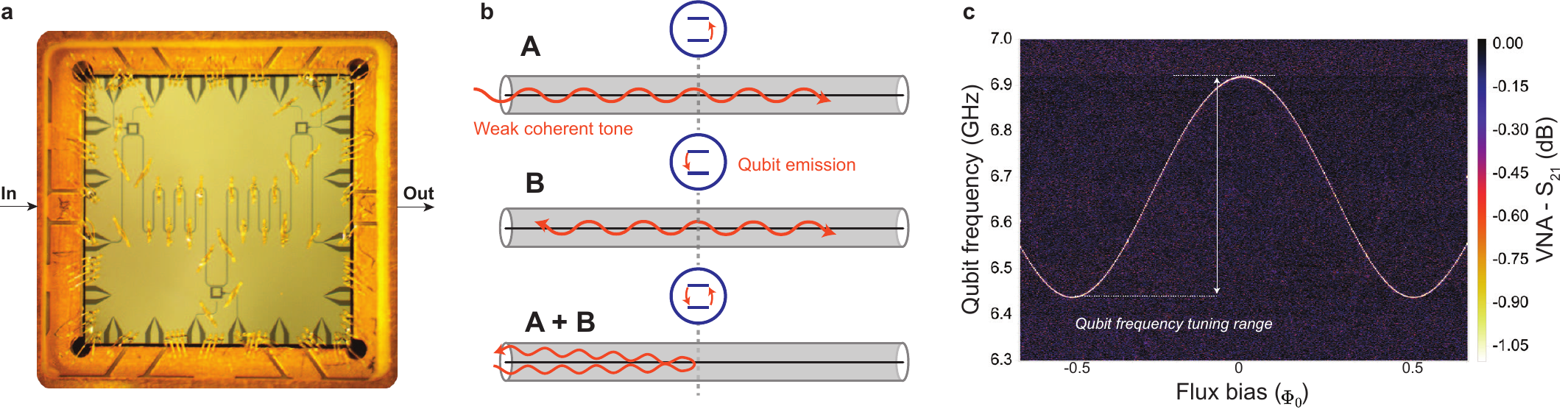}
% \captionsetup{labelformat=empty}
\caption{\textbf{$|$ Waveguide coupled to a qubit as a noise power reference.} \textbf{a.} A microscopic image of the wQED device, where three qubits coupled to a meandering transmission line are visible. In the calibration protocol, we only need to use one of the qubits. \textbf{b.} A schematic illustrating the interference effect between a qubit and a weak coherent tone. Process A: in the limit $\bar{n} < 1$, the qubit absorbs a photon from the coherent driving field. Process B: the qubit emits the photon isotropically in the forward and reverse directions with a $\pi$ phase shift. As a result (A+B), the emission from the qubit causes a destructive interference with the forward propagating field and results in the reflection of the signal~\cite{Kannan2020}.} \textbf{c.} Qubit spectrum measured by scanning DC magnetic flux bias and measuring its transmission profile at large drive. The noise temperature characterization is performed at various qubit frequencies between its two sweet spots (marked between the white dashed lines).
\label{fig:otlsetup}
\end{figure}
\FloatBarrier

In systems with a qubit coupled to a waveguide (\cref{fig:otlsetup}(b)), the qubit will reflect weak incident coherent tones ($\bar{n} = |\alpha|^2 \ll 1$) in the transmission line~\cite{Kannan2020}. In the limit $\bar{n} \ll 1$, the probability of two or more photons is negligible. The qubit absorbs a single photon from the coherent drive and emits the photon isotropically in the forward and reverse directions with a $\pi$ phase shift. As a result, the forward direction destructively interferes with the transmitted driving field, while the reverse direction constructively interferes with the reflected field. Therefore, under ideal conditions, all photons are reflected and no photons are transmitted. This perfect destructive interference is modified by the presence of decoherence, which yields and changes the transmission coefficient \cref{eq:transcoef}. Each qubit can be treated independently as long as they are far-detuned from each other, when $|\omega_i - \omega_j| \gg \Gamma_{i,i}, \Gamma_{j,j}$, where $\omega_i$ is the qubit $i$'s frequency, and $\Gamma_{i,i}$ is its self-decoherence rate due to the transmission line. The vector-network analyzer (VNA) measures the transmission of coherent signals $\langle V_{\rm out} \rangle/ \langle V_{\rm in} \rangle$. The transmission coefficient is \cite{Mirhosseini2019, Kannan2020}
\begin{equation}\label{eq:transcoef}
  t = 1 - \frac{\xi \Gamma_1}{2\Gamma_2}\frac{1-\frac{i\Delta}{\Gamma_2}}{1+\left( \frac{\Delta}{\Gamma_2}\right)^2 + \frac{\Omega^2}{\Gamma_1 \Gamma_2}}.
\end{equation}

\begin{figure}
\centering
\includegraphics[width=0.5\textwidth]{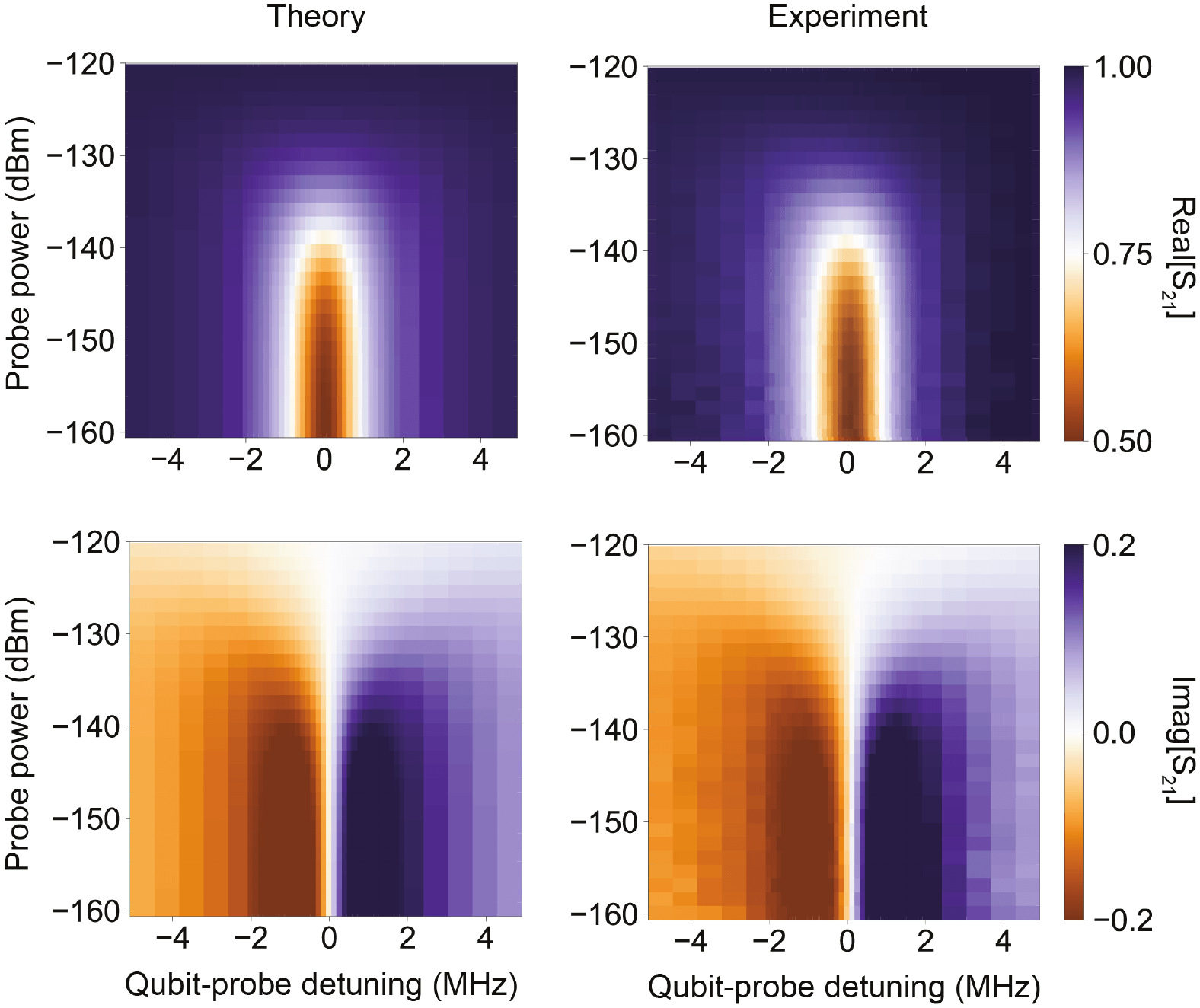}
% \captionsetup{labelformat=empty}
\caption{\textbf{$|$ Transmission scan of a qubit as a function of input power and qubit-probe detuning.} Real and imaginary parts of the experimental data (right) and theory (left) are plotted together as a comparison.}
\label{fig:otl2Dscan}
\end{figure}

Using this equation, we can calibrate the absolute power at the device with independently measured parameters: $\Gamma_1$ is the spontaneous emission rate of the qubit into the transmission line, $\Gamma_2 = \Gamma_1/2 + \Gamma_\phi$ is the transverse decoherence of the qubit, and $\Gamma_\phi$ is the qubit dephasing rate; $\Omega$ is the drive amplitude in the unit of Hz as seen by the qubit. These three are the fitting parameters that can be extracted from the 2D plot shown in \cref{fig:otl2Dscan}. Moreover, $\Delta$ is the qubit-drive detuning; $\xi$ is the ratio of emission to the waveguide compared to all loss channels, and within the SNR of the data, it is assumed to be unity, since the qubit is considered to be strongly coupled to the waveguide so that the decay into the waveguide dominates all the other decay channels. Finally, the drive power at the qubit is given by~\cite{Mirhosseini2019} 
\begin{equation}\label{eq:powerotl}
P = \pi \hbar \omega_i \Omega^2/2\Gamma_1,
\end{equation}
which can be used to plot the transmission coefficient versus power by substituting $\Omega$ in the equation. Note that the transmission coefficient through the qubit is normalized by subtracting the background (without the qubit resonance, determined by detuning the qubit away).

\begin{figure}[!htbp]
\centering
\includegraphics[width=0.9\textwidth]{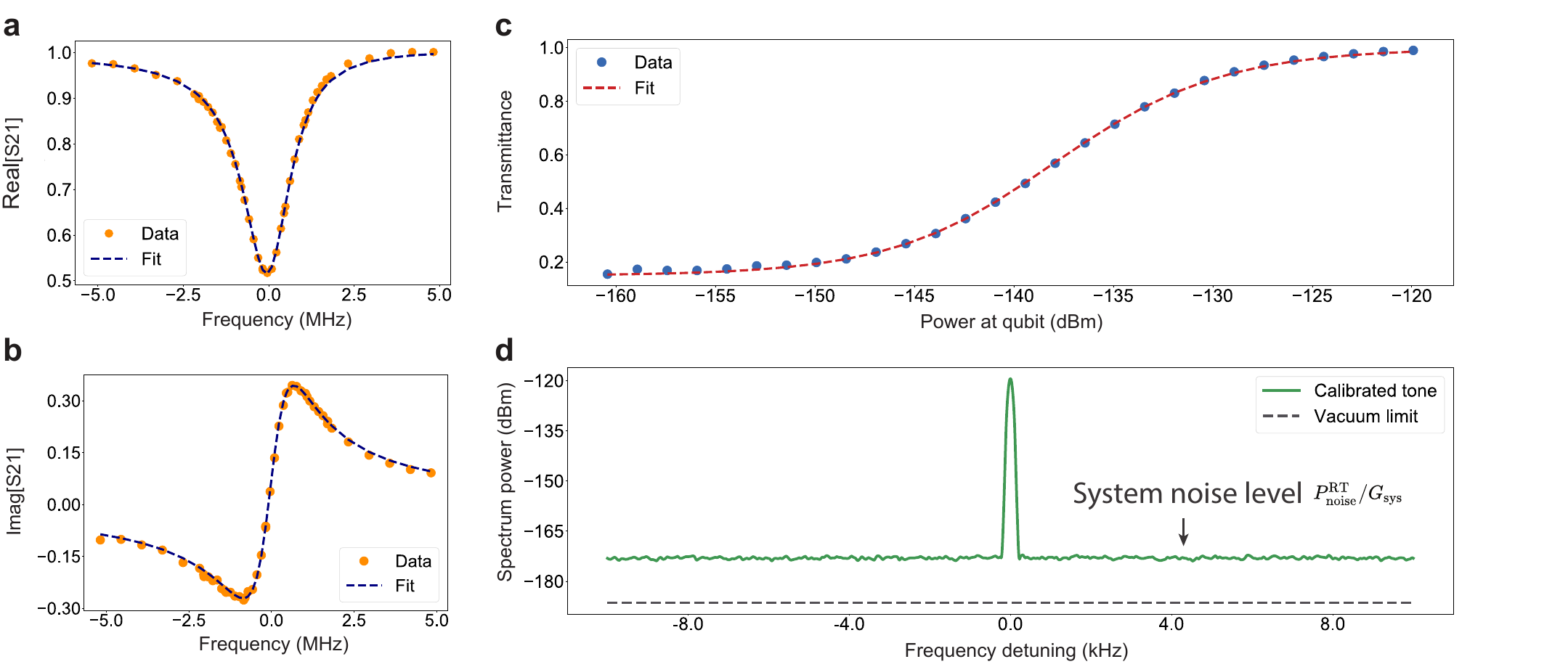}
% \captionsetup{labelformat=empty}
\caption{\textbf{$|$ Transmission profiles and photon number calibration.} Experimental and theory fits are plotted together. \textbf{a. \& b.} Real and imaginary part of the transmission coefficient as a function of input power. \textbf{c.} Resonant transmittivity as a function of input power. \textbf{d.} System noise extracted using a spectrum analyzer. The system noise is shown as a raised noise floor from the vacuum limit.}
\label{fig:otlrealimag}
\end{figure}

As shown in \cref{fig:otlrealimag}(c), we fit the data to equation \cref{eq:transcoef}. Next, we perform the same VNA measurement while also sweeping the input power. The input of a coherent state $|\alpha\rangle$ is mostly reflected at low power ($\bar{n} = |\alpha|^2 \ll 1$) due to interference between the input field and the qubit emission. As we increase the input power, the coherent state $|\alpha\rangle$ will have more contributions from higher number states $|n\rangle $, where $(n>1)$, while the qubit can only perfectly reflect up to a single photon. As a result, the resonant transmission increases and approaches unity at sufficiently high power. The power dependence of the transmission calibrates the absolute power at the qubit, which enables us to further calibrate the noise power.

\FloatBarrier

$S_{21}$ measured by a VNA corresponds to the complex transmission $t$ as defined in~\cref{eq:transcoef}. \cref{fig:otlrealimag} compares the real and imaginary parts  of the data (points) with the theory (line). Fitting is performed over the entire 2D scan, as shown in \cref{fig:otl2Dscan}. In \cref{fig:otlrealimag}(c), we show the transmittance $|t|^2$ as a function of power at zero frequency detuning $\Delta = 0$ from the resonance. Fitting the entire 2D scan enables us to extract $\Omega$ and $\Gamma_1$. Using \cref{eq:powerotl}, we can extract powers at the qubit given the preset powers at refrigerator input at room temperature. As a result, this method also gives us the information for the setup input attenuation from the signal source to the qubit.

\FloatBarrier
To calibrate the system noise level, we first extract the system gain by sending a calibrated input field through the qubit-waveguide system --- $P^{\rm MXC}_{\rm OTL}$ --- power at the wQED reference plane at the mixing chamber (MXC), and measure its output $P^{\rm RT}_{\rm OTL}$ --- power at room temperature using a spectrum analyzer~\cite{Macklin307}. 
The system gain $G_{\rm sys} = P^{\rm RT}_{\rm OTL}/P^{\rm MXC}_{\rm OTL}$ is then used to obtain the system noise temperature
\begin{equation}\label{eq:ntcal}
T_{\rm sys} = P^{\rm RT}_{\rm noise}/G_{\rm sys}k_{\rm B}B,  
\end{equation}
where $P^{\rm RT}_{\rm noise}$ is the noise level measured at the spectrum analyzer. At frequency $\omega = \unit[6.7]{GHz} \times 2\pi$, $P^{\rm RT}_{\rm noise}$ = \unit[-109.63]{dBm}, $G_{\rm sys}$ = \unit[65.06]{dB} and measurement bandwidth $B = \unit[100]{Hz}$, giving a system noise temperature $T_{\rm sys}$ = \unit[2.46]{K} using~\cref{eq:ntcal}, which is equivalent to a measurement efficiency $\eta_{\rm meas} = \hbar \omega/2k_{\rm B}T_{\rm sys} = \unit[6.53]{\%}$.   

\subsection*{Results Comparison Between the wQED (primary) and the SNTJ (secondary) Calibration Methods}
\begin{figure}[!htbp]
\centering
\includegraphics[width=0.99\textwidth]{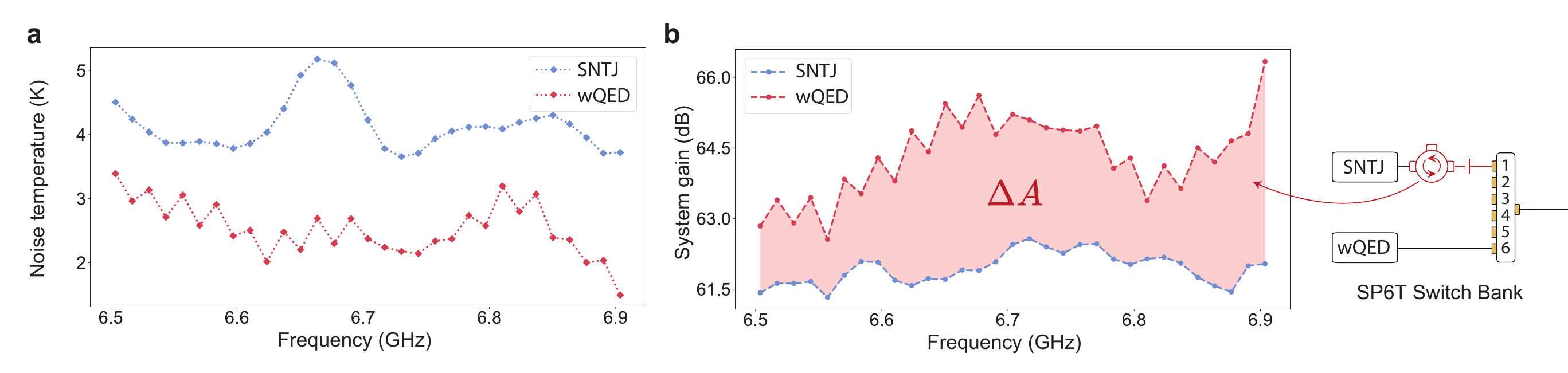}
% \captionsetup{labelformat=empty}
\caption{\textbf{$|$ Uncorrected noise temperature characterization and system gain.} \textbf{a.} Noise temperature as a function of frequency from \unit[6.5]{GHz} to \unit[6.9]{GHz} measured using the SNTJ and the wQED device separately. \textbf{b.} System gain measured using the two methods. Due to the finite difference in their RF transmission $\Delta A$, the extracted system gains are different. In other words, we are using the wQED (primary) to calibrate the SNTJ and isolator loss (secondary). We use this to correct the system noise in panel (a) independently.}
\label{fig:NTuncorrected}
\end{figure}

\begin{figure}[!htbp]
\centering
\includegraphics[width=0.875\textwidth]{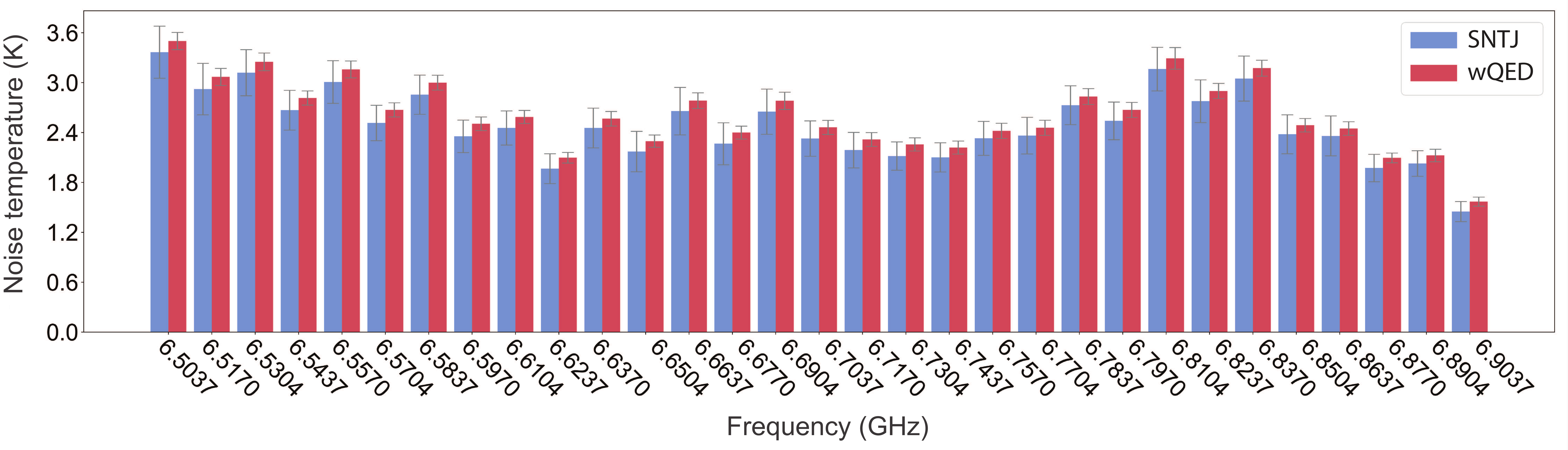}
% \captionsetup{labelformat=empty}
\caption{\textbf{$|$ SNTJ calibrated using wQED.} The SNTJ noise temperature has been corrected here while the wQED measurement values remain unchanged. The SNTJ data are presented as mean values of 20 sets of repeated measurements (each with 2000 samples) +/- standard error of the mean. The wQED data show averaged results for 14 sets of repeated measurements (each with 10000 samples) presented as mean values +/- standard deviation.}
\label{fig:NTcorrected}
\end{figure}

We have employed two different methods --- the primary wQED qubit power calibration technique and the secondary SNTJ method to cross-check the measurement results. We perform the noise temperature characterization using both methods from \unit[6.5]{GHz} to \unit[6.9]{GHz} as shown in \cref{fig:NTuncorrected}. Given the identical setup after the SP6T switch, the difference between the two curves most likely arises from the insertion loss $\Delta A$ imposed by the additional components required to operate the SNTJ (highlighted in red color in the figure). Based on this assumption, the overestimated noise temperature can be corrected by accounting for $\Delta A$ and scaling the noise temperature accordingly. The adjusted results can be seen from \cref{fig:NTcorrected}. In other words, we calibrate the SNTJ using the wQED. The latter, in principle, gives a more accurate system noise characterization for the squeezing measurement without additional circuit components as employed for the former. In this experiment, one drawback of the wQED method is its limited frequency range. However, it can be readily addressed with different qubit designs to fit a particular frequency band.

\FloatBarrier

\newpage
\section{JTWPA Performance}
\subsection*{1-dB Compression Point}
\begin{figure}[!htbp]
\includegraphics[width=0.75\textwidth]{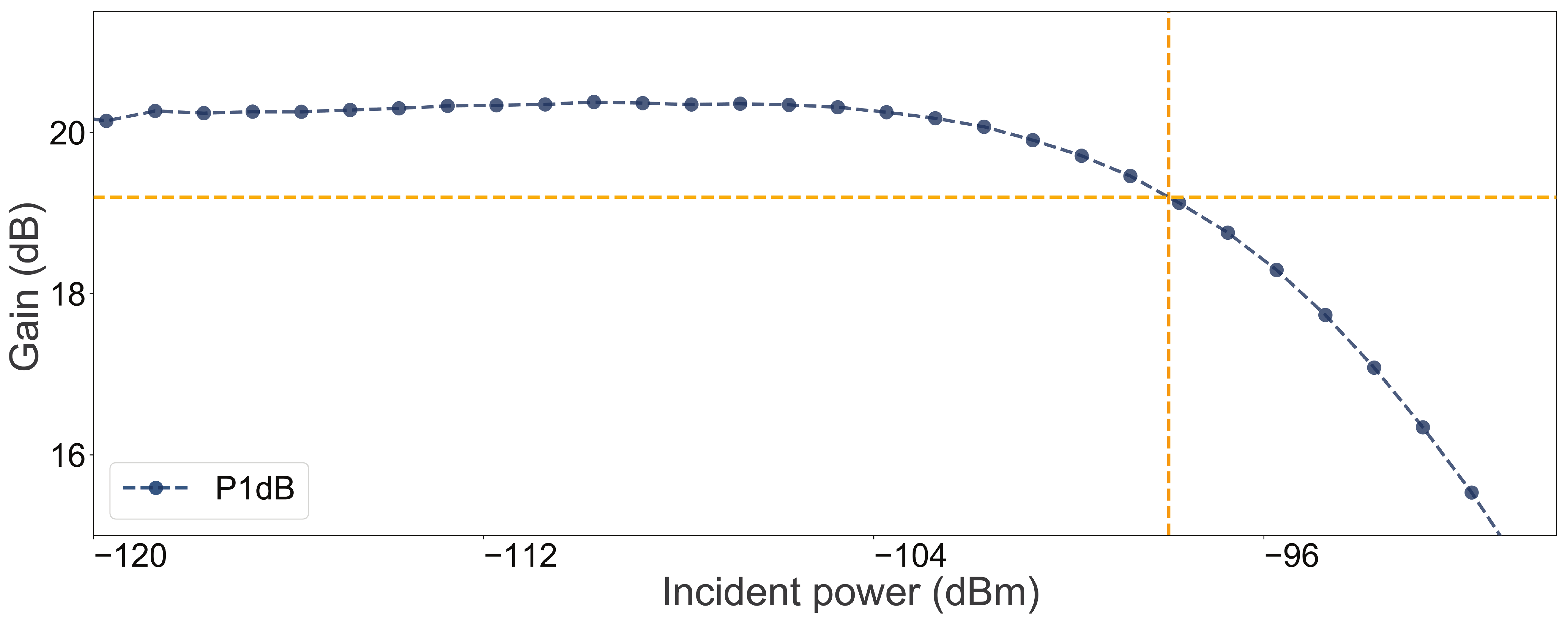}
% \captionsetup{labelformat=empty}
\caption{\textbf{$|$ 1-dB compression point.} JTWPA gain at \unit[6.37]{GHz} as a function of incident power. The yellow crosshair indicates the 1-dB compression point.}
\label{fig:p1dB}
\end{figure}
\FloatBarrier

The 1-dB compression point (P1dB) refers to the incident signal power level that causes the amplifier gain to deviate (decrease) by \unit[1]{dB} from its value at low power. \cref{fig:p1dB} is measured at the signal frequency at \unit[6.70]{GHz}. The input signal power is swept and while the pump powers are fixed. We extract a P1dB value of \unit[-98]{dBm}, on par with the value reported for a single-pump JTWPA~\cite{Macklin307}.

\subsection*{Wavevector}
\begin{figure}[!htbp]
\includegraphics[width=0.725\textwidth]{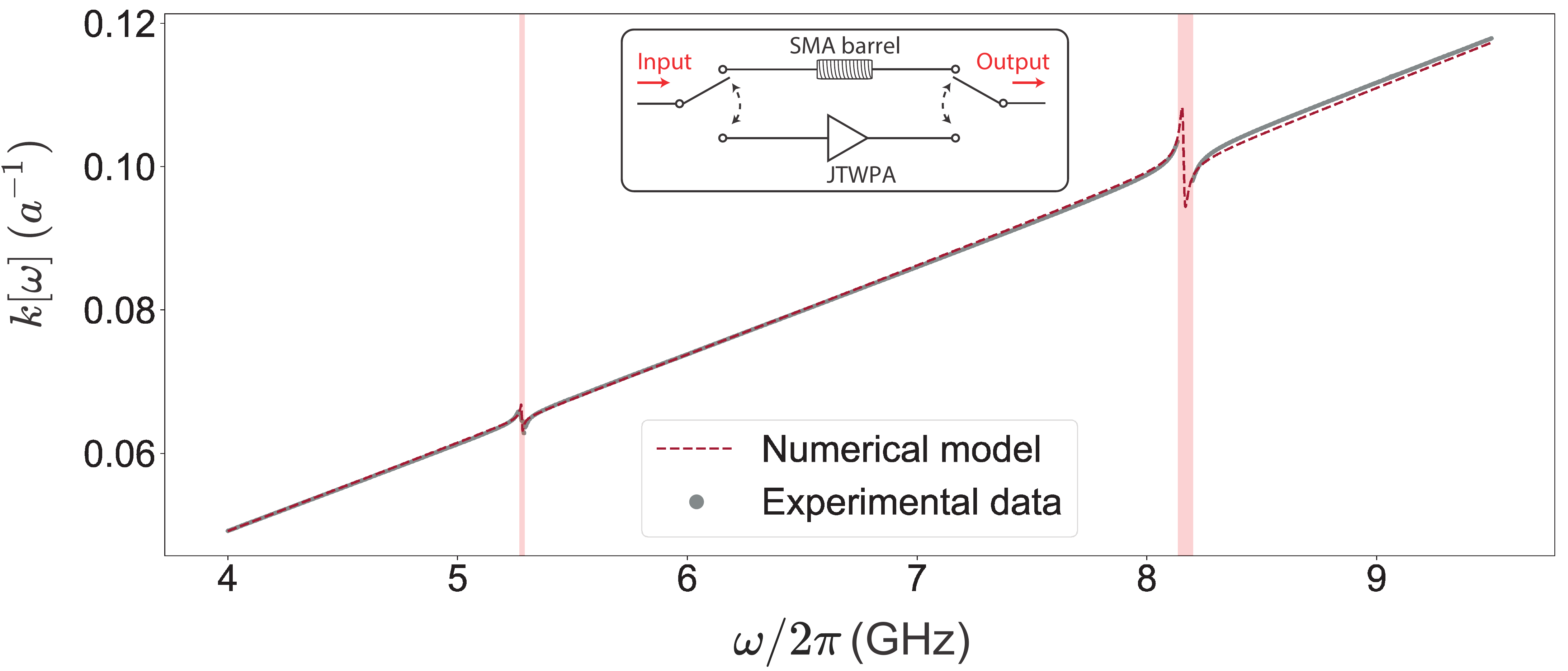}
%\captionsetup{labelformat=empty}
\caption{\textbf{$|$ Experimental data and numerical model for the JTWPA wavevector.} The wavevector is experimentally characterized by measuring the phase of the transmission through the JTWPA when the pumps are turned off. The numerical model is calculated from the dispersion relation (see~\cref{eq:disprel} using fitted circuit parameters to match the experimental data. Crimson-shaded areas are regions where the phase inside the JTWPA cannot be fully resolved because the phase-matching resonators act as stop-band filters and reflect the probe back; the transmission is dominated by instrumental noise within the stopband. The inset shows the measurement schematic using a switch to calibrate the wavevector of a JTWPA.}
\label{fig:wavevector}
\end{figure}

To characterize the phase-matching condition of a nonlinear process, we need to characterize the JTWPA dispersion as a function of frequency. The wavevector of a JTWPA can be extracted relative to that of a SMA coaxial through-line shown in \cref{fig:wavevector}. This measurement scheme aims to single out only the phase change induced by the JTWPA itself. To be more specific, the real part of the wavevector is measured via the phase of the transmitted field. However, due to the presence of other microwave components, the phase response incorporates an additional frequency-dependent phase $\phi_0$, such that

\begin{equation} \label{eq:JTWPAphase}
\phi_{\rm JTWPA} + \phi_0 = k(\omega)L,
\end{equation}
where $k$ and $L$ are the wavevector and length of the device. In contrast, the phase response from the through-line is:
\begin{equation}\label{eq:thruphase}
\phi_{\rm thru} + \phi_0 = k_{\rm thru}L_{\rm thru}
\end{equation}
By subtracting \cref{eq:JTWPAphase} from \cref{eq:thruphase}, we can ideally eliminate the offset $\phi_0$. Furthermore, we can back out $\phi_{\rm JTWPA}$ (which includes the effects from wirebonds, coplanar waveguide boards, etc.) given the through-line phase at DC is zero and the phase is approximately linear in our frequency range of interest \cite{Macklin307}.

\subsection*{JTWPA Insertion Loss}

\begin{tcolorbox}
\tcbsidebyside[sidebyside adapt=both,
enhanced,center,
title= Circuit parameters for a dual-dispersion JTWPA,
attach boxed title to top center={yshift=-2mm}, coltitle=black,boxed title style={colback=orange!25}, segmentation style=solid,colback=orange!5,colframe=gray!50!white
]{\begin{tabular}{LL}
\centering
N (\text{unit cells}) & 3141 \\
C_{\rm g} (\text{capacitance to ground}) & 28.616\, \text{fF} \\
I_{\rm c} (\text{junction critical current}) & 3.14\, \mu\text{A} \\
\hline
\multicolumn{2}{c}{Phase-matching LC resonator (PMLC) 1} \\
\hline\hline
\omega_{\rm r1} (\text{PMLC 1 frequency}) & 5.2815 \times 2\pi\, \text{GHz} \\
C_{\rm r1} (\text{PMLC 1 capacitance}) & 6.653\, \text{pF} \\
C_{\rm c1} (\text{coupling capacitance of PM LC 1})& 28.616\, \text{fF} \\
\hline
\multicolumn{2}{c}{Phase-matching LC resonator (PMLC) 2} \\
\hline\hline
\omega_{\rm r2} (\text{PMLC 2 frequency}) & 8.169 \times 2\pi \, \text{GHz} \\
C_{\rm r1} (\text{PMLC 2 capacitance}) & 2.781\, \text{pF} \\
C_{\rm c2} (\text{coupling capacitance for PM LC 2})& 28.616\, \text{fF} \\
\hline
\multicolumn{2}{c}{Pump frequencies} \\
\hline\hline
\Omega_{\rm 1} (\text{pump 1 frequency}) & 5.2984 \times 2\pi \, \text{GHz} \\
\Omega_{\rm 2} (\text{pump 2 frequency}) & 8.109 \times 2\pi \, \text{GHz} 
\end{tabular} }{\includegraphics[width=0.4\textwidth]{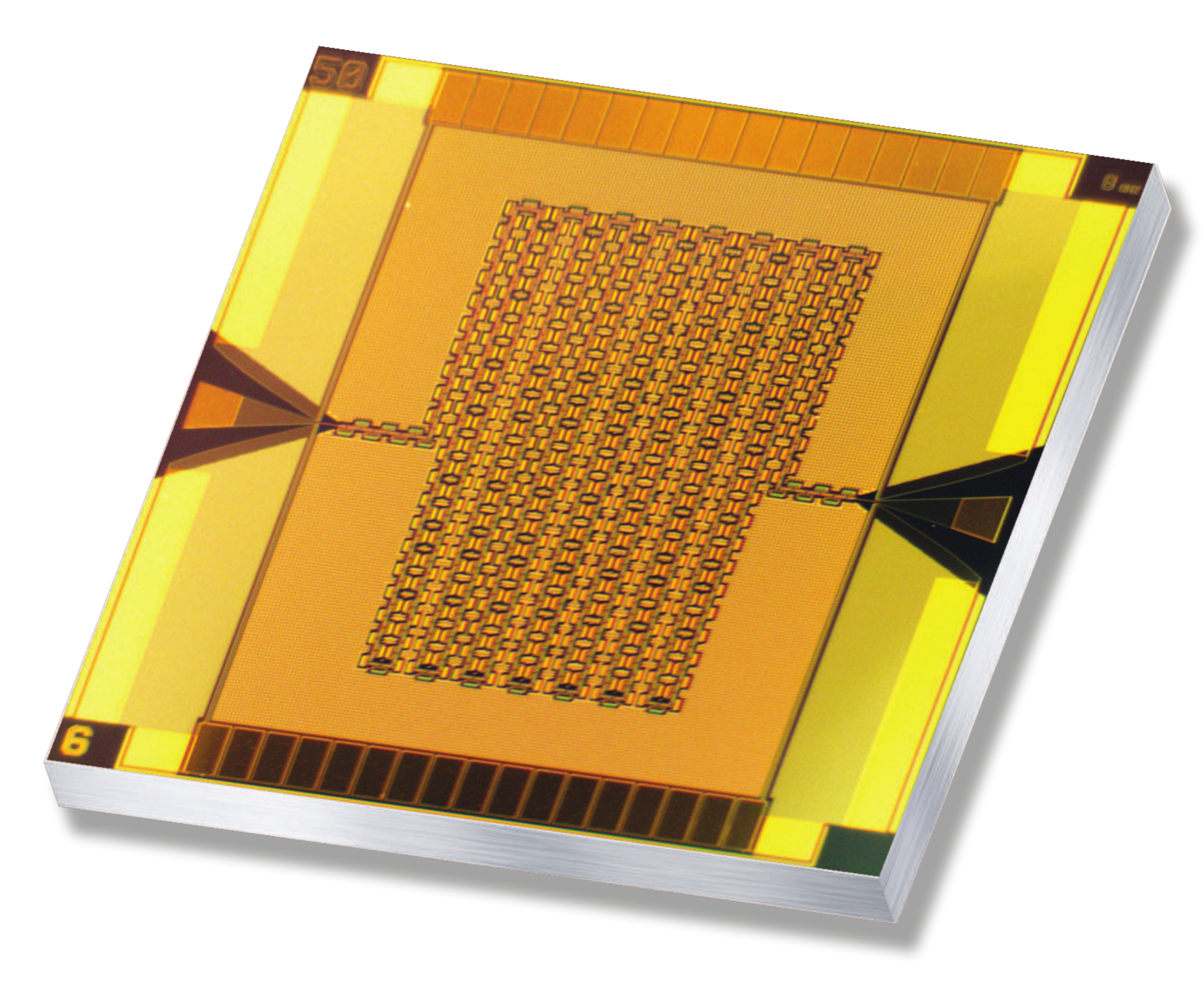}}
\end{tcolorbox}

\begin{figure}[!htbp]
\includegraphics[width=0.835\textwidth]{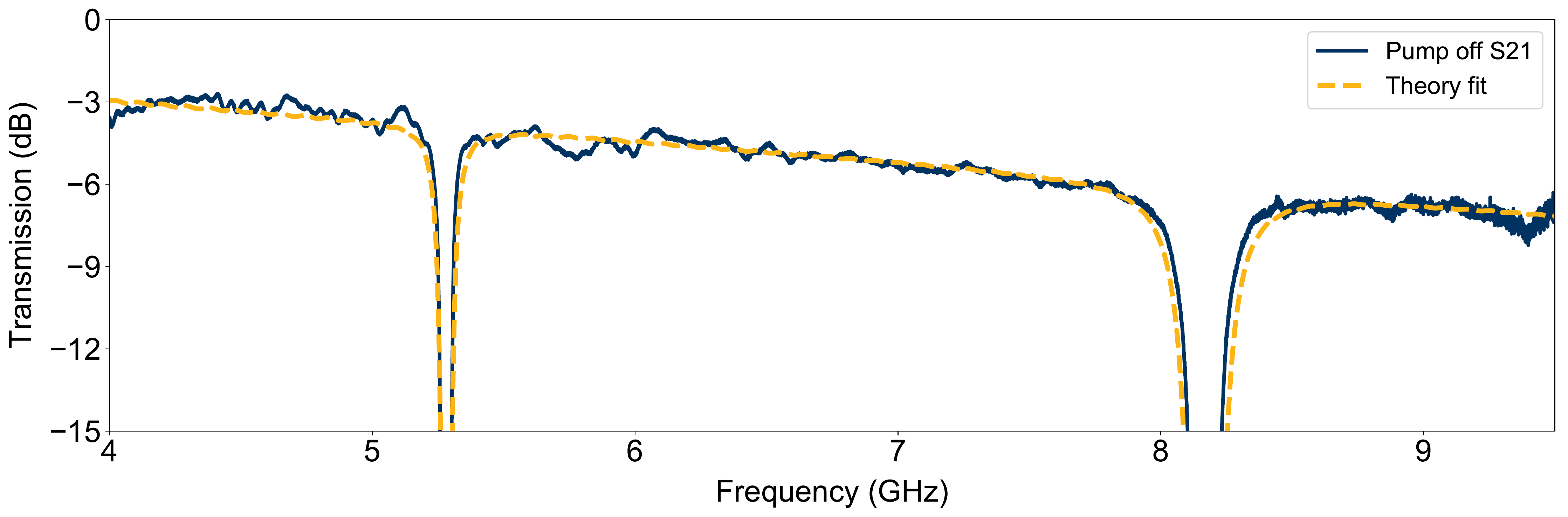}
%\captionsetup{labelformat=empty}
\caption{\textbf{$|$ JTWPA insertion loss and theory fit.} The same plot displayed in Fig. 2a in the main text with a fitted theoretical curve.}
\label{fig:JTWPAloss}
\end{figure}
\FloatBarrier

Utilizing the JTWPA circuit parameters obtained from modeling the measured wavevector in~\cref{fig:wavevector}, we fit the measured insertion loss $|e^{i\tilde{k}(\omega)L}|$ using a single parameter of loss tangent $\tan \delta$ of $4.9\times 10^{-3}$

% \begin{equation} \label{eq:ktheorynew}
% \tilde{k}(\omega) = \frac{\omega \sqrt{L_g C_g (1-i\tan \delta)}}{\sqrt{1-\omega^2 L_J C_J}}

% \begin{align}\label{eq:ktheorynew}

\begin{equation} \label{eq:ktheorynew}
\tilde{k}(\omega) = \frac{\omega \sqrt{L_g C_g (1-i\tan \delta)}}{\sqrt{1-\omega^2 L_J C_J}}\xi,
\end{equation}
where $$\xi = \sqrt{1-\frac{C_{c1}+C_{c2}}{10C_g} + \frac{C_{c1}}{10C_g}\frac{1-\omega^2L_{r1}C_{r1}(1-i\tan \delta)}{1-\omega^2L_{r1}(C_{r1}+C_{c1})(1-i\tan \delta)} +\frac{C_{c2}}{10C_g}\frac{1-\omega^2L_{r2}C_{r2}(1-i\tan \delta)}{1-\omega^2L_{r2}(C_{r2}+C_{c2})(1-i\tan \delta)}},$$ $\omega$ is the frequency, $L_J$ is the junction inductance, and $L_r$ is the phase-matching resonator inductance. The expression of $\tilde{k}(\omega)$ above is obtained from the lossless formula~\cref{eq:disprel} by replacing every capacitor term $C \rightarrow C(1-i\tan \delta)$ to account for the dielectric loss tangent of the capacitors using a parallel RC model. In addition, the 1/10 factors appearing in the resonance terms account for the fact that each set of phase-matching resonators is only inserted once every ten unit cells.

\subsection*{Phase Mismatch for Different Processes}
To understand the phase mismatch quantitatively as a function of pump power $P_1$ and $P_2$, we define an effective power-dependent phase mismatch for the parametric amplification (PA) process,
\begin{equation}\label{eq:pm_13p907}
\Delta k(P_1, P_2) = (2k_{\omega_s} - k_{\Omega_1} -k_{\Omega_2}) + \left[2k_{\omega_s}(P_1) - k_{\Omega_1}(P_1)- k_{\Omega_2}(P_1)\right] + \left[ 2k_{\omega_s}(P_2) - k_{\Omega_1}(P_2)- k_{\Omega_2}(P_2) \right],
\end{equation} 
where the quantity in the first parentheses is the ``bare'' (linear) phase mismatch due to the linear dispersion of the JTWPA measured at low probe power (below single-photon level), the second group of terms in the bracket represents the pump 1 induced Kerr modulation, and the third set is the pump 2 power induced phase shifts. We need to account for all of these terms as we operate the device in the dual-pump scheme. 

\begin{figure}[!htbp]
\includegraphics[width=0.775\textwidth]{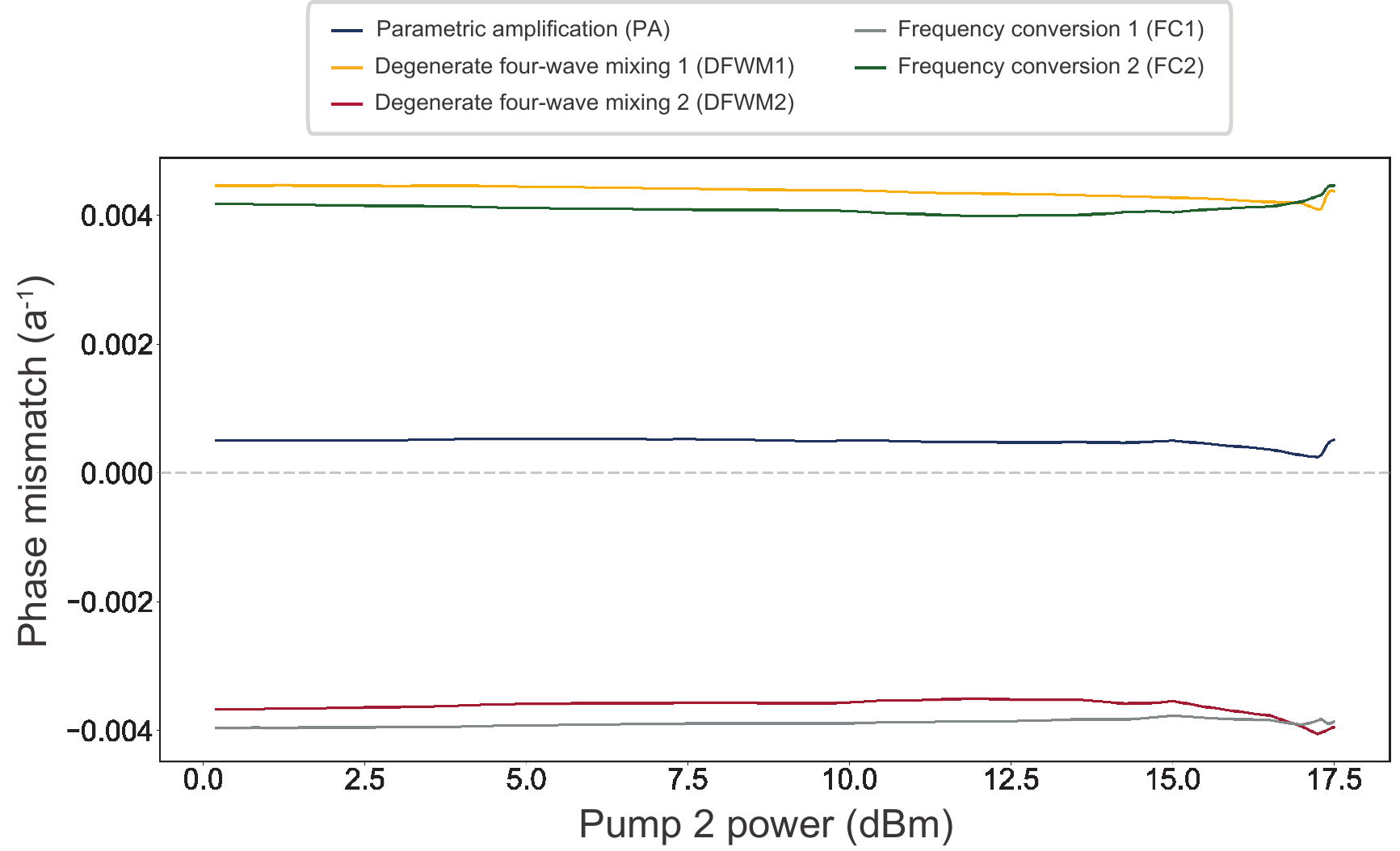}
%\captionsetup{labelformat=empty}
\caption{\textbf{$|$ Phase mismatch for different nonlinear processes.} The phase mismatch is characterized as a function of pump 2 power with a fixed pump 1 power (the same power sweep performed in the squeezing measurement). The unwanted processes are highly phase-mismatched compared to the desired parametric amplification process (blue).}
\label{fig:PM_all_processes}
\end{figure}

\begin{figure}[!htbp]
\includegraphics[width=0.6\textwidth]{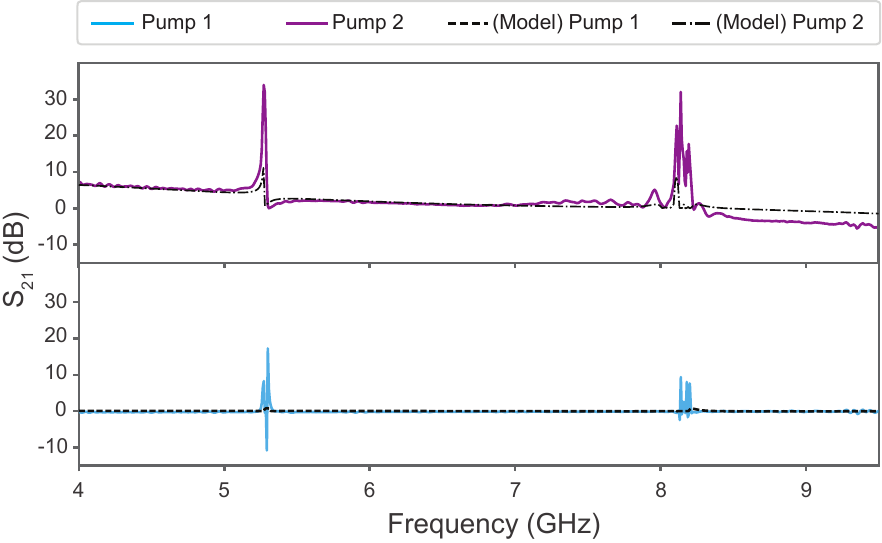}
%\captionsetup{labelformat=empty}
\caption{\textbf{$|$ Single pump gain profiles.} Single pump gain and numerical simulations that fit and predict the gain profiles. The deviation between the model and the experimental data is expected to be caused by the bandwidth constraint of cryogenic components such as the isolator/circulator.}
\label{fig:gainother}
\end{figure}

Similarly, we consider other two-pump-photon nonlinear processes that could lead to degradation in squeezing performance~\cite{peng2022}. There are four major parasitic processes, and their corresponding phase mismatch are power-dependent as well and expressed as the following: 
\setlength{\leftmargini}{1cm}
\begin{itemize}
    \item Degenerate four-wave mixing 1 (DFWM1): $\Delta k _{\rm DFWM1}(P_1, P_2) = k_{\omega_s}(P_1, P_2)+k_{\omega_{i1}}(P_1, P_2)-2k_{\Omega_1}(P_1, P_2)$
    \item Degenerate four-wave mixing 2 (DFWM2): $\Delta k_{\rm DFWM2}(P_1, P_2) = k_{\omega_s}(P_1, P_2)+k_{\omega_{i2}}(P_1, P_2)-2k_{\Omega_2}(P_1, P_2)$
    \item Frequency conversion 1 (FC1): $\Delta k _{\rm FC1}(P_1, P_2) = k_{\omega_s}(P_1, P_2)-k_{\omega_{i1}}(P_1, P_2)-\left(k_{\Omega_{2}}(P_1, P_2)-k_{\Omega_1}(P_1, P_2)\right)$
    \item Frequency conversion 2 (FC2): $\Delta k_{\rm FC2}(P_1, P_2) = k_{\omega_s}(P_1, P_2)-k_{\omega_{i2}}(P_1, P_2)-\left(k_{\Omega_{1}}(P_1, P_2)-k_{\Omega_2}(P_1, P_2)\right)$
\end{itemize}

with frequencies:
%\begin{block}
\setlength{\leftmargini}{6.5cm}
\begin{itemize}
    \item $\Omega_1$: pump 1 frequency
    \item $\Omega_2$: pump 2 frequency
    \item $\omega_s = \frac{\Omega_1 + \Omega_2}{2}$
    \item $\omega_{i1} = 2\Omega_1 - \omega_s$
    \item $\omega_{i2} = 2\Omega_2 - \omega_s$
\end{itemize}
%\end{block}
Here $k_{\omega}$ represents the measured wavevector that accounts for both the linear dispersion and the nonlinear phase modulations from the pumps. It is related to a measurement of a phase $\Delta \phi(\omega)$ using $\Delta \phi(\omega) \sim k(\omega) \times L$. We perform a measurement as shown in \cref{fig:PM_all_processes}, demonstrating the effectiveness of our dispersion engineering technique to suppress undesirable processes while achieving large dual-pump gain. As an example shown in \cref{fig:gainother},  we suppress (highly phase-mismatch) the undesirable DFWM processes (DFWM1 and DFWM2), leading to minimum gain. The results are illustrated by the blue and purple curves, representing the phase-preserving-parametric gain when only a single pump is turned on. They are in stark contrast to the broadband gain, as demonstrated in Fig. 2b with the same pump parameters.

\FloatBarrier

\section*{Numerical modeling of squeezing}

\subsection{Linearized input-output theory}

\begin{figure}
\centering
\includegraphics[width=0.9\textwidth]{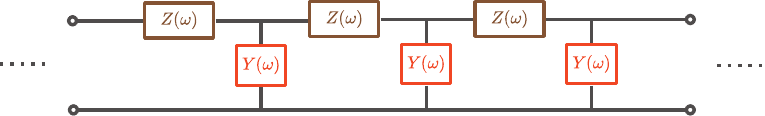}
\caption{\label{fig:theory:line}\textbf{$|$ Illustration of linear impedance $Z(\omega)$ and admittance $Y(\omega)$ for an arbitrary linear transmission line.}}
\end{figure}

In this section, we describe the numerical models of JTWPA squeezing. Following the approach from Ref.~\cite{Grimsmo17}, we derive a Hamiltonian for the JTWPA in the continuum limit where the unit cell distance $a\to 0$ such that the total length $z=Na$ is held constant
\begin{equation}
  \hat H = \hat H_0 + \hat H_1,
\end{equation}
where to fourth order in the Josephson junction potential we have
\begin{align}
    \hat H_0 ={}& \int_{0}^\infty \dd \omega \hbar \omega \hat a_{\omega}^\dagger\hat a_{\omega},\\
    \hat H_1 ={}& -\frac{\gamma}{2} \int_{0}^{z} \dd x [\partial_x\hat \phi(x)]^4.
\end{align}
Here $\hat a_{\omega}\dg$ creates a delocalized right-moving photon of energy $\hbar\omega$,
$\gamma = a^3 E_J \left(\frac{2\pi}{\Phi_0}\right)^4/12$ is a parameter that describes the strength of the non-linearity, and
\begin{equation}
    \hphi(x) = \int_{0}^\infty \dd\omega \sqrt{\frac{\hbar Z_0}{4\pi k_{\omega} v}}\e^{i k_{\omega} x}
    \ha_{\omega} + \text{H.c.},
\end{equation}
is the flux field along the JTWPA. For simplicity, we only consider the right-moving part of the field, under the assumption the input/output transmission lines are well impedance matched and that back-scattering is negligible.
We have moreover introduced a characteristic impedance $Z_0 = \sqrt{l/c}$ and nominal speed of light $v=1/\sqrt{lc}$, with $c = C_0/a$ and $l = \Phi_0/(2\pi I_c a)$, the capacitance to ground and inductance per unit length, respectively.
The dispersion relation for the wavenumber $k_\omega$ is given by the series impedance $Z(\omega)$ and parallel admittance to ground $Y(\omega)$ of each unit cell~\cite{OBrien2014,Grimsmo17} (see~\cref{fig:theory:line}).
\begin{equation}\label{eq:disprel}
  % k_\omega  = \sqrt{\frac{-i\omega y(\omega) l}{1-\omega^2/\omega_p^2}}.
 k_\omega a  = -i\sqrt{Z(\omega) Y(\omega)}.
\end{equation}

We linearize the problem by assuming a strong right-moving classical pump and replace $\hat a_\omega \to \hat a_\omega + b(\omega)$, with $b(\omega)$ the pump amplitude, and neglect terms higher than second order in $\hat a_\omega^{(\dagger)}$, as well as the influence of the quantum fields on the pump. Moreover, dropping fast rotating terms, we have a Hamiltonian
\begin{equation}
\hH = \hH_0 + \hH_{\text{fc}} + \hH_{\text{sq}},
\end{equation}
where
\begin{align}
    \hH_{\fc} ={}& -\frac{\hbar}{2\pi} \int_0^\infty\dd\omega_1\dd\omega_2\Phi_\fc(\omega_1,\omega_2) \ha_{\omega_1}^\dagger \ha_{\omega_2}  + \text{H.c.},\\
    \hH_{\text{sq}} ={}& -\frac{\hbar}{4\pi} \int_0^\infty\dd\omega_1\dd\omega_2\Phi_\text{sq}(\omega_1,\omega_2) \ha_{\omega_1}^\dagger \ha_{\omega_2}^\dagger  + \text{H.c.}
\end{align}
describes frequency conversion and photon pair creation, respectively.
For notational convenience, we have defined phase matching functions
\begin{align}
  &\begin{aligned}
    \Phi_\fc&(\omega_1,\omega_2) =\sqrt{k_{\omega_1}k_{\omega_2}} \int_0^\infty\dd\Omega_1\dd\Omega_2
    \times \int_{0}^{z} \dd x \beta(\Omega_1)^*\beta(\Omega_2) \e^{i\left( k_{\omega_1}-k_{\omega_2}+k_{\Omega_1}-k_{\Omega_2} \right) x}
  \end{aligned}\\
  &\begin{aligned}
    \Phi_\text{sq}&(\omega_1,\omega_2) =\sqrt{k_{\omega_1}k_{\omega_2}} \int_0^\infty\dd\Omega_1\dd\Omega_2
    \times \int_{0}^{z} \dd x \beta(\Omega_1)\beta(\Omega_2) \e^{i\left( k_{\omega_1}+k_{\omega_2}-k_{\Omega_1}-k_{\Omega_2} \right) x},
  \end{aligned}
\end{align}
where $\beta(\omega) = \sqrt{3 \gamma l}\sqrt{\hbar Z_0 k_\omega/4\pi v} b(\omega)$
is a rescaled pump amplitude with units of inverse frequency. The pump amplitude can be related to the pump current as~\cite{Grimsmo17}
\begin{equation}
\beta(\omega) = \frac{I_p(\omega)}{4I_c},
\end{equation}
where the current is defined as 
$I_p(\omega) = \sqrt{\hbar Z_0/4\pi k_\omega v} b(\omega)/l$ and we used that $\gamma l^3 = 1/12 I_c$.

Similarly, the classical pump Hamiltonian can be written
\begin{equation}
  H_p = H_{p0} + H_{p1}
\end{equation}
with
\begin{align}
    &H_{p0} = \frac{16\pi E_J}{a} \int_0^\infty \frac{\dd\omega \omega}{k_\omega} \beta(\omega)^* \beta(\omega),\\
    &\begin{aligned}
    H_{p1} =& -\frac{4E_J}{a} \int_0^\infty\frac{\dd\omega_1\dd\omega_2\Phi_\fc(\omega_1,\omega_2)}{\sqrt{k_{\omega_1}k_{\omega_2}}}
   \beta(\omega_1)^* \beta(\omega_2)
    +\text{c.c.}
    \end{aligned}
\end{align}

To simplify the problem, we consider the steady-state solution by going to an interaction picture with respect to $\hat H_0 + H_{p0}$ and integrating from an initial time $t_0 = -\infty$ to final time $t_1=\infty$~\cite{Grimsmo17,Quesada14}.
Moreover, we take the pump to be a sum of two delta functions in frequency $\beta(\omega) = \sum_{p=1}^2 \beta_p \delta(\omega-\Omega_p)$, with $\beta_p$ a dimensionless pump amplitude.

In the scattering limit, we find position-dependent equations of motion for the pump and the quantum fields~\cite{Grimsmo17}. Specifically,
\begin{equation}\label{eq:sm:eomp}
    \frac{\dd\beta_p}{\dd x} = i k_{p}\left( |\beta_p|^2 + 2|\beta_q|^2 \right)\beta_p,
\end{equation}
with $p=1,2$, $q\neq p$ $k_p \equiv k_{\Omega_p}$ and
\begin{multline}\label{eq:sm:eom}
    \frac{\dd\ha_{\omega}}{\dd x}
    = 2i\sum_{p=1}^2\lambda_{pp}^\fc(\omega)\e^{-i\Delta k_{pp}^\fc(\omega)x} \ha_{\omega}
    +2i\sum_{p\neq q}\lambda_{pq}^\fc(\omega)\e^{-i\Delta k_{pq}^\fc(\omega)x} \ha_{\omega+\Omega_p-\Omega_q}
    +i \sum_{p,q=1}^2 \lambda_{pq}^\text{sq}(\omega)\e^{-i\Delta k_{pq}^\text{sq}(\omega)x} \hat a_{\Omega_p+\Omega_q-\omega}^\dagger,
\end{multline}
where
\begin{subequations}\label{eq:sm:couplings}
\begin{align}
  \lambda_{pq}^\fc(\omega) ={}& \beta_p^*\beta_q \sqrt{k_\omega k_{\omega+\Omega_p-\Omega_q}},\\
  \lambda_{pq}^\text{sq}(\omega) ={}& \beta_p\beta_q \sqrt{k_\omega k_{\Omega_p+\Omega_q-\omega}},\\
  \Delta k_{pq}^\fc(\omega) ={}& -k_\omega + k_{\omega + \Omega_p - \Omega_q} - k_{p} + k_{q},\\
  \Delta k_{pq}^\text{sq}(\omega) ={}& -k_\omega - k_{\Omega_p + \Omega_q  - \omega} + k_{p} + k_{q}.
\end{align}
\end{subequations}

The first term in~\cref{eq:sm:eom} describes cross-phase modulation, which contributes to the phase mismatch.
It is convenient to transform to a rotating frame with respect to this process by defining 
$\hat c_\omega = \hat a_\omega \e^{-2 i \sum_p |\beta_p|^2 k_\omega x}$,
such that we have an equation of motion
\begin{equation}\label{eq:sm:eom2}
  \begin{aligned}
    \frac{\dd\hat c_{\omega}}{\dd x}
    = 2&i\sum_{p \neq q} \lambda_{pq}^\fc(\omega)\e^{-i\tilde \Delta k_{pq}^\fc(\omega)x}  \hat c_{\omega+\Omega_p-\Omega_q}
    +i \sum_{p,q=1}^2  \lambda_{pq}^\text{sq}(\omega)\e^{-i\tilde \Delta k_{pq}^\text{sq}(\omega)x} \hat c_{\Omega_p+\Omega_q-\omega}^\dagger,
  \end{aligned}
\end{equation}
with a non-linear modification to the phase mismatch
\begin{align}
  \tilde\Delta k_{pq}^\fc(\omega) ={}&- \tilde k_\omega + \tilde k_{\omega + \Omega_p - \Omega_q} - \tilde k_{p} + \tilde k_{q},\\
  \tilde\Delta k_{pq}^\text{sq}(\omega) ={}&- \tilde k_\omega - \tilde k_{\Omega_p + \Omega_q  - \omega} + \tilde k_{p} + \tilde k_{q},
\end{align}
where
\begin{subequations}\label{eq:sm:k_nonlin}
\begin{align}
  \tilde k_\omega ={}& \left(1 + 2\sum_p |\beta_p|^2 \right)k_\omega,\\
  \tilde k_{p} ={}& \left(1+|\beta_p|^2 + 2|\beta_q|^2\right)k_{p} \quad (p\neq q).
\end{align}
\end{subequations}

\subsection*{Quantum Loss Model}

We introduce a phenomenological distributed loss model by adding loss terms to~\cref{eq:sm:eom2}~\cite{Caves1987}
\begin{equation}\label{eq:sm:eom_loss}
  \begin{aligned}
    \frac{\dd\hat c_{\omega}}{\dd x}(x)
    = 2&i\sum_{p \neq q} \lambda_{pq}^\fc(\omega)\e^{-i\tilde \Delta k_{pq}^\fc(\omega)x}  \hat c_{\omega+\Omega_p-\Omega_q}(x)
    +i \sum_{p,q=1}^2  \lambda_{pq}^\text{sq}(\omega)\e^{-i\tilde \Delta k_{pq}^\text{sq}(\omega)x} \hat c_{\Omega_p+\Omega_q-\omega}^\dagger(x)\\
    &- \frac{\gamma(\omega)}{2}\hat c_\omega(x) + \sqrt{\gamma(\omega)}\hat c_\text{in}(x),
  \end{aligned}
\end{equation}
where the loss rate $\gamma(\omega)$ has units of inverse length and $\hat c_\text{in}(x)$ describes vacuum input noise coupled to the JTWPA at each position $x$. Similarly, the pump equation of motion is modified to
\begin{equation}\label{eq:sm:eqn30}
    \frac{\dd\beta_p}{\dd x}(x) = i k_{p}\left( |\beta_p(x)|^2 + 2|\beta_q(x)|^2 \right)\beta_p(x)
    - \frac{\gamma_p}{2}\beta_p(x).
\end{equation}

The JTWPA output field is found by integrating the spatial differential equations from $x=0$ to $x=z$, with $\hat c_\omega(0)$ taken to be vacuum input. The pump amplitudes can be solved independently and substituted into~\cref{eq:sm:eom_loss}.
We have the following solution to the pump equation:
\begin{equation}\label{eq:sm:eomp_loss}
    \beta_p(x) = \beta_p(0) \e^{-\frac{\gamma_p x}{2}  -i k_p \left\{ \frac{1}{\gamma_p}\left[|\beta_p(x)|^2 - |\beta_p(0)|^2\right] + \frac{2}{\gamma_q}\left[|\beta_q(x)|^2 - |\beta_q(0)|^2 \right]  \right\}}.
\end{equation}
Note that with this solution, we have
\begin{equation}
    |\beta_p(x)|^2 = |\beta_p(0)|^2 \e^{-\gamma_p x} \Rightarrow \frac{\dd|\beta_p(x)|^2}{\dd x} = -\gamma_p |\beta_p(x)|^2,
\end{equation}
such that we can write
\begin{equation}\label{eq:sm:pumpsol}
    \beta_p(x) = \beta_p(0) \e^{-\frac{\gamma_p x}{2}  +i k_p \left\{ \frac{1}{\gamma_p}(1-\e^{-\gamma_p x})|\beta_p(0)|^2 + \frac{2}{\gamma_q}\left(1-\e^{-\gamma_q x} \right)|\beta_q(0)|^2  \right\}}.
\end{equation}
Differentiating~\cref{eq:sm:pumpsol} gives~\cref{eq:sm:eqn30}. Note that in the limit $\gamma_p\to 0$~\cref{eq:sm:pumpsol} gives
\begin{equation}\label{eq:sm:pumpsol_noloss}
    \beta_p(x) = \beta_p(0) \e^{i k_p x \left\{|\beta_p(0)|^2 + 2|\beta_q(0)|^2 \right\}}.
\end{equation}
\Cref{eq:sm:eom_loss} can now be solved by inserting the solution for $\beta_p(x)$ into the coupling constants $\lambda_{pq}^{\fc,\text{sq}}(\omega)$ and phase mismatch $\Delta k_{pq}^{\fc,\text{sq}}(\omega)$.

\subsection*{Temperature \& Power-Dependent Loss}

\vspace{0pt}
\begin{figure}[!htbp]
\includegraphics[width=0.675\textwidth]{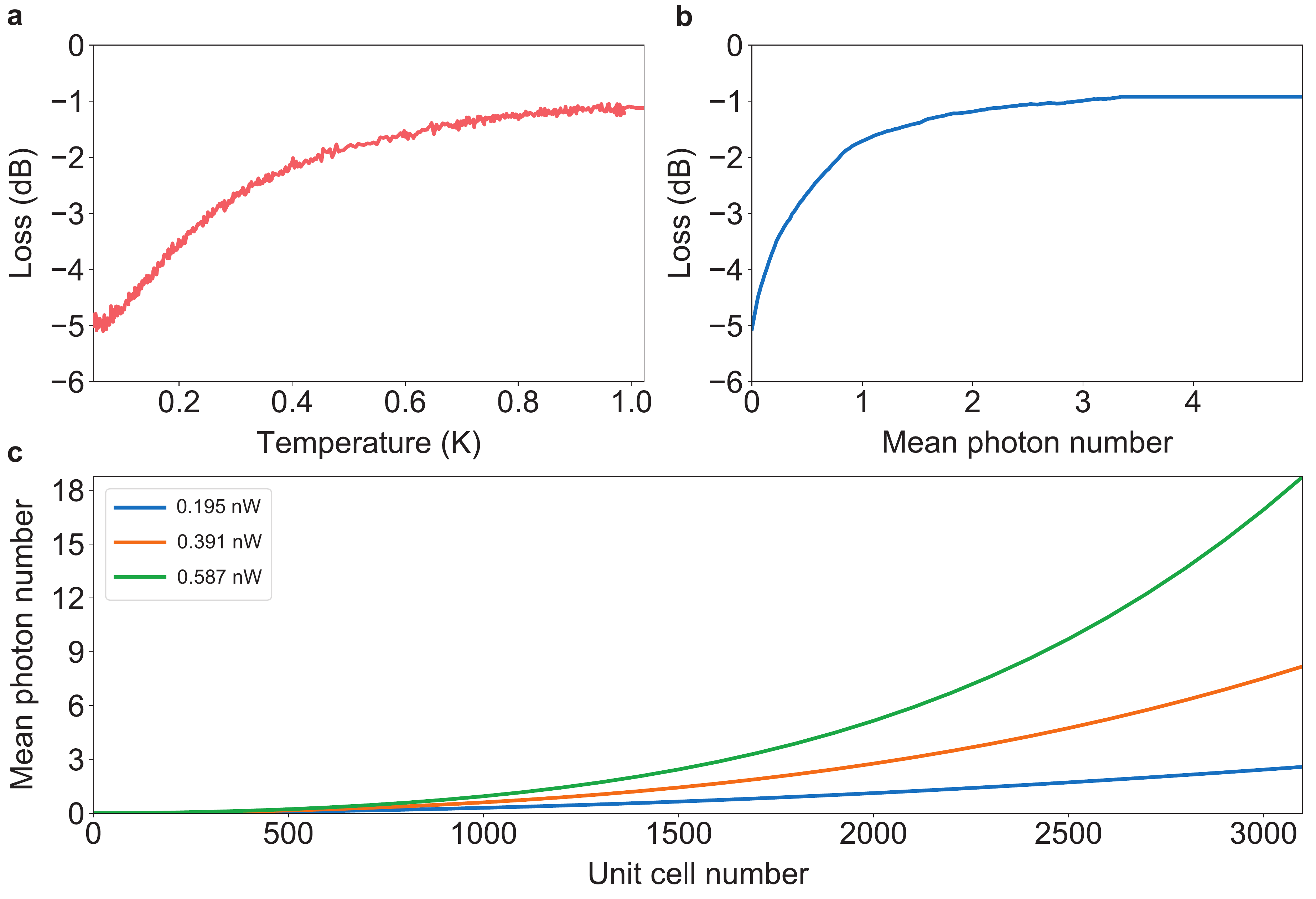}
%\captionsetup{labelformat=empty}
\caption{\textbf{$|$ JTWPA loss saturation.} \textbf{a.} JTWPA insertion loss at \unit[6.7]{GHz} as a function of temperature controlled using a heater at the mixing chamber. \textbf{b.} JTWPA insertion loss plotted versus thermal photon number $n_s$ associated with the temperatures $T_s$, based on Bose-Einstein statistics $n_s = 1/(\e^{h f/k_{\rm B} T_{\rm s}} - 1)$. \textbf{c.} The amplification process of the JTWPA produces an effective thermal state, and its photon number depends on the gain and the position inside the amplifier, as shown in the simulation plot. Each curve corresponds to a specific pump 2 power. Pump 1 power is always fixed at \unit[1.57]{nW} as in Fig. 3 from the main text.}
\label{fig:temploss1D}
\end{figure}
\FloatBarrier

The majority of the loss in JTWPA comes from two-level systems (TLSs) within the dielectric material (silicon dioxide) that constitutes the parallel-plate capacitors in the device~\cite{Sage2011}. We have observed that the loss becomes saturated as the temperature increases. \cref{fig:temploss1D}(a) is a plot of the empirical characterization of the saturation behavior of our device. The temperature is controlled by adjusting the current through a heating element on the mixing chamber of the dilution refrigerator.

In a real device, we expect the pump loss $\gamma_p$ to be dependent on power due to TLS loss saturation, and similarly, the frequency-dependent loss rate per unit length $\gamma(\omega)$ to be dependent on the photon number at 
$\omega$. This means that the loss rates also implicitly depend on position. For the numerical solutions, we have neglected position dependence of $\gamma_p$ and taken $\gamma_p = \gamma_p(P)$ where the loss for a given input power $P$ is a measured quantity. For the loss factors evaluated at any frequency away from the pumps, we use either a constant or photon-number-dependent factor, such that $\gamma(\omega)$ depends on the concomitant value of $\braket{c_\omega\dg c_\omega}(x)$. The photon number dependent loss model is motivated by the well-known observation that the loss rate is temperature dependent, as measured in this experiment, see~\cref{fig:temploss1D}(a) and elsewhere~\cite{Sage2011}.

\begin{figure}
\includegraphics[width=0.75\textwidth]{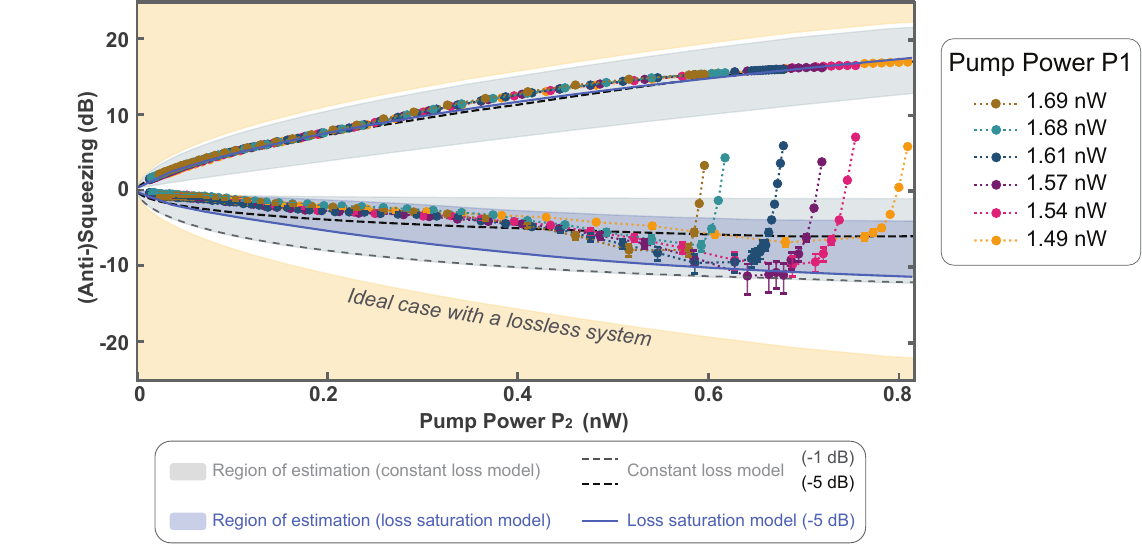}
%\captionsetup{labelformat=empty}
\caption{\textbf{$|$ Measurement of squeezing and anti-squeezing versus $P_2$ with six different $P_1$ configurations.} Single pump gain and numerical simulations that fit and predict the gain profiles. The deviation between the model and the experimental data is expected to be caused by the bandwidth constraint of cryogenic components. Pump power $P_1$ corresponds to the output of its RF source at room temperature. The data are presented as mean values of 3 sets of repeated measurement (each with $6\times 10^6$ sample points). Their statistical variation is almost entirely due to the uncertainty in estimating the noise temperature, which dominates the error bars shown in the plot as an estimation range for the squeezing/anti-squeezing levels.}
\label{fig:sqztrend}
\end{figure}

In \cref{fig:sqztrend}, the boundary of the beige region corresponds to the ideal-case squeezing achievable for a lossless JTWPA. The gray-shaded areas represent regions of estimation for squeezing and anti-squeezing levels; we define a lower bound that corresponds to numerical results assuming all of the loss (\unit[-5]{dB}) is at the end of the device (worst case), while the upper bounds are obtained using Caves and Crouch's distributed beamsplitter loss model~\cite{Caves1987} with \unit[-1]{dB} distributed loss (best case). The black dashed lines represent the numerical model with a uniformly distributed \unit[-5]{dB}-loss across the device. These numerical models confine the possible squeezing and anti-squeezing levels given the loss of the JTWPA.

We estimate the loss saturation effect on squeezing using a distributed loss saturation model plotted as a blue line in~\cref{fig:sqztrend}. In this model, the loss rate $\gamma(x)$ at position $x$ is determined by an effective temperature $T_{\rm e}$ extracted from the instantaneous photon number in the numerical simulation shown in~\cref{fig:temploss1D}(c). The lower bound is given by the loss saturation model with all of the loss towards the end of the device, while the upper bound is provided by the same model with a more realistic distributed \unit[-5]{dB} loss model. Together, they form a refined region of estimation as displayed in the blue-shaded region. The discrepancy between the measured behavior of squeezing at moderate pump powers prior to saturation and the numerical simulation could be due to more complicated pump dynamics and multimode interactions~\cite{peng2022} mixing in un-squeezed vacuum, which are not captured by the input-output model used here. 

As mentioned in the main text, there are two major approaches to improve the JTWPA squeezing performance based on our current architecture. Through Floquet engineering and its potential benefit of suppressing spurious nonlinear processes such as sideband generation, the squeezing level is expected to approach the performance dominated by loss, and the squeezing purity will improve in the low-to-mid power region. Moreover, we can further decrease the JTWPA loss from the dielectrics by using a high-Q fabrication process. In the limit of near-lossless performance, the maximum squeezing level limit will approach \unit[-20]{dB} --- an almost \unit[10]{dB} improvement --- assuming the device performance is soley constrained by loss at this point.

\section*{Numerical method}

To solve for the output fields numerically, we first have to choose a finite set of frequencies
\begin{equation}
    \mathbb{K} \equiv \{\omega_0,\omega_1,\dots,\omega_n\},
\end{equation}
and set $\hat c_{\omega} = 0$ for $\omega \not\in \mathbb{K}$ in~\cref{eq:sm:eom_loss}. For a given ``signal'' frequency $\omega_0$ we construct the set $\mathbb{K}$ in an iterative manner. For the first ``level'' we add the two frequencies
\begin{equation}
    \mathbb{K}_0 = \{\omega_0, \Omega_1 + \Omega_2 - \omega_0\}.
\end{equation}
Then we construct the next level as follows:
\begin{equation}
    \mathbb{K}_{l+1} = \{\Omega_p + \Omega_q - \omega_l, \omega_l \pm (\Omega_2 - \Omega_1) \mid \omega_l \in \mathbb{K}_l, p,q \in \{1, 2\} \},
\end{equation}
but remove from $\mathbb{K}_{l+1}$ any $\omega_l < 0$, any $\omega_l \simeq \Omega_1,\Omega_2$ and any $\omega_l$ already in $\mathbb{K}_{l}$. Finally, $\mathbb{K} = \cup_{l=0}^{k} \mathbb{K}_l$ up to some truncation $k$.
The first two levels thus include
\begin{subequations}\label{eq:sm:K}
\begin{align}
\mathbb{K}_0 ={}& \{\omega_0, \Omega_1 + \Omega_2 - \omega_0\},\\
\mathbb{K}_1 ={}& \{2\Omega_1 - \omega_0, 2\Omega_2 - \omega_0, \omega_0 + \Omega_1 - \Omega_2, \omega_0 + \Omega_2 - \Omega_1\}.
\end{align}
\end{subequations}
In practice, we have found after extensive numerical testing that including frequencies beyond the first level $\mathbb K_0$ does not improve the fit to the experimental squeezing data. 

Once a finite set of frequencies has been chosen, we can use~\cref{eq:sm:eom_loss} to compute expectation values. For numerical purposes, it is convenient to introduce a matrix-vector notation
\begin{equation}
    \vec c = [\hat c_1, \dots, \hat c_n, \hat c_1\dg, \dots, \hat c_n\dg]^T,
\end{equation}
and write
\begin{equation}\label{eq:sm:eom:avg}
    \frac{\dd}{\dd x} \braket{\vec c} = i K H \braket{\vec c} - \frac{\Gamma}{2} \braket{\vec c},
\end{equation}
where $K$, $H$ and $\Gamma$ are $2n \times 2n$ matrices, with $K = \text{diag}[I_n,-I_n]$, $$\Gamma = \text{diag}[\gamma_1,\dots,\gamma_n,\gamma_1,\dots,\gamma_n],$$ and $H$ is a Hermitian matrix that can be written in the block form
\begin{equation}
  H = \left[\begin{array}{cc}
      \Lambda_1 & \Lambda_2 \\
       \Lambda_2^* & \Lambda_1^*
  \end{array}\right],
\end{equation}
with $\Lambda_1\dg = \Lambda_1$ Hermitian and $\Lambda_2^T = \Lambda_2$ symmetric.

From~\cref{eq:sm:eom:avg} we can compute the gain using as initial condition $$\braket{\vec c(0)} = [\alpha,0,\dots,0,\alpha^*,0,\dots,0]^T,$$ and define the gain to be $g = \braket{\hat c_0}/\alpha$, and power gain $G = |g|^2$.

To compute squeezing, we also need to solve for all second order moments, $\Big \langle \hat c_i \hat c_j \Big \rangle$, $\braket{\hat c_i \hat c_j\dg}$, etc. For this purpose it is convenient to define a ``correlation matrix''
\begin{equation}
  C = [\braket{\vec c_i \vec c_j}] =
  \left[\begin{array}{cc}
  \Big \langle \hat c_i \hat c_j \Big \rangle & \braket{\hat c_i \hat c_j\dg} \\
  \braket{\hat c_i\dg \hat c_j} & \braket{\hat c_i\dg \hat c_j\dg}
  \end{array}\right],
\end{equation}
where each block is $n \times n$. An equation of motion can be derived from~\cref{eq:sm:eomp_loss} by using that
\begin{equation}
    \begin{aligned}
    \frac{\dd}{\dd x}
    \braket{\hat c_i \hat c_j} ={}& \braket{(\partial_x \hat c_i) \hat c_j} + \braket{\hat c_i (\partial_x \hat c_j)},
    \end{aligned}
\end{equation}
etc. We find
\begin{equation}\label{eq:sm:eom:corr}
  \frac{\dd}{\dd x} C = i KH C + i C (KH)^T
  + \left(\begin{array}{cc}
  0 & \Gamma \\
  0 & 0 \\
  \end{array}\right),
\end{equation}
where we have assumed a vacuum input field $\hat c_\text{in}(x)$.

To compute squeezing we first define a ``squeezing matrix''
\begin{equation}
  \begin{aligned}
  \Delta Y_{ij}^2(\theta) ={}& \half \braket{\left(\hat Y_i^\theta + \hat Y_j^\theta\right)^2} - \frac{1}{4}\braket{\hat Y_i + \hat Y_j}^2\\
  ={}& \frac{1}{4} \sum_{\stackrel{i'=i,j}{j'=i,j}} \big(
  \braket{\hat c_{i'}^\dagger \hat c_{j'}} + \braket{\hat c_{i'} \hat c_{j'}^\dagger}
  -\e^{i\theta} \braket{\hat c_{i'}^\dagger \hat c_{j'}^\dagger} - \e^{-i\theta}\braket{\hat c_{i'} \hat c_{j'}} \big),
  \end{aligned}
\end{equation}
where $\hat Y_i^\theta = \frac{i}{\sqrt 2}\left(\e^{i\theta/2}\hat c_i^\dagger - \e^{-i\theta/2}\hat c_i\right)$, and in the second line we have used $\braket{\hat Y_i + \hat Y_j} = 0$ for vacuum input.

The squeezing matrix is here defined such that high squeezing level means that $\Delta Y_{ij}^2(\theta)$ is small. Squeezing is thus maximized between modes $i$ and $j$ ($i=j$ for single-mode squeezing) by choosing $\theta$ such that $\e^{i\theta}\braket{\hat c_i^\dagger \hat c_j^\dagger} = \left|\braket{\hat c_i^\dagger \hat c_j^\dagger}\right|$. Note that the $\theta$ that maximizes squeezing might in general be different for different $ij$.

The squeezing in dB is defined as
\begin{equation}
    \mathcal S_{ij} = 10 \log_{10} \left( \frac{\Delta Y_{ij}^2}{1/2} \right),
\end{equation}
where the $1/2$ is the vacuum fluctuations. To compute the squeezing, \cref{eq:sm:eom:corr} is integrated numerically with initial condition
\begin{equation}
  C(0) = \left(\begin{array}{cc}
  0 & I_n \\
  0 & 0 \\
  \end{array}\right),
\end{equation}
corresponding to vacuum input.

\subsection*{Calibrating pump power at the device}

Matching the numerical results to experimental data requires knowing the dimensionless pump strength $\beta_p = I_p/4I_c$ at the device for a given input power $P$. One approach to determine $\beta_p$ is to measure the power dependent phase shift $\Delta \phi_p$ at the pump frequency in the presence of a single pump. From~\cref{eq:sm:eomp_loss} we have that
\begin{equation}
    |\beta_p(x=0)|^2 = \frac{1}{k_p}\frac{\gamma_p(P)}{1-\e^{-\gamma_p(P) z}} \Delta \phi_p(P),
\end{equation}
where we have included the power dependence of the pump loss rate $\gamma_p(P)$.
This procedure is, however, complicated by the fact that we do not observe a linear relationship between $\Delta \phi_p$ and $P$ in the experiment. This could be, amongst other factors, due to the non-trivial dependence of the dispersion feature on power: As the pump power increases, the dispersion feature is observed to become significantly more narrow in frequency, likely due to saturation of two-level systems in the LC oscillators.

Nevertheless, we have found reasonable agreement with experiments by assuming a power dependence of the form
\begin{equation}
    |\beta_p(0)|^2 = \frac{1}{k_p}\frac{\gamma_p(P)}{1-\e^{-\gamma_p(P) z}} \times c_p P,
\end{equation}
where $c_p$ is a power-independent fit parameter. In practice, we first vary $c_p$ to fit the numerical results to the gain curve and subsequently use the same value of $c_p$ to extract squeezing and anti-squeezing.

Since the gain curve has been fitted, the theory does not directly predict the gain at a given input power $P$. Nevertheless, it is noteworthy that an excellent fit to the overall shape of the gain curve can be found using this method, as shown in Fig. 3c in the main text. Most importantly, this method allows us to predict the squeezing and anti-squeezing at a given gain.

\bibliography{main}

\begin{thebibliography}{57}%
\makeatletter
\providecommand \@ifxundefined [1]{%
 \@ifx{#1\undefined}
}%
\providecommand \@ifnum [1]{%
 \ifnum #1\expandafter \@firstoftwo
 \else \expandafter \@secondoftwo
 \fi
}%
\providecommand \@ifx [1]{%
 \ifx #1\expandafter \@firstoftwo
 \else \expandafter \@secondoftwo
 \fi
}%
\providecommand \natexlab [1]{#1}%
\providecommand \enquote  [1]{``#1''}%
\providecommand \bibnamefont  [1]{#1}%
\providecommand \bibfnamefont [1]{#1}%
\providecommand \citenamefont [1]{#1}%
\providecommand \href@noop [0]{\@secondoftwo}%
\providecommand \href [0]{\begingroup \@sanitize@url \@href}%
\providecommand \@href[1]{\@@startlink{#1}\@@href}%
\providecommand \@@href[1]{\endgroup#1\@@endlink}%
\providecommand \@sanitize@url [0]{\catcode `\\12\catcode `\$12\catcode
  `\&12\catcode `\#12\catcode `\^12\catcode `\_12\catcode `\%12\relax}%
\providecommand \@@startlink[1]{}%
\providecommand \@@endlink[0]{}%
\providecommand \url  [0]{\begingroup\@sanitize@url \@url }%
\providecommand \@url [1]{\endgroup\@href {#1}{\urlprefix }}%
\providecommand \urlprefix  [0]{URL }%
\providecommand \Eprint [0]{\href }%
\providecommand \doibase [0]{http://dx.doi.org/}%
\providecommand \selectlanguage [0]{\@gobble}%
\providecommand \bibinfo  [0]{\@secondoftwo}%
\providecommand \bibfield  [0]{\@secondoftwo}%
\providecommand \translation [1]{[#1]}%
\providecommand \BibitemOpen [0]{}%
\providecommand \bibitemStop [0]{}%
\providecommand \bibitemNoStop [0]{.\EOS\space}%
\providecommand \EOS [0]{\spacefactor3000\relax}%
\providecommand \BibitemShut  [1]{\csname bibitem#1\endcsname}%
\let\auto@bib@innerbib\@empty
%</preamble>
\bibitem [{\citenamefont {Wallraff}\ \emph {et~al.}(2004)\citenamefont
  {Wallraff}, \citenamefont {Schuster}, \citenamefont {Blais}, \citenamefont
  {Frunzio}, \citenamefont {Huang}, \citenamefont {Majer}, \citenamefont
  {Kumar}, \citenamefont {Girvin},\ and\ \citenamefont
  {Schoelkopf}}]{Wallraff2004}%
  \BibitemOpen
  \bibfield  {author} {\bibinfo {author} {\bibfnamefont {A.}~\bibnamefont
  {Wallraff}}, \bibinfo {author} {\bibfnamefont {D.~I.}\ \bibnamefont
  {Schuster}}, \bibinfo {author} {\bibfnamefont {A.}~\bibnamefont {Blais}},
  \bibinfo {author} {\bibfnamefont {L.}~\bibnamefont {Frunzio}}, \bibinfo
  {author} {\bibfnamefont {R.-S.}\ \bibnamefont {Huang}}, \bibinfo {author}
  {\bibfnamefont {J.}~\bibnamefont {Majer}}, \bibinfo {author} {\bibfnamefont
  {S.}~\bibnamefont {Kumar}}, \bibinfo {author} {\bibfnamefont {S.~M.}\
  \bibnamefont {Girvin}}, \ and\ \bibinfo {author} {\bibfnamefont {R.~J.}\
  \bibnamefont {Schoelkopf}},\ }\href {http://dx.doi.org/10.1038/nature02851}
  {\bibfield  {journal} {\bibinfo  {journal} {Nature}\ }\textbf {\bibinfo
  {volume} {431}},\ \bibinfo {pages} {162} (\bibinfo {year}
  {2004})}\BibitemShut {NoStop}%
\bibitem [{\citenamefont {Caves}(1981)}]{Caves}%
  \BibitemOpen
  \bibfield  {author} {\bibinfo {author} {\bibfnamefont {C.~M.}\ \bibnamefont
  {Caves}},\ }\href {\doibase 10.1103/PhysRevD.23.1693} {\bibfield  {journal}
  {\bibinfo  {journal} {Phys. Rev. D}\ }\textbf {\bibinfo {volume} {23}},\
  \bibinfo {pages} {1693} (\bibinfo {year} {1981})}\BibitemShut {NoStop}%
\bibitem [{\citenamefont {Bienfait}\ \emph {et~al.}(2016)\citenamefont
  {Bienfait}, \citenamefont {Pla}, \citenamefont {Kubo}, \citenamefont {Stern},
  \citenamefont {Zhou}, \citenamefont {Lo}, \citenamefont {Weis}, \citenamefont
  {Schenkel}, \citenamefont {Thewalt}, \citenamefont {Vion}, \citenamefont
  {Esteve}, \citenamefont {Julsgaard}, \citenamefont {M{\o}lmer}, \citenamefont
  {Morton},\ and\ \citenamefont {Bertet}}]{Bienfait2016}%
  \BibitemOpen
  \bibfield  {author} {\bibinfo {author} {\bibfnamefont {A.}~\bibnamefont
  {Bienfait}}, \bibinfo {author} {\bibfnamefont {J.~J.}\ \bibnamefont {Pla}},
  \bibinfo {author} {\bibfnamefont {Y.}~\bibnamefont {Kubo}}, \bibinfo {author}
  {\bibfnamefont {M.}~\bibnamefont {Stern}}, \bibinfo {author} {\bibfnamefont
  {X.}~\bibnamefont {Zhou}}, \bibinfo {author} {\bibfnamefont {C.~C.}\
  \bibnamefont {Lo}}, \bibinfo {author} {\bibfnamefont {C.~D.}\ \bibnamefont
  {Weis}}, \bibinfo {author} {\bibfnamefont {T.}~\bibnamefont {Schenkel}},
  \bibinfo {author} {\bibfnamefont {M.~L.~W.}\ \bibnamefont {Thewalt}},
  \bibinfo {author} {\bibfnamefont {D.}~\bibnamefont {Vion}}, \bibinfo {author}
  {\bibfnamefont {D.}~\bibnamefont {Esteve}}, \bibinfo {author} {\bibfnamefont
  {B.}~\bibnamefont {Julsgaard}}, \bibinfo {author} {\bibfnamefont
  {K.}~\bibnamefont {M{\o}lmer}}, \bibinfo {author} {\bibfnamefont {J.~J.~L.}\
  \bibnamefont {Morton}}, \ and\ \bibinfo {author} {\bibfnamefont
  {P.}~\bibnamefont {Bertet}},\ }\href {\doibase 10.1038/nnano.2015.282}
  {\bibfield  {journal} {\bibinfo  {journal} {Nature Nanotechnology}\ }\textbf
  {\bibinfo {volume} {11}},\ \bibinfo {pages} {253} (\bibinfo {year}
  {2016})}\BibitemShut {NoStop}%
\bibitem [{\citenamefont {Slusher}\ \emph {et~al.}(1985)\citenamefont
  {Slusher}, \citenamefont {Hollberg}, \citenamefont {Yurke}, \citenamefont
  {Mertz},\ and\ \citenamefont {Valley}}]{Slusher1985}%
  \BibitemOpen
  \bibfield  {author} {\bibinfo {author} {\bibfnamefont {R.~E.}\ \bibnamefont
  {Slusher}}, \bibinfo {author} {\bibfnamefont {L.~W.}\ \bibnamefont
  {Hollberg}}, \bibinfo {author} {\bibfnamefont {B.}~\bibnamefont {Yurke}},
  \bibinfo {author} {\bibfnamefont {J.~C.}\ \bibnamefont {Mertz}}, \ and\
  \bibinfo {author} {\bibfnamefont {J.~F.}\ \bibnamefont {Valley}},\
  }\href@noop {} {\bibfield  {journal} {\bibinfo  {journal} {Phys. Rev. Lett.}\
  }\textbf {\bibinfo {volume} {55}},\ \bibinfo {pages} {2409} (\bibinfo {year}
  {1985})}\BibitemShut {NoStop}%
\bibitem [{\citenamefont {Toyli}\ \emph {et~al.}(2016)\citenamefont {Toyli},
  \citenamefont {Eddins}, \citenamefont {Boutin}, \citenamefont {Puri},
  \citenamefont {Hover}, \citenamefont {Bolkhovsky}, \citenamefont {Oliver},
  \citenamefont {Blais},\ and\ \citenamefont {Siddiqi}}]{Toyli2016}%
  \BibitemOpen
  \bibfield  {author} {\bibinfo {author} {\bibfnamefont {D.~M.}\ \bibnamefont
  {Toyli}}, \bibinfo {author} {\bibfnamefont {A.~W.}\ \bibnamefont {Eddins}},
  \bibinfo {author} {\bibfnamefont {S.}~\bibnamefont {Boutin}}, \bibinfo
  {author} {\bibfnamefont {S.}~\bibnamefont {Puri}}, \bibinfo {author}
  {\bibfnamefont {D.}~\bibnamefont {Hover}}, \bibinfo {author} {\bibfnamefont
  {V.}~\bibnamefont {Bolkhovsky}}, \bibinfo {author} {\bibfnamefont {W.~D.}\
  \bibnamefont {Oliver}}, \bibinfo {author} {\bibfnamefont {A.}~\bibnamefont
  {Blais}}, \ and\ \bibinfo {author} {\bibfnamefont {I.}~\bibnamefont
  {Siddiqi}},\ }\href {\doibase 10.1103/PhysRevX.6.031004} {\bibfield
  {journal} {\bibinfo  {journal} {Phys. Rev. X}\ }\textbf {\bibinfo {volume}
  {6}},\ \bibinfo {pages} {031004} (\bibinfo {year} {2016})}\BibitemShut
  {NoStop}%
\bibitem [{\citenamefont {Aoki}\ \emph {et~al.}(2009)\citenamefont {Aoki},
  \citenamefont {Takahashi}, \citenamefont {Kajiya}, \citenamefont {Yoshikawa},
  \citenamefont {Braunstein}, \citenamefont {van Loock},\ and\ \citenamefont
  {Furusawa}}]{Aoki2009}%
  \BibitemOpen
  \bibfield  {author} {\bibinfo {author} {\bibfnamefont {T.}~\bibnamefont
  {Aoki}}, \bibinfo {author} {\bibfnamefont {G.}~\bibnamefont {Takahashi}},
  \bibinfo {author} {\bibfnamefont {T.}~\bibnamefont {Kajiya}}, \bibinfo
  {author} {\bibfnamefont {J.-i.}\ \bibnamefont {Yoshikawa}}, \bibinfo {author}
  {\bibfnamefont {S.~L.}\ \bibnamefont {Braunstein}}, \bibinfo {author}
  {\bibfnamefont {P.}~\bibnamefont {van Loock}}, \ and\ \bibinfo {author}
  {\bibfnamefont {A.}~\bibnamefont {Furusawa}},\ }\href@noop {} {\bibfield
  {journal} {\bibinfo  {journal} {Nature Physics}\ }\textbf {\bibinfo {volume}
  {5}},\ \bibinfo {pages} {541} (\bibinfo {year} {2009})}\BibitemShut {NoStop}%
\bibitem [{\citenamefont {{The LIGO Scientific
  Collaboration}}(2011)}]{LIGO2011}%
  \BibitemOpen
  \bibfield  {author} {\bibinfo {author} {\bibnamefont {{The LIGO Scientific
  Collaboration}}},\ }\href@noop {} {\bibfield  {journal} {\bibinfo  {journal}
  {Nature Physics}\ ,\ \bibinfo {pages} {962 }} (\bibinfo {year}
  {2011})}\BibitemShut {NoStop}%
\bibitem [{\citenamefont {Boutin}\ \emph {et~al.}(2017)\citenamefont {Boutin},
  \citenamefont {Toyli}, \citenamefont {Venkatramani}, \citenamefont {Eddins},
  \citenamefont {Siddiqi},\ and\ \citenamefont {Blais}}]{Boutin2017}%
  \BibitemOpen
  \bibfield  {author} {\bibinfo {author} {\bibfnamefont {S.}~\bibnamefont
  {Boutin}}, \bibinfo {author} {\bibfnamefont {D.~M.}\ \bibnamefont {Toyli}},
  \bibinfo {author} {\bibfnamefont {A.~V.}\ \bibnamefont {Venkatramani}},
  \bibinfo {author} {\bibfnamefont {A.~W.}\ \bibnamefont {Eddins}}, \bibinfo
  {author} {\bibfnamefont {I.}~\bibnamefont {Siddiqi}}, \ and\ \bibinfo
  {author} {\bibfnamefont {A.}~\bibnamefont {Blais}},\ }\href@noop {}
  {\bibfield  {journal} {\bibinfo  {journal} {Phys. Rev. Applied}\ }\textbf
  {\bibinfo {volume} {8}},\ \bibinfo {pages} {054030} (\bibinfo {year}
  {2017})}\BibitemShut {NoStop}%
\bibitem [{\citenamefont {Malnou}\ \emph {et~al.}(2018)\citenamefont {Malnou},
  \citenamefont {Palken}, \citenamefont {Vale}, \citenamefont {Hilton},\ and\
  \citenamefont {Lehnert}}]{Malnou_2018}%
  \BibitemOpen
  \bibfield  {author} {\bibinfo {author} {\bibfnamefont {M.}~\bibnamefont
  {Malnou}}, \bibinfo {author} {\bibfnamefont {D.~A.}\ \bibnamefont {Palken}},
  \bibinfo {author} {\bibfnamefont {L.~R.}\ \bibnamefont {Vale}}, \bibinfo
  {author} {\bibfnamefont {G.~C.}\ \bibnamefont {Hilton}}, \ and\ \bibinfo
  {author} {\bibfnamefont {K.~W.}\ \bibnamefont {Lehnert}},\ }\href {\doibase
  10.1103/PhysRevApplied.9.044023} {\bibfield  {journal} {\bibinfo  {journal}
  {Phys. Rev. Applied}\ }\textbf {\bibinfo {volume} {9}},\ \bibinfo {pages}
  {044023} (\bibinfo {year} {2018})}\BibitemShut {NoStop}%
\bibitem [{\citenamefont {Murch}\ \emph {et~al.}(2013)\citenamefont {Murch},
  \citenamefont {Weber}, \citenamefont {Beck}, \citenamefont {Ginossar},\ and\
  \citenamefont {Siddiqi}}]{Murch2013}%
  \BibitemOpen
  \bibfield  {author} {\bibinfo {author} {\bibfnamefont {K.~W.}\ \bibnamefont
  {Murch}}, \bibinfo {author} {\bibfnamefont {S.~J.}\ \bibnamefont {Weber}},
  \bibinfo {author} {\bibfnamefont {K.~M.}\ \bibnamefont {Beck}}, \bibinfo
  {author} {\bibfnamefont {E.}~\bibnamefont {Ginossar}}, \ and\ \bibinfo
  {author} {\bibfnamefont {I.}~\bibnamefont {Siddiqi}},\ }\href@noop {}
  {\bibfield  {journal} {\bibinfo  {journal} {Nature}\ }\textbf {\bibinfo
  {volume} {499}},\ \bibinfo {pages} {62 } (\bibinfo {year}
  {2013})}\BibitemShut {NoStop}%
\bibitem [{\citenamefont {Menzel}\ \emph {et~al.}(2012)\citenamefont {Menzel},
  \citenamefont {Di~Candia}, \citenamefont {Deppe}, \citenamefont {Eder},
  \citenamefont {Zhong}, \citenamefont {Ihmig}, \citenamefont {Haeberlein},
  \citenamefont {Baust}, \citenamefont {Hoffmann}, \citenamefont {Ballester},
  \citenamefont {Inomata}, \citenamefont {Yamamoto}, \citenamefont {Nakamura},
  \citenamefont {Solano}, \citenamefont {Marx},\ and\ \citenamefont
  {Gross}}]{Menzel_2012}%
  \BibitemOpen
  \bibfield  {author} {\bibinfo {author} {\bibfnamefont {E.~P.}\ \bibnamefont
  {Menzel}}, \bibinfo {author} {\bibfnamefont {R.}~\bibnamefont {Di~Candia}},
  \bibinfo {author} {\bibfnamefont {F.}~\bibnamefont {Deppe}}, \bibinfo
  {author} {\bibfnamefont {P.}~\bibnamefont {Eder}}, \bibinfo {author}
  {\bibfnamefont {L.}~\bibnamefont {Zhong}}, \bibinfo {author} {\bibfnamefont
  {M.}~\bibnamefont {Ihmig}}, \bibinfo {author} {\bibfnamefont
  {M.}~\bibnamefont {Haeberlein}}, \bibinfo {author} {\bibfnamefont
  {A.}~\bibnamefont {Baust}}, \bibinfo {author} {\bibfnamefont
  {E.}~\bibnamefont {Hoffmann}}, \bibinfo {author} {\bibfnamefont
  {D.}~\bibnamefont {Ballester}}, \bibinfo {author} {\bibfnamefont
  {K.}~\bibnamefont {Inomata}}, \bibinfo {author} {\bibfnamefont
  {T.}~\bibnamefont {Yamamoto}}, \bibinfo {author} {\bibfnamefont
  {Y.}~\bibnamefont {Nakamura}}, \bibinfo {author} {\bibfnamefont
  {E.}~\bibnamefont {Solano}}, \bibinfo {author} {\bibfnamefont
  {A.}~\bibnamefont {Marx}}, \ and\ \bibinfo {author} {\bibfnamefont
  {R.}~\bibnamefont {Gross}},\ }\href {\doibase 10.1103/PhysRevLett.109.250502}
  {\bibfield  {journal} {\bibinfo  {journal} {Phys. Rev. Lett.}\ }\textbf
  {\bibinfo {volume} {109}},\ \bibinfo {pages} {250502} (\bibinfo {year}
  {2012})}\BibitemShut {NoStop}%
\bibitem [{\citenamefont {Bienfait}\ \emph {et~al.}(2017)\citenamefont
  {Bienfait}, \citenamefont {Campagne-Ibarcq}, \citenamefont {Kiilerich},
  \citenamefont {Zhou}, \citenamefont {Probst}, \citenamefont {Pla},
  \citenamefont {Schenkel}, \citenamefont {Vion}, \citenamefont {Esteve},
  \citenamefont {Morton}, \citenamefont {Moelmer},\ and\ \citenamefont
  {Bertet}}]{Bienfait_2017}%
  \BibitemOpen
  \bibfield  {author} {\bibinfo {author} {\bibfnamefont {A.}~\bibnamefont
  {Bienfait}}, \bibinfo {author} {\bibfnamefont {P.}~\bibnamefont
  {Campagne-Ibarcq}}, \bibinfo {author} {\bibfnamefont {A.~H.}\ \bibnamefont
  {Kiilerich}}, \bibinfo {author} {\bibfnamefont {X.}~\bibnamefont {Zhou}},
  \bibinfo {author} {\bibfnamefont {S.}~\bibnamefont {Probst}}, \bibinfo
  {author} {\bibfnamefont {J.~J.}\ \bibnamefont {Pla}}, \bibinfo {author}
  {\bibfnamefont {T.}~\bibnamefont {Schenkel}}, \bibinfo {author}
  {\bibfnamefont {D.}~\bibnamefont {Vion}}, \bibinfo {author} {\bibfnamefont
  {D.}~\bibnamefont {Esteve}}, \bibinfo {author} {\bibfnamefont {J.~J.~L.}\
  \bibnamefont {Morton}}, \bibinfo {author} {\bibfnamefont {K.}~\bibnamefont
  {Moelmer}}, \ and\ \bibinfo {author} {\bibfnamefont {P.}~\bibnamefont
  {Bertet}},\ }\href {\doibase 10.1103/PhysRevX.7.041011} {\bibfield  {journal}
  {\bibinfo  {journal} {Phys. Rev. X}\ }\textbf {\bibinfo {volume} {7}},\
  \bibinfo {pages} {041011} (\bibinfo {year} {2017})}\BibitemShut {NoStop}%
\bibitem [{\citenamefont {Krantz}\ \emph {et~al.}(2013)\citenamefont {Krantz},
  \citenamefont {Reshitnyk}, \citenamefont {Wustmann}, \citenamefont
  {Bylander}, \citenamefont {Gustavsson}, \citenamefont {Oliver}, \citenamefont
  {Duty}, \citenamefont {Shumeiko},\ and\ \citenamefont
  {Delsing}}]{Krantz_2013}%
  \BibitemOpen
  \bibfield  {author} {\bibinfo {author} {\bibfnamefont {P.}~\bibnamefont
  {Krantz}}, \bibinfo {author} {\bibfnamefont {Y.}~\bibnamefont {Reshitnyk}},
  \bibinfo {author} {\bibfnamefont {W.}~\bibnamefont {Wustmann}}, \bibinfo
  {author} {\bibfnamefont {J.}~\bibnamefont {Bylander}}, \bibinfo {author}
  {\bibfnamefont {S.}~\bibnamefont {Gustavsson}}, \bibinfo {author}
  {\bibfnamefont {W.~D.}\ \bibnamefont {Oliver}}, \bibinfo {author}
  {\bibfnamefont {T.}~\bibnamefont {Duty}}, \bibinfo {author} {\bibfnamefont
  {V.}~\bibnamefont {Shumeiko}}, \ and\ \bibinfo {author} {\bibfnamefont
  {P.}~\bibnamefont {Delsing}},\ }\href {\doibase
  10.1088/1367-2630/15/10/105002} {\bibfield  {journal} {\bibinfo  {journal}
  {New Journal of Physics}\ }\textbf {\bibinfo {volume} {15}},\ \bibinfo
  {pages} {105002} (\bibinfo {year} {2013})}\BibitemShut {NoStop}%
\bibitem [{\citenamefont {Renger}\ \emph {et~al.}(2021)\citenamefont {Renger},
  \citenamefont {Pogorzalek}, \citenamefont {Chen}, \citenamefont {Nojiri},
  \citenamefont {Inomata}, \citenamefont {Nakamura}, \citenamefont {Partanen},
  \citenamefont {Marx}, \citenamefont {Gross}, \citenamefont {Deppe},\ and\
  \citenamefont {Fedorov}}]{Renger2021}%
  \BibitemOpen
  \bibfield  {author} {\bibinfo {author} {\bibfnamefont {M.}~\bibnamefont
  {Renger}}, \bibinfo {author} {\bibfnamefont {S.}~\bibnamefont {Pogorzalek}},
  \bibinfo {author} {\bibfnamefont {Q.}~\bibnamefont {Chen}}, \bibinfo {author}
  {\bibfnamefont {Y.}~\bibnamefont {Nojiri}}, \bibinfo {author} {\bibfnamefont
  {K.}~\bibnamefont {Inomata}}, \bibinfo {author} {\bibfnamefont
  {Y.}~\bibnamefont {Nakamura}}, \bibinfo {author} {\bibfnamefont
  {M.}~\bibnamefont {Partanen}}, \bibinfo {author} {\bibfnamefont
  {A.}~\bibnamefont {Marx}}, \bibinfo {author} {\bibfnamefont {R.}~\bibnamefont
  {Gross}}, \bibinfo {author} {\bibfnamefont {F.}~\bibnamefont {Deppe}}, \ and\
  \bibinfo {author} {\bibfnamefont {K.~G.}\ \bibnamefont {Fedorov}},\ }\href
  {\doibase 10.1038/s41534-021-00495-y} {\bibfield  {journal} {\bibinfo
  {journal} {npj Quantum Information}\ }\textbf {\bibinfo {volume} {7}},\
  \bibinfo {pages} {160} (\bibinfo {year} {2021})}\BibitemShut {NoStop}%
\bibitem [{\citenamefont {Roy}\ \emph {et~al.}(2015)\citenamefont {Roy},
  \citenamefont {Kundu}, \citenamefont {Chand}, \citenamefont {Vadiraj},
  \citenamefont {Ranadive}, \citenamefont {Nehra}, \citenamefont {Patankar},
  \citenamefont {Aumentado}, \citenamefont {Clerk},\ and\ \citenamefont
  {Vijay}}]{Roy2015}%
  \BibitemOpen
  \bibfield  {author} {\bibinfo {author} {\bibfnamefont {T.}~\bibnamefont
  {Roy}}, \bibinfo {author} {\bibfnamefont {S.}~\bibnamefont {Kundu}}, \bibinfo
  {author} {\bibfnamefont {M.}~\bibnamefont {Chand}}, \bibinfo {author}
  {\bibfnamefont {A.~M.}\ \bibnamefont {Vadiraj}}, \bibinfo {author}
  {\bibfnamefont {A.}~\bibnamefont {Ranadive}}, \bibinfo {author}
  {\bibfnamefont {N.}~\bibnamefont {Nehra}}, \bibinfo {author} {\bibfnamefont
  {M.~P.}\ \bibnamefont {Patankar}}, \bibinfo {author} {\bibfnamefont
  {J.}~\bibnamefont {Aumentado}}, \bibinfo {author} {\bibfnamefont {A.~A.}\
  \bibnamefont {Clerk}}, \ and\ \bibinfo {author} {\bibfnamefont
  {R.}~\bibnamefont {Vijay}},\ }\href {\doibase 10.1063/1.4939148} {\bibfield
  {journal} {\bibinfo  {journal} {Applied Physics Letters}\ }\textbf {\bibinfo
  {volume} {107}},\ \bibinfo {pages} {262601} (\bibinfo {year}
  {2015})}\BibitemShut {NoStop}%
\bibitem [{\citenamefont {Mutus}\ \emph {et~al.}(2014)\citenamefont {Mutus},
  \citenamefont {White}, \citenamefont {Barends}, \citenamefont {Chen},
  \citenamefont {Chen}, \citenamefont {Chiaro}, \citenamefont {Dunsworth},
  \citenamefont {Jeffrey}, \citenamefont {Kelly}, \citenamefont {Megrant},
  \citenamefont {Neill}, \citenamefont {O'Malley}, \citenamefont {Roushan},
  \citenamefont {Sank}, \citenamefont {Vainsencher}, \citenamefont {Wenner},
  \citenamefont {Sundqvist}, \citenamefont {Cleland},\ and\ \citenamefont
  {Martinis}}]{Mutus2014}%
  \BibitemOpen
  \bibfield  {author} {\bibinfo {author} {\bibfnamefont {J.~Y.}\ \bibnamefont
  {Mutus}}, \bibinfo {author} {\bibfnamefont {T.~C.}\ \bibnamefont {White}},
  \bibinfo {author} {\bibfnamefont {R.}~\bibnamefont {Barends}}, \bibinfo
  {author} {\bibfnamefont {Y.}~\bibnamefont {Chen}}, \bibinfo {author}
  {\bibfnamefont {Z.}~\bibnamefont {Chen}}, \bibinfo {author} {\bibfnamefont
  {B.}~\bibnamefont {Chiaro}}, \bibinfo {author} {\bibfnamefont
  {A.}~\bibnamefont {Dunsworth}}, \bibinfo {author} {\bibfnamefont
  {E.}~\bibnamefont {Jeffrey}}, \bibinfo {author} {\bibfnamefont
  {J.}~\bibnamefont {Kelly}}, \bibinfo {author} {\bibfnamefont
  {A.}~\bibnamefont {Megrant}}, \bibinfo {author} {\bibfnamefont
  {C.}~\bibnamefont {Neill}}, \bibinfo {author} {\bibfnamefont {P.~J.~J.}\
  \bibnamefont {O'Malley}}, \bibinfo {author} {\bibfnamefont {P.}~\bibnamefont
  {Roushan}}, \bibinfo {author} {\bibfnamefont {D.}~\bibnamefont {Sank}},
  \bibinfo {author} {\bibfnamefont {A.}~\bibnamefont {Vainsencher}}, \bibinfo
  {author} {\bibfnamefont {J.}~\bibnamefont {Wenner}}, \bibinfo {author}
  {\bibfnamefont {K.~M.}\ \bibnamefont {Sundqvist}}, \bibinfo {author}
  {\bibfnamefont {A.~N.}\ \bibnamefont {Cleland}}, \ and\ \bibinfo {author}
  {\bibfnamefont {J.~M.}\ \bibnamefont {Martinis}},\ }\href {\doibase
  10.1063/1.4886408} {\bibfield  {journal} {\bibinfo  {journal} {Applied
  Physics Letters}\ }\textbf {\bibinfo {volume} {104}},\ \bibinfo {pages}
  {263513} (\bibinfo {year} {2014})}\BibitemShut {NoStop}%
\bibitem [{\citenamefont {Sivak}\ \emph {et~al.}(2019)\citenamefont {Sivak},
  \citenamefont {Frattini}, \citenamefont {Joshi}, \citenamefont
  {Lingenfelter}, \citenamefont {Shankar},\ and\ \citenamefont
  {Devoret}}]{Sivak2019}%
  \BibitemOpen
  \bibfield  {author} {\bibinfo {author} {\bibfnamefont {V.}~\bibnamefont
  {Sivak}}, \bibinfo {author} {\bibfnamefont {N.}~\bibnamefont {Frattini}},
  \bibinfo {author} {\bibfnamefont {V.}~\bibnamefont {Joshi}}, \bibinfo
  {author} {\bibfnamefont {A.}~\bibnamefont {Lingenfelter}}, \bibinfo {author}
  {\bibfnamefont {S.}~\bibnamefont {Shankar}}, \ and\ \bibinfo {author}
  {\bibfnamefont {M.}~\bibnamefont {Devoret}},\ }\href@noop {} {\bibfield
  {journal} {\bibinfo  {journal} {Phys. Rev. Applied}\ }\textbf {\bibinfo
  {volume} {11}},\ \bibinfo {pages} {054060} (\bibinfo {year}
  {2019})}\BibitemShut {NoStop}%
\bibitem [{\citenamefont {Frattini}\ \emph {et~al.}(2018)\citenamefont
  {Frattini}, \citenamefont {Sivak}, \citenamefont {Lingenfelter},
  \citenamefont {Shankar},\ and\ \citenamefont {Devoret}}]{Frattini2018}%
  \BibitemOpen
  \bibfield  {author} {\bibinfo {author} {\bibfnamefont {N.~E.}\ \bibnamefont
  {Frattini}}, \bibinfo {author} {\bibfnamefont {V.~V.}\ \bibnamefont {Sivak}},
  \bibinfo {author} {\bibfnamefont {A.}~\bibnamefont {Lingenfelter}}, \bibinfo
  {author} {\bibfnamefont {S.}~\bibnamefont {Shankar}}, \ and\ \bibinfo
  {author} {\bibfnamefont {M.~H.}\ \bibnamefont {Devoret}},\ }\href@noop {}
  {\bibfield  {journal} {\bibinfo  {journal} {Phys. Rev. Applied}\ }\textbf
  {\bibinfo {volume} {10}},\ \bibinfo {pages} {054020} (\bibinfo {year}
  {2018})}\BibitemShut {NoStop}%
\bibitem [{\citenamefont {Sivak}\ \emph {et~al.}(2020)\citenamefont {Sivak},
  \citenamefont {Shankar}, \citenamefont {Liu}, \citenamefont {Aumentado},\
  and\ \citenamefont {Devoret}}]{Sivak2020}%
  \BibitemOpen
  \bibfield  {author} {\bibinfo {author} {\bibfnamefont {V.~V.}\ \bibnamefont
  {Sivak}}, \bibinfo {author} {\bibfnamefont {S.}~\bibnamefont {Shankar}},
  \bibinfo {author} {\bibfnamefont {G.}~\bibnamefont {Liu}}, \bibinfo {author}
  {\bibfnamefont {J.}~\bibnamefont {Aumentado}}, \ and\ \bibinfo {author}
  {\bibfnamefont {M.~H.}\ \bibnamefont {Devoret}},\ }\href@noop {} {\bibfield
  {journal} {\bibinfo  {journal} {Phys. Rev. Applied}\ }\textbf {\bibinfo
  {volume} {13}},\ \bibinfo {pages} {024014} (\bibinfo {year}
  {2020})}\BibitemShut {NoStop}%
\bibitem [{\citenamefont {Esposito}\ \emph {et~al.}(2021)\citenamefont
  {Esposito}, \citenamefont {Ranadive}, \citenamefont {Planat}, \citenamefont
  {Leger}, \citenamefont {Fraudet}, \citenamefont {Jouanny}, \citenamefont
  {Buisson}, \citenamefont {Guichard}, \citenamefont {Naud}, \citenamefont
  {Aumentado}, \citenamefont {Lecocq},\ and\ \citenamefont
  {Roch}}]{esposito2021}%
  \BibitemOpen
  \bibfield  {author} {\bibinfo {author} {\bibfnamefont {M.}~\bibnamefont
  {Esposito}}, \bibinfo {author} {\bibfnamefont {A.}~\bibnamefont {Ranadive}},
  \bibinfo {author} {\bibfnamefont {L.}~\bibnamefont {Planat}}, \bibinfo
  {author} {\bibfnamefont {S.}~\bibnamefont {Leger}}, \bibinfo {author}
  {\bibfnamefont {D.}~\bibnamefont {Fraudet}}, \bibinfo {author} {\bibfnamefont
  {V.}~\bibnamefont {Jouanny}}, \bibinfo {author} {\bibfnamefont
  {O.}~\bibnamefont {Buisson}}, \bibinfo {author} {\bibfnamefont
  {W.}~\bibnamefont {Guichard}}, \bibinfo {author} {\bibfnamefont
  {C.}~\bibnamefont {Naud}}, \bibinfo {author} {\bibfnamefont {J.}~\bibnamefont
  {Aumentado}}, \bibinfo {author} {\bibfnamefont {F.}~\bibnamefont {Lecocq}}, \
  and\ \bibinfo {author} {\bibfnamefont {N.}~\bibnamefont {Roch}},\ }\href@noop
  {} {\  (\bibinfo {year} {2021})},\ \Eprint {http://arxiv.org/abs/2111.03696}
  {arXiv:2111.03696 [quant-ph]} \BibitemShut {NoStop}%
\bibitem [{\citenamefont {Perelshtein}\ \emph {et~al.}(2021)\citenamefont
  {Perelshtein}, \citenamefont {Petrovnin}, \citenamefont {Vesterinen},
  \citenamefont {Raja}, \citenamefont {Lilja}, \citenamefont {Will},
  \citenamefont {Savin}, \citenamefont {Simbierowicz}, \citenamefont
  {Jabdaraghi}, \citenamefont {Lehtinen}, \citenamefont {Grönberg},
  \citenamefont {Hassel}, \citenamefont {Prunnila}, \citenamefont {Govenius},
  \citenamefont {Paraoanu},\ and\ \citenamefont {Hakonen}}]{perelshtein2021}%
  \BibitemOpen
  \bibfield  {author} {\bibinfo {author} {\bibfnamefont {M.}~\bibnamefont
  {Perelshtein}}, \bibinfo {author} {\bibfnamefont {K.}~\bibnamefont
  {Petrovnin}}, \bibinfo {author} {\bibfnamefont {V.}~\bibnamefont
  {Vesterinen}}, \bibinfo {author} {\bibfnamefont {S.~H.}\ \bibnamefont
  {Raja}}, \bibinfo {author} {\bibfnamefont {I.}~\bibnamefont {Lilja}},
  \bibinfo {author} {\bibfnamefont {M.}~\bibnamefont {Will}}, \bibinfo {author}
  {\bibfnamefont {A.}~\bibnamefont {Savin}}, \bibinfo {author} {\bibfnamefont
  {S.}~\bibnamefont {Simbierowicz}}, \bibinfo {author} {\bibfnamefont
  {R.}~\bibnamefont {Jabdaraghi}}, \bibinfo {author} {\bibfnamefont
  {J.}~\bibnamefont {Lehtinen}}, \bibinfo {author} {\bibfnamefont
  {L.}~\bibnamefont {Grönberg}}, \bibinfo {author} {\bibfnamefont
  {J.}~\bibnamefont {Hassel}}, \bibinfo {author} {\bibfnamefont
  {M.}~\bibnamefont {Prunnila}}, \bibinfo {author} {\bibfnamefont
  {J.}~\bibnamefont {Govenius}}, \bibinfo {author} {\bibfnamefont
  {S.}~\bibnamefont {Paraoanu}}, \ and\ \bibinfo {author} {\bibfnamefont
  {P.}~\bibnamefont {Hakonen}},\ }\href@noop {} {\  (\bibinfo {year} {2021})},\
  \Eprint {http://arxiv.org/abs/2111.06145} {arXiv:2111.06145 [quant-ph]}
  \BibitemShut {NoStop}%
\bibitem [{\citenamefont {Parker}\ \emph {et~al.}(2021)\citenamefont {Parker},
  \citenamefont {Savytskyi}, \citenamefont {Vine}, \citenamefont {Laucht},
  \citenamefont {Duty}, \citenamefont {Morello}, \citenamefont {Grimsmo},\ and\
  \citenamefont {Pla}}]{parker2021}%
  \BibitemOpen
  \bibfield  {author} {\bibinfo {author} {\bibfnamefont {D.~J.}\ \bibnamefont
  {Parker}}, \bibinfo {author} {\bibfnamefont {M.}~\bibnamefont {Savytskyi}},
  \bibinfo {author} {\bibfnamefont {W.}~\bibnamefont {Vine}}, \bibinfo {author}
  {\bibfnamefont {A.}~\bibnamefont {Laucht}}, \bibinfo {author} {\bibfnamefont
  {T.}~\bibnamefont {Duty}}, \bibinfo {author} {\bibfnamefont {A.}~\bibnamefont
  {Morello}}, \bibinfo {author} {\bibfnamefont {A.~L.}\ \bibnamefont
  {Grimsmo}}, \ and\ \bibinfo {author} {\bibfnamefont {J.~J.}\ \bibnamefont
  {Pla}},\ }\href@noop {} {\  (\bibinfo {year} {2021})},\ \Eprint
  {http://arxiv.org/abs/2108.10471} {arXiv:2108.10471 [quant-ph]} \BibitemShut
  {NoStop}%
\bibitem [{\citenamefont {Malnou}\ \emph {et~al.}(2021)\citenamefont {Malnou},
  \citenamefont {Vissers}, \citenamefont {Wheeler}, \citenamefont {Aumentado},
  \citenamefont {Hubmayr}, \citenamefont {Ullom},\ and\ \citenamefont
  {Gao}}]{Malnou2021}%
  \BibitemOpen
  \bibfield  {author} {\bibinfo {author} {\bibfnamefont {M.}~\bibnamefont
  {Malnou}}, \bibinfo {author} {\bibfnamefont {M.}~\bibnamefont {Vissers}},
  \bibinfo {author} {\bibfnamefont {J.}~\bibnamefont {Wheeler}}, \bibinfo
  {author} {\bibfnamefont {J.}~\bibnamefont {Aumentado}}, \bibinfo {author}
  {\bibfnamefont {J.}~\bibnamefont {Hubmayr}}, \bibinfo {author} {\bibfnamefont
  {J.}~\bibnamefont {Ullom}}, \ and\ \bibinfo {author} {\bibfnamefont
  {J.}~\bibnamefont {Gao}},\ }\href@noop {} {\bibfield  {journal} {\bibinfo
  {journal} {PRX Quantum}\ }\textbf {\bibinfo {volume} {2}},\ \bibinfo {pages}
  {010302} (\bibinfo {year} {2021})}\BibitemShut {NoStop}%
\bibitem [{\citenamefont {Bockstiegel}\ \emph {et~al.}(2014)\citenamefont
  {Bockstiegel}, \citenamefont {Gao}, \citenamefont {Vissers}, \citenamefont
  {Sandberg}, \citenamefont {Chaudhuri}, \citenamefont {Sanders}, \citenamefont
  {Vale}, \citenamefont {Irwin},\ and\ \citenamefont
  {Pappas}}]{Bockstiegel2014}%
  \BibitemOpen
  \bibfield  {author} {\bibinfo {author} {\bibfnamefont {C.}~\bibnamefont
  {Bockstiegel}}, \bibinfo {author} {\bibfnamefont {J.}~\bibnamefont {Gao}},
  \bibinfo {author} {\bibfnamefont {M.~R.}\ \bibnamefont {Vissers}}, \bibinfo
  {author} {\bibfnamefont {M.}~\bibnamefont {Sandberg}}, \bibinfo {author}
  {\bibfnamefont {S.}~\bibnamefont {Chaudhuri}}, \bibinfo {author}
  {\bibfnamefont {A.}~\bibnamefont {Sanders}}, \bibinfo {author} {\bibfnamefont
  {L.~R.}\ \bibnamefont {Vale}}, \bibinfo {author} {\bibfnamefont {K.~D.}\
  \bibnamefont {Irwin}}, \ and\ \bibinfo {author} {\bibfnamefont {D.~P.}\
  \bibnamefont {Pappas}},\ }\href@noop {} {\bibfield  {journal} {\bibinfo
  {journal} {Journal of Low Temperature Physics}\ }\textbf {\bibinfo {volume}
  {176}},\ \bibinfo {pages} {476} (\bibinfo {year} {2014})}\BibitemShut
  {NoStop}%
\bibitem [{\citenamefont {Macklin}\ \emph {et~al.}(2015)\citenamefont
  {Macklin}, \citenamefont {O{\textquoteright}Brien}, \citenamefont {Hover},
  \citenamefont {Schwartz}, \citenamefont {Bolkhovsky}, \citenamefont {Zhang},
  \citenamefont {Oliver},\ and\ \citenamefont {Siddiqi}}]{Macklin307}%
  \BibitemOpen
  \bibfield  {author} {\bibinfo {author} {\bibfnamefont {C.}~\bibnamefont
  {Macklin}}, \bibinfo {author} {\bibfnamefont {K.}~\bibnamefont
  {O{\textquoteright}Brien}}, \bibinfo {author} {\bibfnamefont
  {D.}~\bibnamefont {Hover}}, \bibinfo {author} {\bibfnamefont {M.~E.}\
  \bibnamefont {Schwartz}}, \bibinfo {author} {\bibfnamefont {V.}~\bibnamefont
  {Bolkhovsky}}, \bibinfo {author} {\bibfnamefont {X.}~\bibnamefont {Zhang}},
  \bibinfo {author} {\bibfnamefont {W.~D.}\ \bibnamefont {Oliver}}, \ and\
  \bibinfo {author} {\bibfnamefont {I.}~\bibnamefont {Siddiqi}},\ }\href@noop
  {} {\bibfield  {journal} {\bibinfo  {journal} {Science}\ }\textbf {\bibinfo
  {volume} {350}},\ \bibinfo {pages} {307} (\bibinfo {year}
  {2015})}\BibitemShut {NoStop}%
\bibitem [{\citenamefont {O'Brien}\ \emph {et~al.}(2014)\citenamefont
  {O'Brien}, \citenamefont {Macklin}, \citenamefont {Siddiqi},\ and\
  \citenamefont {Zhang}}]{OBrien2014}%
  \BibitemOpen
  \bibfield  {author} {\bibinfo {author} {\bibfnamefont {K.}~\bibnamefont
  {O'Brien}}, \bibinfo {author} {\bibfnamefont {C.}~\bibnamefont {Macklin}},
  \bibinfo {author} {\bibfnamefont {I.}~\bibnamefont {Siddiqi}}, \ and\
  \bibinfo {author} {\bibfnamefont {X.}~\bibnamefont {Zhang}},\ }\href@noop {}
  {\bibfield  {journal} {\bibinfo  {journal} {Phys. Rev. Lett.}\ }\textbf
  {\bibinfo {volume} {113}},\ \bibinfo {pages} {157001} (\bibinfo {year}
  {2014})}\BibitemShut {NoStop}%
\bibitem [{\citenamefont {Thol{\'{e}}n}\ \emph {et~al.}(2009)\citenamefont
  {Thol{\'{e}}n}, \citenamefont {Ergül}, \citenamefont {Stannigel},
  \citenamefont {Hutter},\ and\ \citenamefont {Haviland}}]{Thol_2009}%
  \BibitemOpen
  \bibfield  {author} {\bibinfo {author} {\bibfnamefont {E.~A.}\ \bibnamefont
  {Thol{\'{e}}n}}, \bibinfo {author} {\bibfnamefont {A.}~\bibnamefont
  {Ergül}}, \bibinfo {author} {\bibfnamefont {K.}~\bibnamefont {Stannigel}},
  \bibinfo {author} {\bibfnamefont {C.}~\bibnamefont {Hutter}}, \ and\ \bibinfo
  {author} {\bibfnamefont {D.~B.}\ \bibnamefont {Haviland}},\ }\href {\doibase
  10.1088/0031-8949/2009/t137/014019} {\bibfield  {journal} {\bibinfo
  {journal} {Physica Scripta}\ }\textbf {\bibinfo {volume} {T137}},\ \bibinfo
  {pages} {014019} (\bibinfo {year} {2009})}\BibitemShut {NoStop}%
\bibitem [{\citenamefont {Castellanos-Beltran}\ \emph
  {et~al.}(2008)\citenamefont {Castellanos-Beltran}, \citenamefont {Irwin},
  \citenamefont {Hilton}, \citenamefont {Vale},\ and\ \citenamefont
  {Lehnert}}]{Castellanos_2008}%
  \BibitemOpen
  \bibfield  {author} {\bibinfo {author} {\bibfnamefont {M.~A.}\ \bibnamefont
  {Castellanos-Beltran}}, \bibinfo {author} {\bibfnamefont {K.}~\bibnamefont
  {Irwin}}, \bibinfo {author} {\bibfnamefont {G.}~\bibnamefont {Hilton}},
  \bibinfo {author} {\bibfnamefont {L.}~\bibnamefont {Vale}}, \ and\ \bibinfo
  {author} {\bibfnamefont {K.}~\bibnamefont {Lehnert}},\ }\href@noop {}
  {\bibfield  {journal} {\bibinfo  {journal} {Nature Physics}\ }\textbf
  {\bibinfo {volume} {4}},\ \bibinfo {pages} {929} (\bibinfo {year}
  {2008})}\BibitemShut {NoStop}%
\bibitem [{\citenamefont {Zhong}\ \emph {et~al.}(2013)\citenamefont {Zhong},
  \citenamefont {Menzel}, \citenamefont {Candia}, \citenamefont {Eder},
  \citenamefont {Ihmig}, \citenamefont {Baust}, \citenamefont {Haeberlein},
  \citenamefont {Hoffmann}, \citenamefont {Inomata}, \citenamefont {Yamamoto},
  \citenamefont {Nakamura}, \citenamefont {Solano}, \citenamefont {Deppe},
  \citenamefont {Marx},\ and\ \citenamefont {Gross}}]{Zhong_2013}%
  \BibitemOpen
  \bibfield  {author} {\bibinfo {author} {\bibfnamefont {L.}~\bibnamefont
  {Zhong}}, \bibinfo {author} {\bibfnamefont {E.~P.}\ \bibnamefont {Menzel}},
  \bibinfo {author} {\bibfnamefont {R.~D.}\ \bibnamefont {Candia}}, \bibinfo
  {author} {\bibfnamefont {P.}~\bibnamefont {Eder}}, \bibinfo {author}
  {\bibfnamefont {M.}~\bibnamefont {Ihmig}}, \bibinfo {author} {\bibfnamefont
  {A.}~\bibnamefont {Baust}}, \bibinfo {author} {\bibfnamefont
  {M.}~\bibnamefont {Haeberlein}}, \bibinfo {author} {\bibfnamefont
  {E.}~\bibnamefont {Hoffmann}}, \bibinfo {author} {\bibfnamefont
  {K.}~\bibnamefont {Inomata}}, \bibinfo {author} {\bibfnamefont
  {T.}~\bibnamefont {Yamamoto}}, \bibinfo {author} {\bibfnamefont
  {Y.}~\bibnamefont {Nakamura}}, \bibinfo {author} {\bibfnamefont
  {E.}~\bibnamefont {Solano}}, \bibinfo {author} {\bibfnamefont
  {F.}~\bibnamefont {Deppe}}, \bibinfo {author} {\bibfnamefont
  {A.}~\bibnamefont {Marx}}, \ and\ \bibinfo {author} {\bibfnamefont
  {R.}~\bibnamefont {Gross}},\ }\href {\doibase 10.1088/1367-2630/15/12/125013}
  {\bibfield  {journal} {\bibinfo  {journal} {New Journal of Physics}\ }\textbf
  {\bibinfo {volume} {15}},\ \bibinfo {pages} {125013} (\bibinfo {year}
  {2013})}\BibitemShut {NoStop}%
\bibitem [{\citenamefont {{Zorin}}\ \emph {et~al.}(2017)\citenamefont
  {{Zorin}}, \citenamefont {{Khabipov}}, \citenamefont {{Dietel}},\ and\
  \citenamefont {{Dolata}}}]{Zorin2017}%
  \BibitemOpen
  \bibfield  {author} {\bibinfo {author} {\bibfnamefont {A.~B.}\ \bibnamefont
  {{Zorin}}}, \bibinfo {author} {\bibfnamefont {M.}~\bibnamefont {{Khabipov}}},
  \bibinfo {author} {\bibfnamefont {J.}~\bibnamefont {{Dietel}}}, \ and\
  \bibinfo {author} {\bibfnamefont {R.}~\bibnamefont {{Dolata}}},\ }in\ \href
  {\doibase 10.1109/ISEC.2017.8314196} {\emph {\bibinfo {booktitle} {2017 16th
  International Superconductive Electronics Conference (ISEC)}}}\ (\bibinfo
  {year} {2017})\ pp.\ \bibinfo {pages} {1--3}\BibitemShut {NoStop}%
\bibitem [{\citenamefont {Eichler}\ \emph {et~al.}(2014)\citenamefont
  {Eichler}, \citenamefont {Salathe}, \citenamefont {Mlynek}, \citenamefont
  {Schmidt},\ and\ \citenamefont {Wallraff}}]{Eicher2014}%
  \BibitemOpen
  \bibfield  {author} {\bibinfo {author} {\bibfnamefont {C.}~\bibnamefont
  {Eichler}}, \bibinfo {author} {\bibfnamefont {Y.}~\bibnamefont {Salathe}},
  \bibinfo {author} {\bibfnamefont {J.}~\bibnamefont {Mlynek}}, \bibinfo
  {author} {\bibfnamefont {S.}~\bibnamefont {Schmidt}}, \ and\ \bibinfo
  {author} {\bibfnamefont {A.}~\bibnamefont {Wallraff}},\ }\href {\doibase
  10.1103/PhysRevLett.113.110502} {\bibfield  {journal} {\bibinfo  {journal}
  {Phys. Rev. Lett.}\ }\textbf {\bibinfo {volume} {113}},\ \bibinfo {pages}
  {110502} (\bibinfo {year} {2014})}\BibitemShut {NoStop}%
\bibitem [{\citenamefont {Kannan}\ \emph {et~al.}(2020)\citenamefont {Kannan},
  \citenamefont {Campbell}, \citenamefont {Vasconcelos}, \citenamefont {Winik},
  \citenamefont {Kim}, \citenamefont {Kjaergaard}, \citenamefont {Krantz},
  \citenamefont {Melville}, \citenamefont {Niedzielski}, \citenamefont {Yoder},
  \citenamefont {Orlando}, \citenamefont {Gustavsson},\ and\ \citenamefont
  {Oliver}}]{Kannan2020}%
  \BibitemOpen
  \bibfield  {author} {\bibinfo {author} {\bibfnamefont {B.}~\bibnamefont
  {Kannan}}, \bibinfo {author} {\bibfnamefont {D.~L.}\ \bibnamefont
  {Campbell}}, \bibinfo {author} {\bibfnamefont {F.}~\bibnamefont
  {Vasconcelos}}, \bibinfo {author} {\bibfnamefont {R.}~\bibnamefont {Winik}},
  \bibinfo {author} {\bibfnamefont {D.~K.}\ \bibnamefont {Kim}}, \bibinfo
  {author} {\bibfnamefont {M.}~\bibnamefont {Kjaergaard}}, \bibinfo {author}
  {\bibfnamefont {P.}~\bibnamefont {Krantz}}, \bibinfo {author} {\bibfnamefont
  {A.}~\bibnamefont {Melville}}, \bibinfo {author} {\bibfnamefont {B.~M.}\
  \bibnamefont {Niedzielski}}, \bibinfo {author} {\bibfnamefont {J.~L.}\
  \bibnamefont {Yoder}}, \bibinfo {author} {\bibfnamefont {T.~P.}\ \bibnamefont
  {Orlando}}, \bibinfo {author} {\bibfnamefont {S.}~\bibnamefont {Gustavsson}},
  \ and\ \bibinfo {author} {\bibfnamefont {W.~D.}\ \bibnamefont {Oliver}},\
  }\href@noop {} {\bibfield  {journal} {\bibinfo  {journal} {Science Advances}\
  }\textbf {\bibinfo {volume} {6}} (\bibinfo {year} {2020})}\BibitemShut
  {NoStop}%
\bibitem [{\citenamefont {Spietz}\ \emph {et~al.}(2003)\citenamefont {Spietz},
  \citenamefont {Lehnert}, \citenamefont {Siddiqi},\ and\ \citenamefont
  {Schoelkopf}}]{Spietz2003}%
  \BibitemOpen
  \bibfield  {author} {\bibinfo {author} {\bibfnamefont {L.}~\bibnamefont
  {Spietz}}, \bibinfo {author} {\bibfnamefont {K.~W.}\ \bibnamefont {Lehnert}},
  \bibinfo {author} {\bibfnamefont {I.}~\bibnamefont {Siddiqi}}, \ and\
  \bibinfo {author} {\bibfnamefont {R.~J.}\ \bibnamefont {Schoelkopf}},\
  }\href@noop {} {\ \textbf {\bibinfo {volume} {300}},\ \bibinfo {pages} {1929}
  (\bibinfo {year} {2003})}\BibitemShut {NoStop}%
\bibitem [{\citenamefont {Mallet}\ \emph {et~al.}(2011)\citenamefont {Mallet},
  \citenamefont {Castellanos-Beltran}, \citenamefont {Ku}, \citenamefont
  {Glancy}, \citenamefont {Knill}, \citenamefont {Irwin}, \citenamefont
  {Hilton}, \citenamefont {Vale},\ and\ \citenamefont {Lehnert}}]{Mallet_2011}%
  \BibitemOpen
  \bibfield  {author} {\bibinfo {author} {\bibfnamefont {F.}~\bibnamefont
  {Mallet}}, \bibinfo {author} {\bibfnamefont {M.~A.}\ \bibnamefont
  {Castellanos-Beltran}}, \bibinfo {author} {\bibfnamefont {H.~S.}\
  \bibnamefont {Ku}}, \bibinfo {author} {\bibfnamefont {S.}~\bibnamefont
  {Glancy}}, \bibinfo {author} {\bibfnamefont {E.}~\bibnamefont {Knill}},
  \bibinfo {author} {\bibfnamefont {K.~D.}\ \bibnamefont {Irwin}}, \bibinfo
  {author} {\bibfnamefont {G.~C.}\ \bibnamefont {Hilton}}, \bibinfo {author}
  {\bibfnamefont {L.~R.}\ \bibnamefont {Vale}}, \ and\ \bibinfo {author}
  {\bibfnamefont {K.~W.}\ \bibnamefont {Lehnert}},\ }\href {\doibase
  10.1103/PhysRevLett.106.220502} {\bibfield  {journal} {\bibinfo  {journal}
  {Phys. Rev. Lett.}\ }\textbf {\bibinfo {volume} {106}},\ \bibinfo {pages}
  {220502} (\bibinfo {year} {2011})}\BibitemShut {NoStop}%
\bibitem [{\citenamefont {Movshovich}\ \emph {et~al.}(1990)\citenamefont
  {Movshovich}, \citenamefont {Yurke}, \citenamefont {Kaminsky}, \citenamefont
  {Smith}, \citenamefont {Silver}, \citenamefont {Simon},\ and\ \citenamefont
  {Schneider}}]{Movshovich1990}%
  \BibitemOpen
  \bibfield  {author} {\bibinfo {author} {\bibfnamefont {R.}~\bibnamefont
  {Movshovich}}, \bibinfo {author} {\bibfnamefont {B.}~\bibnamefont {Yurke}},
  \bibinfo {author} {\bibfnamefont {P.~G.}\ \bibnamefont {Kaminsky}}, \bibinfo
  {author} {\bibfnamefont {A.~D.}\ \bibnamefont {Smith}}, \bibinfo {author}
  {\bibfnamefont {A.~H.}\ \bibnamefont {Silver}}, \bibinfo {author}
  {\bibfnamefont {R.~W.}\ \bibnamefont {Simon}}, \ and\ \bibinfo {author}
  {\bibfnamefont {M.~V.}\ \bibnamefont {Schneider}},\ }\href {\doibase
  10.1103/PhysRevLett.65.1419} {\bibfield  {journal} {\bibinfo  {journal}
  {Phys. Rev. Lett.}\ }\textbf {\bibinfo {volume} {65}},\ \bibinfo {pages}
  {1419} (\bibinfo {year} {1990})}\BibitemShut {NoStop}%
\bibitem [{\citenamefont {Clark}\ \emph {et~al.}(2017)\citenamefont {Clark},
  \citenamefont {Lecocq}, \citenamefont {Simmonds}, \citenamefont {Aumentado},\
  and\ \citenamefont {Teufel}}]{Clark2017}%
  \BibitemOpen
  \bibfield  {author} {\bibinfo {author} {\bibfnamefont {J.~B.}\ \bibnamefont
  {Clark}}, \bibinfo {author} {\bibfnamefont {F.}~\bibnamefont {Lecocq}},
  \bibinfo {author} {\bibfnamefont {R.}~\bibnamefont {Simmonds}}, \bibinfo
  {author} {\bibfnamefont {J.}~\bibnamefont {Aumentado}}, \ and\ \bibinfo
  {author} {\bibfnamefont {J.}~\bibnamefont {Teufel}},\ }\href@noop {}
  {\bibfield  {journal} {\bibinfo  {journal} {Nature}\ }\textbf {\bibinfo
  {volume} {541}},\ \bibinfo {pages} {191} (\bibinfo {year}
  {2017})}\BibitemShut {NoStop}%
\bibitem [{\citenamefont {Sage}\ \emph {et~al.}(2011)\citenamefont {Sage},
  \citenamefont {Bolkhovsky}, \citenamefont {Oliver}, \citenamefont {Turek},\
  and\ \citenamefont {Welander}}]{Sage2011}%
  \BibitemOpen
  \bibfield  {author} {\bibinfo {author} {\bibfnamefont {J.~M.}\ \bibnamefont
  {Sage}}, \bibinfo {author} {\bibfnamefont {V.}~\bibnamefont {Bolkhovsky}},
  \bibinfo {author} {\bibfnamefont {W.~D.}\ \bibnamefont {Oliver}}, \bibinfo
  {author} {\bibfnamefont {B.}~\bibnamefont {Turek}}, \ and\ \bibinfo {author}
  {\bibfnamefont {P.~B.}\ \bibnamefont {Welander}},\ }\href@noop {} {\bibfield
  {journal} {\bibinfo  {journal} {Journal of Applied Physics}\ }\textbf
  {\bibinfo {volume} {109}},\ \bibinfo {pages} {063915} (\bibinfo {year}
  {2011})}\BibitemShut {NoStop}%
\bibitem [{\citenamefont {Houde}\ \emph {et~al.}(2019)\citenamefont {Houde},
  \citenamefont {Govia},\ and\ \citenamefont {Clerk}}]{Govia2019}%
  \BibitemOpen
  \bibfield  {author} {\bibinfo {author} {\bibfnamefont {M.}~\bibnamefont
  {Houde}}, \bibinfo {author} {\bibfnamefont {L.}~\bibnamefont {Govia}}, \ and\
  \bibinfo {author} {\bibfnamefont {A.}~\bibnamefont {Clerk}},\ }\href
  {\doibase 10.1103/PhysRevApplied.12.034054} {\bibfield  {journal} {\bibinfo
  {journal} {Phys. Rev. Applied}\ }\textbf {\bibinfo {volume} {12}},\ \bibinfo
  {pages} {034054} (\bibinfo {year} {2019})}\BibitemShut {NoStop}%
\bibitem [{\citenamefont {Peng}\ \emph {et~al.}(2022)\citenamefont {Peng},
  \citenamefont {Naghiloo}, \citenamefont {Wang}, \citenamefont {Cunningham},
  \citenamefont {Ye},\ and\ \citenamefont {O'Brien}}]{peng2022}%
  \BibitemOpen
  \bibfield  {author} {\bibinfo {author} {\bibfnamefont {K.}~\bibnamefont
  {Peng}}, \bibinfo {author} {\bibfnamefont {M.}~\bibnamefont {Naghiloo}},
  \bibinfo {author} {\bibfnamefont {J.}~\bibnamefont {Wang}}, \bibinfo {author}
  {\bibfnamefont {G.~D.}\ \bibnamefont {Cunningham}}, \bibinfo {author}
  {\bibfnamefont {Y.}~\bibnamefont {Ye}}, \ and\ \bibinfo {author}
  {\bibfnamefont {K.~P.}\ \bibnamefont {O'Brien}},\ }\href@noop {} {\bibfield
  {journal} {\bibinfo  {journal} {PRX Quantum}\ }\textbf {\bibinfo {volume}
  {3}},\ \bibinfo {pages} {020306} (\bibinfo {year} {2022})}\BibitemShut
  {NoStop}%
\bibitem [{\citenamefont {Grimsmo}\ and\ \citenamefont
  {Blais}(2017)}]{Grimsmo17}%
  \BibitemOpen
  \bibfield  {author} {\bibinfo {author} {\bibfnamefont {A.~L.}\ \bibnamefont
  {Grimsmo}}\ and\ \bibinfo {author} {\bibfnamefont {A.}~\bibnamefont
  {Blais}},\ }\href {\doibase 10.1038/s41534-017-0020-8} {\bibfield  {journal}
  {\bibinfo  {journal} {npj Quantum Information}\ }\textbf {\bibinfo {volume}
  {3}},\ \bibinfo {pages} {20} (\bibinfo {year} {2017})}\BibitemShut {NoStop}%
\bibitem [{\citenamefont {Eichler}\ \emph {et~al.}(2011)\citenamefont
  {Eichler}, \citenamefont {Bozyigit}, \citenamefont {Lang}, \citenamefont
  {Baur}, \citenamefont {Steffen}, \citenamefont {Fink}, \citenamefont
  {Filipp},\ and\ \citenamefont {Wallraff}}]{Eichler2011}%
  \BibitemOpen
  \bibfield  {author} {\bibinfo {author} {\bibfnamefont {C.}~\bibnamefont
  {Eichler}}, \bibinfo {author} {\bibfnamefont {D.}~\bibnamefont {Bozyigit}},
  \bibinfo {author} {\bibfnamefont {C.}~\bibnamefont {Lang}}, \bibinfo {author}
  {\bibfnamefont {M.}~\bibnamefont {Baur}}, \bibinfo {author} {\bibfnamefont
  {L.}~\bibnamefont {Steffen}}, \bibinfo {author} {\bibfnamefont {J.~M.}\
  \bibnamefont {Fink}}, \bibinfo {author} {\bibfnamefont {S.}~\bibnamefont
  {Filipp}}, \ and\ \bibinfo {author} {\bibfnamefont {A.}~\bibnamefont
  {Wallraff}},\ }\href {\doibase 10.1103/PhysRevLett.107.113601} {\bibfield
  {journal} {\bibinfo  {journal} {Phys. Rev. Lett.}\ }\textbf {\bibinfo
  {volume} {107}},\ \bibinfo {pages} {113601} (\bibinfo {year}
  {2011})}\BibitemShut {NoStop}%
\bibitem [{\citenamefont {Flurin}\ \emph {et~al.}(2012)\citenamefont {Flurin},
  \citenamefont {Roch}, \citenamefont {Mallet}, \citenamefont {Devoret},\ and\
  \citenamefont {Huard}}]{Flurin_2012}%
  \BibitemOpen
  \bibfield  {author} {\bibinfo {author} {\bibfnamefont {E.}~\bibnamefont
  {Flurin}}, \bibinfo {author} {\bibfnamefont {N.}~\bibnamefont {Roch}},
  \bibinfo {author} {\bibfnamefont {F.}~\bibnamefont {Mallet}}, \bibinfo
  {author} {\bibfnamefont {M.~H.}\ \bibnamefont {Devoret}}, \ and\ \bibinfo
  {author} {\bibfnamefont {B.}~\bibnamefont {Huard}},\ }\href {\doibase
  10.1103/PhysRevLett.109.183901} {\bibfield  {journal} {\bibinfo  {journal}
  {Phys. Rev. Lett.}\ }\textbf {\bibinfo {volume} {109}},\ \bibinfo {pages}
  {183901} (\bibinfo {year} {2012})}\BibitemShut {NoStop}%
\bibitem [{\citenamefont {Schneider}\ \emph {et~al.}(2020)\citenamefont
  {Schneider}, \citenamefont {Bengtsson}, \citenamefont {Svensson},
  \citenamefont {Aref}, \citenamefont {Johansson}, \citenamefont {Bylander},\
  and\ \citenamefont {Delsing}}]{Schneider2020}%
  \BibitemOpen
  \bibfield  {author} {\bibinfo {author} {\bibfnamefont {B.~H.}\ \bibnamefont
  {Schneider}}, \bibinfo {author} {\bibfnamefont {A.}~\bibnamefont
  {Bengtsson}}, \bibinfo {author} {\bibfnamefont {I.~M.}\ \bibnamefont
  {Svensson}}, \bibinfo {author} {\bibfnamefont {T.}~\bibnamefont {Aref}},
  \bibinfo {author} {\bibfnamefont {G.}~\bibnamefont {Johansson}}, \bibinfo
  {author} {\bibfnamefont {J.}~\bibnamefont {Bylander}}, \ and\ \bibinfo
  {author} {\bibfnamefont {P.}~\bibnamefont {Delsing}},\ }\href@noop {}
  {\bibfield  {journal} {\bibinfo  {journal} {Phys. Rev. Lett.}\ }\textbf
  {\bibinfo {volume} {124}},\ \bibinfo {pages} {140503} (\bibinfo {year}
  {2020})}\BibitemShut {NoStop}%
\bibitem [{\citenamefont {Heinsoo}\ \emph {et~al.}(2018)\citenamefont
  {Heinsoo}, \citenamefont {Andersen}, \citenamefont {Remm}, \citenamefont
  {Krinner}, \citenamefont {Walter}, \citenamefont {Salath\'e}, \citenamefont
  {Gasparinetti}, \citenamefont {Besse}, \citenamefont
  {Poto\ifmmode~\check{c}\else \v{c}\fi{}nik}, \citenamefont {Wallraff},\ and\
  \citenamefont {Eichler}}]{Heinsoo2018}%
  \BibitemOpen
  \bibfield  {author} {\bibinfo {author} {\bibfnamefont {J.}~\bibnamefont
  {Heinsoo}}, \bibinfo {author} {\bibfnamefont {C.~K.}\ \bibnamefont
  {Andersen}}, \bibinfo {author} {\bibfnamefont {A.}~\bibnamefont {Remm}},
  \bibinfo {author} {\bibfnamefont {S.}~\bibnamefont {Krinner}}, \bibinfo
  {author} {\bibfnamefont {T.}~\bibnamefont {Walter}}, \bibinfo {author}
  {\bibfnamefont {Y.}~\bibnamefont {Salath\'e}}, \bibinfo {author}
  {\bibfnamefont {S.}~\bibnamefont {Gasparinetti}}, \bibinfo {author}
  {\bibfnamefont {J.-C.}\ \bibnamefont {Besse}}, \bibinfo {author}
  {\bibfnamefont {A.}~\bibnamefont {Poto\ifmmode~\check{c}\else
  \v{c}\fi{}nik}}, \bibinfo {author} {\bibfnamefont {A.}~\bibnamefont
  {Wallraff}}, \ and\ \bibinfo {author} {\bibfnamefont {C.}~\bibnamefont
  {Eichler}},\ }\href@noop {} {\bibfield  {journal} {\bibinfo  {journal} {Phys.
  Rev. Applied}\ }\textbf {\bibinfo {volume} {10}},\ \bibinfo {pages} {034040}
  (\bibinfo {year} {2018})}\BibitemShut {NoStop}%
\bibitem [{\citenamefont {Backes}\ \emph {et~al.}(2021)\citenamefont {Backes},
  \citenamefont {Palken}, \citenamefont {Kenany}, \citenamefont {Brubaker},
  \citenamefont {Cahn}, \citenamefont {Droster}, \citenamefont {Hilton},
  \citenamefont {Ghosh}, \citenamefont {Jackson}, \citenamefont {Lamoreaux},
  \citenamefont {Leder}, \citenamefont {Lehnert}, \citenamefont {Lewis},
  \citenamefont {Malnou}, \citenamefont {Maruyama}, \citenamefont {Rapidis},
  \citenamefont {Simanovskaia}, \citenamefont {Singh}, \citenamefont {Speller},
  \citenamefont {Urdinaran}, \citenamefont {Vale}, \citenamefont {van
  Assendelft}, \citenamefont {van Bibber},\ and\ \citenamefont
  {Wang}}]{Backes2021}%
  \BibitemOpen
  \bibfield  {author} {\bibinfo {author} {\bibfnamefont {K.~M.}\ \bibnamefont
  {Backes}}, \bibinfo {author} {\bibfnamefont {D.~A.}\ \bibnamefont {Palken}},
  \bibinfo {author} {\bibfnamefont {S.~A.}\ \bibnamefont {Kenany}}, \bibinfo
  {author} {\bibfnamefont {B.~M.}\ \bibnamefont {Brubaker}}, \bibinfo {author}
  {\bibfnamefont {S.~B.}\ \bibnamefont {Cahn}}, \bibinfo {author}
  {\bibfnamefont {A.}~\bibnamefont {Droster}}, \bibinfo {author} {\bibfnamefont
  {G.~C.}\ \bibnamefont {Hilton}}, \bibinfo {author} {\bibfnamefont
  {S.}~\bibnamefont {Ghosh}}, \bibinfo {author} {\bibfnamefont
  {H.}~\bibnamefont {Jackson}}, \bibinfo {author} {\bibfnamefont {S.~K.}\
  \bibnamefont {Lamoreaux}}, \bibinfo {author} {\bibfnamefont {A.~F.}\
  \bibnamefont {Leder}}, \bibinfo {author} {\bibfnamefont {K.~W.}\ \bibnamefont
  {Lehnert}}, \bibinfo {author} {\bibfnamefont {S.~M.}\ \bibnamefont {Lewis}},
  \bibinfo {author} {\bibfnamefont {M.}~\bibnamefont {Malnou}}, \bibinfo
  {author} {\bibfnamefont {R.~H.}\ \bibnamefont {Maruyama}}, \bibinfo {author}
  {\bibfnamefont {N.~M.}\ \bibnamefont {Rapidis}}, \bibinfo {author}
  {\bibfnamefont {M.}~\bibnamefont {Simanovskaia}}, \bibinfo {author}
  {\bibfnamefont {S.}~\bibnamefont {Singh}}, \bibinfo {author} {\bibfnamefont
  {D.~H.}\ \bibnamefont {Speller}}, \bibinfo {author} {\bibfnamefont
  {I.}~\bibnamefont {Urdinaran}}, \bibinfo {author} {\bibfnamefont {L.~R.}\
  \bibnamefont {Vale}}, \bibinfo {author} {\bibfnamefont {E.~C.}\ \bibnamefont
  {van Assendelft}}, \bibinfo {author} {\bibfnamefont {K.}~\bibnamefont {van
  Bibber}}, \ and\ \bibinfo {author} {\bibfnamefont {H.}~\bibnamefont {Wang}},\
  }\href@noop {} {\bibfield  {journal} {\bibinfo  {journal} {Nature}\ }\textbf
  {\bibinfo {volume} {590}},\ \bibinfo {pages} {238} (\bibinfo {year}
  {2021})}\BibitemShut {NoStop}%
\bibitem [{\citenamefont {Barzanjeh}\ \emph {et~al.}(2014)\citenamefont
  {Barzanjeh}, \citenamefont {DiVincenzo},\ and\ \citenamefont
  {Terhal}}]{Barzanjeh2014}%
  \BibitemOpen
  \bibfield  {author} {\bibinfo {author} {\bibfnamefont {S.}~\bibnamefont
  {Barzanjeh}}, \bibinfo {author} {\bibfnamefont {D.~P.}\ \bibnamefont
  {DiVincenzo}}, \ and\ \bibinfo {author} {\bibfnamefont {B.~M.}\ \bibnamefont
  {Terhal}},\ }\href@noop {} {\bibfield  {journal} {\bibinfo  {journal} {Phys.
  Rev. B}\ }\textbf {\bibinfo {volume} {90}},\ \bibinfo {pages} {134515}
  (\bibinfo {year} {2014})}\BibitemShut {NoStop}%
\bibitem [{\citenamefont {Didier}\ \emph {et~al.}(2015)\citenamefont {Didier},
  \citenamefont {Kamal}, \citenamefont {Oliver}, \citenamefont {Blais},\ and\
  \citenamefont {Clerk}}]{didier}%
  \BibitemOpen
  \bibfield  {author} {\bibinfo {author} {\bibfnamefont {N.}~\bibnamefont
  {Didier}}, \bibinfo {author} {\bibfnamefont {A.}~\bibnamefont {Kamal}},
  \bibinfo {author} {\bibfnamefont {W.~D.}\ \bibnamefont {Oliver}}, \bibinfo
  {author} {\bibfnamefont {A.}~\bibnamefont {Blais}}, \ and\ \bibinfo {author}
  {\bibfnamefont {A.~A.}\ \bibnamefont {Clerk}},\ }\href@noop {} {\bibfield
  {journal} {\bibinfo  {journal} {Phys. Rev. Lett.}\ }\textbf {\bibinfo
  {volume} {115}},\ \bibinfo {pages} {093604} (\bibinfo {year}
  {2015})}\BibitemShut {NoStop}%
\bibitem [{\citenamefont {Barzanjeh}\ \emph {et~al.}(2020)\citenamefont
  {Barzanjeh}, \citenamefont {Pirandola}, \citenamefont {Vitali},\ and\
  \citenamefont {Fink}}]{Barzanjeheabb0451}%
  \BibitemOpen
  \bibfield  {author} {\bibinfo {author} {\bibfnamefont {S.}~\bibnamefont
  {Barzanjeh}}, \bibinfo {author} {\bibfnamefont {S.}~\bibnamefont
  {Pirandola}}, \bibinfo {author} {\bibfnamefont {D.}~\bibnamefont {Vitali}}, \
  and\ \bibinfo {author} {\bibfnamefont {J.~M.}\ \bibnamefont {Fink}},\
  }\href@noop {} {\bibfield  {journal} {\bibinfo  {journal} {Science Advances}\
  }\textbf {\bibinfo {volume} {6}} (\bibinfo {year} {2020})}\BibitemShut
  {NoStop}%
\bibitem [{\citenamefont {Las~Heras}\ \emph {et~al.}(2017)\citenamefont
  {Las~Heras}, \citenamefont {Di~Candia}, \citenamefont {Fedorov},
  \citenamefont {Deppe}, \citenamefont {Sanz},\ and\ \citenamefont
  {Solano}}]{LasHeras2017}%
  \BibitemOpen
  \bibfield  {author} {\bibinfo {author} {\bibfnamefont {U.}~\bibnamefont
  {Las~Heras}}, \bibinfo {author} {\bibfnamefont {R.}~\bibnamefont
  {Di~Candia}}, \bibinfo {author} {\bibfnamefont {K.~G.}\ \bibnamefont
  {Fedorov}}, \bibinfo {author} {\bibfnamefont {F.}~\bibnamefont {Deppe}},
  \bibinfo {author} {\bibfnamefont {M.}~\bibnamefont {Sanz}}, \ and\ \bibinfo
  {author} {\bibfnamefont {E.}~\bibnamefont {Solano}},\ }\href@noop {}
  {\bibfield  {journal} {\bibinfo  {journal} {Scientific Reports}\ }\textbf
  {\bibinfo {volume} {7}},\ \bibinfo {pages} {9333} (\bibinfo {year}
  {2017})}\BibitemShut {NoStop}%
\bibitem [{\citenamefont {Fedorov}\ \emph {et~al.}(2021)\citenamefont
  {Fedorov}, \citenamefont {Renger}, \citenamefont {Pogorzalek}, \citenamefont
  {Candia}, \citenamefont {Chen}, \citenamefont {Nojiri}, \citenamefont
  {Inomata}, \citenamefont {Nakamura}, \citenamefont {Partanen}, \citenamefont
  {Marx}, \citenamefont {Gross},\ and\ \citenamefont {Deppe}}]{Fedorov2021}%
  \BibitemOpen
  \bibfield  {author} {\bibinfo {author} {\bibfnamefont {K.~G.}\ \bibnamefont
  {Fedorov}}, \bibinfo {author} {\bibfnamefont {M.}~\bibnamefont {Renger}},
  \bibinfo {author} {\bibfnamefont {S.}~\bibnamefont {Pogorzalek}}, \bibinfo
  {author} {\bibfnamefont {R.~D.}\ \bibnamefont {Candia}}, \bibinfo {author}
  {\bibfnamefont {Q.}~\bibnamefont {Chen}}, \bibinfo {author} {\bibfnamefont
  {Y.}~\bibnamefont {Nojiri}}, \bibinfo {author} {\bibfnamefont
  {K.}~\bibnamefont {Inomata}}, \bibinfo {author} {\bibfnamefont
  {Y.}~\bibnamefont {Nakamura}}, \bibinfo {author} {\bibfnamefont
  {M.}~\bibnamefont {Partanen}}, \bibinfo {author} {\bibfnamefont
  {A.}~\bibnamefont {Marx}}, \bibinfo {author} {\bibfnamefont {R.}~\bibnamefont
  {Gross}}, \ and\ \bibinfo {author} {\bibfnamefont {F.}~\bibnamefont
  {Deppe}},\ }\href {\doibase 10.1126/sciadv.abk0891} {\bibfield  {journal}
  {\bibinfo  {journal} {Science Advances}\ }\textbf {\bibinfo {volume} {7}},\
  \bibinfo {pages} {eabk0891} (\bibinfo {year} {2021})}\BibitemShut {NoStop}%
\bibitem [{\citenamefont {Fedorov}\ \emph {et~al.}(2016)\citenamefont
  {Fedorov}, \citenamefont {Zhong}, \citenamefont {Pogorzalek}, \citenamefont
  {Eder}, \citenamefont {Fischer}, \citenamefont {Goetz}, \citenamefont {Xie},
  \citenamefont {Wulschner}, \citenamefont {Inomata}, \citenamefont {Yamamoto},
  \citenamefont {Nakamura}, \citenamefont {Di~Candia}, \citenamefont
  {Las~Heras}, \citenamefont {Sanz}, \citenamefont {Solano}, \citenamefont
  {Menzel}, \citenamefont {Deppe}, \citenamefont {Marx},\ and\ \citenamefont
  {Gross}}]{Fedorov2016}%
  \BibitemOpen
  \bibfield  {author} {\bibinfo {author} {\bibfnamefont {K.~G.}\ \bibnamefont
  {Fedorov}}, \bibinfo {author} {\bibfnamefont {L.}~\bibnamefont {Zhong}},
  \bibinfo {author} {\bibfnamefont {S.}~\bibnamefont {Pogorzalek}}, \bibinfo
  {author} {\bibfnamefont {P.}~\bibnamefont {Eder}}, \bibinfo {author}
  {\bibfnamefont {M.}~\bibnamefont {Fischer}}, \bibinfo {author} {\bibfnamefont
  {J.}~\bibnamefont {Goetz}}, \bibinfo {author} {\bibfnamefont
  {E.}~\bibnamefont {Xie}}, \bibinfo {author} {\bibfnamefont {F.}~\bibnamefont
  {Wulschner}}, \bibinfo {author} {\bibfnamefont {K.}~\bibnamefont {Inomata}},
  \bibinfo {author} {\bibfnamefont {T.}~\bibnamefont {Yamamoto}}, \bibinfo
  {author} {\bibfnamefont {Y.}~\bibnamefont {Nakamura}}, \bibinfo {author}
  {\bibfnamefont {R.}~\bibnamefont {Di~Candia}}, \bibinfo {author}
  {\bibfnamefont {U.}~\bibnamefont {Las~Heras}}, \bibinfo {author}
  {\bibfnamefont {M.}~\bibnamefont {Sanz}}, \bibinfo {author} {\bibfnamefont
  {E.}~\bibnamefont {Solano}}, \bibinfo {author} {\bibfnamefont {E.~P.}\
  \bibnamefont {Menzel}}, \bibinfo {author} {\bibfnamefont {F.}~\bibnamefont
  {Deppe}}, \bibinfo {author} {\bibfnamefont {A.}~\bibnamefont {Marx}}, \ and\
  \bibinfo {author} {\bibfnamefont {R.}~\bibnamefont {Gross}},\ }\href@noop {}
  {\bibfield  {journal} {\bibinfo  {journal} {Phys. Rev. Lett.}\ }\textbf
  {\bibinfo {volume} {117}},\ \bibinfo {pages} {020502} (\bibinfo {year}
  {2016})}\BibitemShut {NoStop}%
\bibitem [{\citenamefont {Yan}\ \emph {et~al.}(2018)\citenamefont {Yan},
  \citenamefont {Campbell}, \citenamefont {Krantz}, \citenamefont {Kjaergaard},
  \citenamefont {Kim}, \citenamefont {Yoder}, \citenamefont {Hover},
  \citenamefont {Sears}, \citenamefont {Kerman}, \citenamefont {Orlando},
  \citenamefont {Gustavsson},\ and\ \citenamefont {Oliver}}]{Fei2018}%
  \BibitemOpen
  \bibfield  {author} {\bibinfo {author} {\bibfnamefont {F.}~\bibnamefont
  {Yan}}, \bibinfo {author} {\bibfnamefont {D.}~\bibnamefont {Campbell}},
  \bibinfo {author} {\bibfnamefont {P.}~\bibnamefont {Krantz}}, \bibinfo
  {author} {\bibfnamefont {M.}~\bibnamefont {Kjaergaard}}, \bibinfo {author}
  {\bibfnamefont {D.}~\bibnamefont {Kim}}, \bibinfo {author} {\bibfnamefont
  {J.~L.}\ \bibnamefont {Yoder}}, \bibinfo {author} {\bibfnamefont
  {D.}~\bibnamefont {Hover}}, \bibinfo {author} {\bibfnamefont
  {A.}~\bibnamefont {Sears}}, \bibinfo {author} {\bibfnamefont {A.~J.}\
  \bibnamefont {Kerman}}, \bibinfo {author} {\bibfnamefont {T.~P.}\
  \bibnamefont {Orlando}}, \bibinfo {author} {\bibfnamefont {S.}~\bibnamefont
  {Gustavsson}}, \ and\ \bibinfo {author} {\bibfnamefont {W.~D.}\ \bibnamefont
  {Oliver}},\ }\href@noop {} {\bibfield  {journal} {\bibinfo  {journal} {Phys.
  Rev. Lett.}\ }\textbf {\bibinfo {volume} {120}},\ \bibinfo {pages} {260504}
  (\bibinfo {year} {2018})}\BibitemShut {NoStop}%
\bibitem [{\citenamefont {Jin}\ \emph {et~al.}(2015)\citenamefont {Jin},
  \citenamefont {Kamal}, \citenamefont {Sears}, \citenamefont {Gudmundsen},
  \citenamefont {Hover}, \citenamefont {Miloshi}, \citenamefont {Slattery},
  \citenamefont {Yan}, \citenamefont {Yoder}, \citenamefont {Orlando},
  \citenamefont {Gustavsson},\ and\ \citenamefont {Oliver}}]{Jin2015}%
  \BibitemOpen
  \bibfield  {author} {\bibinfo {author} {\bibfnamefont {X.~Y.}\ \bibnamefont
  {Jin}}, \bibinfo {author} {\bibfnamefont {A.}~\bibnamefont {Kamal}}, \bibinfo
  {author} {\bibfnamefont {A.~P.}\ \bibnamefont {Sears}}, \bibinfo {author}
  {\bibfnamefont {T.}~\bibnamefont {Gudmundsen}}, \bibinfo {author}
  {\bibfnamefont {D.}~\bibnamefont {Hover}}, \bibinfo {author} {\bibfnamefont
  {J.}~\bibnamefont {Miloshi}}, \bibinfo {author} {\bibfnamefont
  {R.}~\bibnamefont {Slattery}}, \bibinfo {author} {\bibfnamefont
  {F.}~\bibnamefont {Yan}}, \bibinfo {author} {\bibfnamefont {J.}~\bibnamefont
  {Yoder}}, \bibinfo {author} {\bibfnamefont {T.~P.}\ \bibnamefont {Orlando}},
  \bibinfo {author} {\bibfnamefont {S.}~\bibnamefont {Gustavsson}}, \ and\
  \bibinfo {author} {\bibfnamefont {W.~D.}\ \bibnamefont {Oliver}},\
  }\href@noop {} {\bibfield  {journal} {\bibinfo  {journal} {Phys. Rev. Lett.}\
  }\textbf {\bibinfo {volume} {114}},\ \bibinfo {pages} {240501} (\bibinfo
  {year} {2015})}\BibitemShut {NoStop}%
\bibitem [{\citenamefont {Dassonneville}\ \emph {et~al.}(2021)\citenamefont
  {Dassonneville}, \citenamefont {Assouly}, \citenamefont {Peronnin},
  \citenamefont {Clerk}, \citenamefont {Bienfait},\ and\ \citenamefont
  {Huard}}]{Dassonneville2021}%
  \BibitemOpen
  \bibfield  {author} {\bibinfo {author} {\bibfnamefont {R.}~\bibnamefont
  {Dassonneville}}, \bibinfo {author} {\bibfnamefont {R.}~\bibnamefont
  {Assouly}}, \bibinfo {author} {\bibfnamefont {T.}~\bibnamefont {Peronnin}},
  \bibinfo {author} {\bibfnamefont {A.}~\bibnamefont {Clerk}}, \bibinfo
  {author} {\bibfnamefont {A.}~\bibnamefont {Bienfait}}, \ and\ \bibinfo
  {author} {\bibfnamefont {B.}~\bibnamefont {Huard}},\ }\href {\doibase
  10.1103/PRXQuantum.2.020323} {\bibfield  {journal} {\bibinfo  {journal} {PRX
  Quantum}\ }\textbf {\bibinfo {volume} {2}},\ \bibinfo {pages} {020323}
  (\bibinfo {year} {2021})}\BibitemShut {NoStop}%
\bibitem [{\citenamefont {Mirhosseini}\ \emph {et~al.}(2019)\citenamefont
  {Mirhosseini}, \citenamefont {Kim}, \citenamefont {Zhang}, \citenamefont
  {Sipahigil}, \citenamefont {Dieterle}, \citenamefont {Keller}, \citenamefont
  {Asenjo-Garcia}, \citenamefont {Chang},\ and\ \citenamefont
  {Painter}}]{Mirhosseini2019}%
  \BibitemOpen
  \bibfield  {author} {\bibinfo {author} {\bibfnamefont {M.}~\bibnamefont
  {Mirhosseini}}, \bibinfo {author} {\bibfnamefont {E.}~\bibnamefont {Kim}},
  \bibinfo {author} {\bibfnamefont {X.}~\bibnamefont {Zhang}}, \bibinfo
  {author} {\bibfnamefont {A.}~\bibnamefont {Sipahigil}}, \bibinfo {author}
  {\bibfnamefont {P.~B.}\ \bibnamefont {Dieterle}}, \bibinfo {author}
  {\bibfnamefont {A.~J.}\ \bibnamefont {Keller}}, \bibinfo {author}
  {\bibfnamefont {A.}~\bibnamefont {Asenjo-Garcia}}, \bibinfo {author}
  {\bibfnamefont {D.~E.}\ \bibnamefont {Chang}}, \ and\ \bibinfo {author}
  {\bibfnamefont {O.}~\bibnamefont {Painter}},\ }\href {\doibase
  10.1038/s41586-019-1196-1} {\bibfield  {journal} {\bibinfo  {journal}
  {Nature}\ }\textbf {\bibinfo {volume} {569}},\ \bibinfo {pages} {692}
  (\bibinfo {year} {2019})}\BibitemShut {NoStop}%
\bibitem [{\citenamefont {Quesada}\ and\ \citenamefont
  {Sipe}(2014)}]{Quesada14}%
  \BibitemOpen
  \bibfield  {author} {\bibinfo {author} {\bibfnamefont {N.}~\bibnamefont
  {Quesada}}\ and\ \bibinfo {author} {\bibfnamefont {J.~E.}\ \bibnamefont
  {Sipe}},\ }\href@noop {} {\bibfield  {journal} {\bibinfo  {journal} {Phys.
  Rev. A}\ }\textbf {\bibinfo {volume} {90}},\ \bibinfo {pages} {063840}
  (\bibinfo {year} {2014})}\BibitemShut {NoStop}%
\bibitem [{\citenamefont {Caves}\ and\ \citenamefont
  {Crouch}(1987)}]{Caves1987}%
  \BibitemOpen
  \bibfield  {author} {\bibinfo {author} {\bibfnamefont {C.~M.}\ \bibnamefont
  {Caves}}\ and\ \bibinfo {author} {\bibfnamefont {D.~D.}\ \bibnamefont
  {Crouch}},\ }\href@noop {} {\bibfield  {journal} {\bibinfo  {journal} {JOSA
  B}\ }\textbf {\bibinfo {volume} {4}},\ \bibinfo {pages} {1535} (\bibinfo
  {year} {1987})}\BibitemShut {NoStop}%
\end{thebibliography}%
\end{document}